\documentclass[prb,twocolumn,amsmath,amssymb,10pt,aps,longbibliography,superscriptaddress,reprint]{revtex4-2}
\usepackage{graphicx} 
\usepackage{dcolumn}
\usepackage{bm}
\usepackage{graphicx,color,xcolor}
\usepackage{comment}
\usepackage{ulem}
\graphicspath{ {./Figures/} }
\usepackage{textcomp}
\usepackage[utf8]{inputenc}
\usepackage[T1]{fontenc}
\usepackage{amsmath}
\usepackage{amssymb}
\usepackage{amsthm}
\usepackage{graphicx}
\usepackage{subfigure}
\usepackage{tabularx}
\usepackage{booktabs}
\usepackage{multirow}
\usepackage[pdftex,colorlinks=true,linkcolor=blue,citecolor=blue]{hyperref} 

\usepackage{xcolor}

\usepackage{ulem}

\begin{document}

\title{\texorpdfstring{$\mathbb{Z}_N$}{Z\_N} generalizations of three-dimensional stabilizer codes}

\author{Chanbeen Lee}
\thanks{These two authors contributed equally.}
\affiliation{Department of Physics, Pohang University of Science and Technology (POSTECH), Pohang 37673, Republic of Korea}
\affiliation{Center for Artificial Low Dimensional Electronic Systems, Institute for Basic Science (IBS), Pohang 37673, Republic of Korea}

\author{Yaozong Hu}
\thanks{These two authors contributed equally.}
\affiliation{Department of Applied Physics, University of Tokyo, Tokyo 113-8656, Japan}

\author{Gil Young Cho}
\thanks{Electronic Address: gilyoungcho@kaist.ac.kr}
\affiliation{Department of Physics, Korea Advanced Institute of Science and Technology, Daejeon 34141, South Korea}
\affiliation{Center for Artificial Low Dimensional Electronic Systems, Institute for Basic Science Pohang, 37673, South Korea}
\affiliation{Asia-Pacific Center for Theoretical Physics, Pohang, Gyeongbuk 37673, South Korea}

\author{Haruki Watanabe}
\thanks{Electronic Address: hwatanabe@g.ecc.u-tokyo.ac.jp}
\affiliation{Department of Applied Physics, University of Tokyo, Tokyo 113-8656, Japan}
 
\date{\today}

\begin{abstract} 
In this work, we generalize several three-dimensional $\mathbb{Z}_2$ stabilizer models—including the X-cube model, the three-dimensional toric code, and Haah’s code—to their $\mathbb{Z}_N$ counterparts. Under periodic boundary conditions, we analyze their ground state degeneracies and topological excitations and uncover behaviors that strongly depend on system size. For the X-cube model, we identify excitations with mobility restricted under local operations but relaxed under nonlocal ones derived from global topology. These excitations, previously confined to open boundaries in the $\mathbb{Z}_2$ model, now appear even under periodic boundaries. In the toric code, we observe nontrivial braiding between string and point excitations despite the absence of ground state degeneracy, indicating long-range entanglement independent of topological degeneracy. Again, this effect extends from open to periodic boundaries in the generalized models. For Haah’s code, we find new excitations—fracton tripoles and monopoles—that remain globally constrained, along with a relaxation of immobility giving rise to lineons and planons. These results reveal new forms of topological order and suggest a broader framework for understanding fracton phases beyond the conventional $\mathbb{Z}_2$ setting.
\end{abstract}
\maketitle

\section{Introduction} 

Topological order is a major theme in theoretical condensed matter physics~\cite{wen1990topological,wen2019choreographed}. Microscopically, long-range entanglement~\cite{kitaev2006topological,levin2006detecting,chen2010local} gives rise to nontrivial topological features such as ground state degeneracy (GSD) that is independent of spontaneous symmetry breaking, and anyonic braiding—phenomena essential for robust quantum memory and fault-tolerant quantum computation even in the presence of decoherence~\cite{RevModPhys.80.1083}. Translation-invariant stabilizer models provide an efficient theoretical framework for studying topological order, since their Hamiltonians are composed of mutually commuted local stabilizer terms. These models are exactly solvable and faithfully capture the key features of topologically ordered phases.

The prototypical example is the two-dimensional toric code introduced by Kitaev~\cite{KITAEV20032}, which employs $\mathbb{Z}_2$ spins and exhibits topological ground state degeneracy and nontrivial braiding between particle excitations. In $3d$ case, braiding between point-like excitations becomes trivial; however, new types of braiding emerge involving particles and loop-like excitations~\cite{PhysRevB.72.035307,Kong2020}. Furthermore, some $3d$ models realize fracton phases, where certain excitations become immobile due to the absence of local operators that can create or move them~\cite{chamon2005quantum,bravyi2011topological,PhysRevA.83.042330,yoshida2013exotic,vijay2015new,vijay2016fracton,pretko2017,lake2021subdimensional}.

By extending the toric code in various directions, richer physical phenomena can be uncovered. In the presence of global symmetries, such systems may enter symmetry-enriched topological (SET) phases, which feature more intricate structures than conventional topological orders~\cite{wen2002quantum,kou2009translation,essin2013classifying,mesaros2013classification,barkeshli2013theory,teo2015theory,stephen2020subsystem, PhysRevB.100.155146,lam2024classification,williamson2019fractonic}. For example, the rank-2 toric code, constructed from rank-2 gauge fields, exhibits unique patterns of symmetry fractionalization~\cite{Pace_2022,oh2022effective,oh2022rank,kim2025unveilinguvirmixingsymmetry}. Likewise, the introduction of modulated symmetries has led to a variety of novel physical effects~\cite{you2024intrinsicmixedstatesptmodulated,delfino20232d,PhysRevB.109.125121,pace2024gauging}.

Recently, a generalized family of two-dimensional $\mathbb{Z}_N$ toric codes with modulated exponential symmetries has been proposed~\cite{watanabe2023ground}. Unlike the original $\mathbb{Z}_2$ model - whose GSD depends solely on spatial topology (e.g., genus) - these generalized models exhibit GSDs that depend on the size of the system, the spin level $N$, and additional Hamiltonian parameters. Moreover, under certain parameter regimes, local string operators can create single excitations, in contrast to the $\mathbb{Z}_2$ toric code where excitations must be created in pairs. These generalizations have been extensively studied in one- and two-dimensional systems~\cite{delfino20232d,delfino2023effective,PhysRevB.109.125121,PhysRevB.107.195139}, raising the natural question: what novel phenomena may arise in three dimensions under similar generalizations?

In this work, we introduce $\mathbb{Z}_N$ generalizations of several three-dimensional stabilizer models, including the X-cube model, the $3d$ toric code, and Haah's code, and investigate their ground state degeneracies and elementary excitations. Each generalized model preserves key features of its $\mathbb{Z}_2$ counterpart while also exhibiting entirely new and exotic behaviors.

Among our results, the most prominent arise in the $\mathbb{Z}_N$ X-cube model, which exhibits either a fracton phase or a conventional topologically ordered phase depending on parameter choices. Traditionally, a defining feature of the fracton phase is the presence of excitations with restricted mobility. However, in our model, the nature of these excitations oscillates continuously with changes in system size. Remarkably, even in the thermodynamic limit, there exist specific large system sizes in which all excitations become fully mobile, even under periodic boundary conditions (PBCs).

To better characterize this behavior, we classify a new type of 'mobile' excitation: one that is strictly immobile under all local operations but becomes mobile under certain non-local operations, even when PBCs are imposed. These excitations exist under open boundary conditions(OBCs) in the original $\mathbb{Z}_2$ model, but are absent under PBCs. In contrast, our generalized models allow them to persist under PBCs. Although these excitations are not strictly immobile, they functionally behave as fractons and maintain the system in the fracton phase. This suggests that the conventional understanding of fractons and fracton phases may require significant refinement.

The remainder of this paper is structured as follows. In Sec.II, we review mathematical preliminaries and discuss the generalized $2d$ toric code as a pedagogical example. Sections III, IV, and V present our main results for the generalized $3d$ models. We conclude in Sec.~VI with a summary and outlook.

\begin{widetext}

\begin{table*}
\centering
\caption{X-cube Model}
\label{table:x-cube}
\begin{tabular}{|l|l|l|l|l|}
\hline
Phase & Parameter condition & Cube Excitation & Vertex excitation & GSD \\ 
\hline
Trivial Phase & $N_{xyz}=1$ & 
Planon or Boson & 
Planon or Boson & 
${d^{2L_x}_{yz}}{d^{2L_y}_{zx}}{d^{2L_z}_{xy}}$ \\ 
\hline
\multirow{2}{*}{Fracton Phase} 
& $N_{xyz}\neq 1, d_{xyz}\neq 1$ & 
\begin{tabular}[c]{@{}l@{}}Fracton, Quasi-Fracton, \\Planon and Boson\end{tabular} & 
\begin{tabular}[c]{@{}l@{}}Lineon, Quasi-Lineon, \\Planon and Boson\end{tabular} & 
$\frac{{d^{2L_x}_{yz}}{d^{2L_y}_{zx}}{d^{2L_z}_{xy}}}{d_{xyz}^3}$ \\
\cline{2-5}
& $N_{xyz}\neq 1, d_{xyz}=1$ & 
\begin{tabular}[c]{@{}l@{}}Quasi-Fracton, Planon\\and Boson\end{tabular} & 
\begin{tabular}[c]{@{}l@{}}Quasi-Lineon, Planon\\and Boson\end{tabular} & 
${d^{2L_x}_{yz}}{d^{2L_y}_{zx}}{d^{2L_z}_{xy}}$ \\  
\hline
\end{tabular}
\\
\caption{$3d$-toric Code}
\label{table:3d toric code}
\begin{tabular}{|l|l|l|l|l|}
\hline
Phase & Parameter condition & Plaquette excitation & Vertex excitation & GSD \\ 
\hline
Trivial Phase & $N_{xyz}=1$ & 
Open string & 
Boson & 
$1$ \\ 
\hline
\multirow{2}{*}{Topological Ordered Phase} 
& $N_{xyz}\neq 1, d_{xyz}\neq 1$ & 
\begin{tabular}[c]{@{}l@{}}Open String, \\Closed string\end{tabular} & 

\begin{tabular}[c]{@{}l@{}}Pair of Topological Point Excitaion, \\Single Topological Point Excitation, \\and Boson\end{tabular} & 
$d_{xyz}^3$ \\
\cline{2-5}
& $N_{xyz}\neq 1, d_{xyz}=1$ & 
\begin{tabular}[c]{@{}l@{}}Open String, \\Closed string\end{tabular} & 

\begin{tabular}[c]{@{}l@{}}Single Topological Point Excitation, \\and Boson\end{tabular} & 
$1$ \\  
\hline
\end{tabular}
\\\caption{Haah's code}
\label{table:Haah's code}
\begin{tabular}{|l|c|l|c|}
\hline
Phase         & \multicolumn{1}{l|}{Parameter condition}                                                                                                                & Cube excitation                                                                                                                                            & \multicolumn{1}{l|}{GSD}                                                                                                                                  \\ \hline
Trivial phase & \begin{tabular}[c]{@{}c@{}}Two or more in $\{a_x,a_y,a_z \}$ are zero\\ or\\ Two or more in $\{a_{xy}',a_{yz}',a_{zx}'\}$ are zero \end{tabular}         & \begin{tabular}[c]{@{}l@{}}Free particle\\  Lineon or Planon\end{tabular}                                                                                 & \multirow{2}{*}{\begin{tabular}[c]{@{}c@{}} \\ \\ Fluctuate under $a_x,a_y,a_z,a_{xy}',a_{yz}',a_{zx}'$,\\ \\ and system sizes $L_x$, $L_y$, $L_z$\end{tabular}} \\ \cline{1-3}
Fracton phase & \begin{tabular}[c]{@{}c@{}}Two or more in $\{a_x,a_y,a_z\}$ are nonzero \\ and\\ Two or more in $\{a_{xy}',a_{yz}',a_{zx}' \}$ are nonzero\end{tabular} & \begin{tabular}[c]{@{}l@{}}Fracton quadruopole,\\ Fracton tripole,\\ Fracton dipole,\\ Quasi-Fracton (monopole),\\ Fracton surface,\\ Fracton line\end{tabular} &                                                                                                                                                           \\ \hline
\end{tabular}

\caption{\label{table} Summary of the results on the $\mathbb{Z}_N$ generalization of the three $3d$ models. For each model, parameters, excitations of different local stabilizer terms, and ground state degeneracies are listed. We define $N_{xyz}$ as the largest factor of $N$ that is coprime with $a_xa_ya_z$, and $d_{xyz}=\text{gcd}(a_{x}^{L_x}-1,a_{y}^{L_y}-1,a_{z}^{L_z}-1,N)$, $d_{ij}=\text{gcd}(a_{i}^{L_i}-1,a_{j}^{L_j}-1,N)$ for $i,j=x,y,z$. For Haah's code, we consider only prime integer $N$ for the technical simplicity, and the phases were classified based on whether each parameter is 0 modulo $N$.
}
\end{table*}
\end{widetext}

\section{Preliminaries} 
In this section, we introduce some basic mathematical background on generalized Pauli matrices and number theory, which will help readers better understand the present work. We also review previous results on the $\mathbb{Z}_N$ generalization of the $2d$ toric code~\cite{watanabe2023ground}, as a warm-up for the $3d$ models discussed later.

\subsection{Mathematical Background} 
\subsubsection{\texorpdfstring{$\mathbb{Z}_N$}{Z\_N} Pauli matrices}
We use  generalized Pauli matrices defined as
\begin{equation}
    X=\begin{pmatrix}
 &  &  &1  \\
 1&  &  &  \\
 & \ddots &  &  \\
 &  &  1&  \\
\end{pmatrix}, \quad
 Z=\begin{pmatrix}
 1&  &  &  \\
 & \omega &  &  \\
 &  & \ddots &  \\
 &  &  & \omega^{N-1} \\
\end{pmatrix},
\end{equation}
which act on an $N$-dimensional Hilbert space at each site. Here and throughout this paper, $N$ denotes a positive integer, and $\omega = e^{2\pi i/N}$ is a primitive $N$-th root of unity. In this work, we focus on the case $N > 2$. These operators satisfy the commutation relation $ZX = \omega XZ$. Note also that $Z^N = X^N = 1$, where $1$ denotes the identity operator on the $N$-dimensional Hilbert space. All the mutually commuting local stabilizer terms in the Hamiltonians introduced below are constructed from $X$ and $Z$.

\subsubsection{Some Number Theory} \label{subsec : number theory}

Here we summarize definitions and facts from number theory that will be used in this work. First, for a given positive integer $N$ and an integer $a$, the multiplicative order of $a$ modulo $N$ is defined as the smallest positive integer $l$ such that
\begin{equation}
    a^l \equiv 1 \pmod{N}.
\end{equation}
We denote this quantity by $M_N(a)$. 

Euler's totient function $\varphi(N)$ counts the number of positive integers less than $N$ that are coprime to $N$. Euler’s theorem states that $a^{\varphi(N)} \equiv 1 \pmod{N}$ for any $a$ coprime to $N$~\cite{stein2008elementary}, implying that $M_N(a)$ divides $\varphi(N)$. When $N$ is prime, it follows from the definition that $\varphi(N) = N - 1$.

Finally, the radical of $N$, denoted $\mathrm{rad}(N)$, is defined as the product of all distinct prime factors of $N$. Namely, if $N$ is factorized as
\begin{equation}
    N = \prod_{j=1}^{n} p_j^{r_j},
\end{equation}
then
\begin{equation}
    \mathrm{rad}(N) \equiv \prod_{j=1}^{n} p_j.
\end{equation}
Let $r$ be the largest exponent among $\{r_1, r_2, \dots, r_n\}$. Then, for any integer $x$ divisible by $\mathrm{rad}(N)$, we have
\begin{equation}
    x^r \equiv 0 \pmod{N}.
\end{equation}

\subsubsection{Definitions of Useful Functions} \label{App_B_function}

We define a class of arithmetic functions based on the exponents $(a_x, a_y, a_z)$ and the spin level $N$. For each $i \in \{x, y, z\}$, we define
\begin{align}
    N_i \equiv \inf_{n \in \mathbb{N}} \left( \frac{N}{\gcd(a_i^n, N)} \right),
\end{align}
where $\inf\{\cdots\}$ denotes the infimum of the set $\{\cdots\}$. In other words, $N_i$ is the largest divisor of $N$ that is coprime to $a_i$.

Note that for any sufficiently large $\ell_i$, we have $N_i a_i^{\ell_i} \equiv 0 \pmod{N}$. In particular, if $a_i \equiv 0 \pmod{\mathrm{rad}(N)}$, then $N_i = 1$.

We also define the analogous joint functions:
\begin{align}
    N_{ij} &\equiv \inf_{n \in \mathbb{N}} \left( \frac{N}{\gcd(a_i^n a_j^n, N)} \right), \\
    N_{ijk} &\equiv \inf_{n \in \mathbb{N}} \left( \frac{N}{\gcd(a_i^n a_j^n a_k^n, N)} \right),
\end{align}
where again $i,j,k \in \{x,y,z\}$. Here, $N_{ij}$ is the largest divisor of $N$ that is coprime to both $a_i$ and $a_j$, and $N_{ijk}$ is the largest divisor of $N$ that is coprime to all three: $a_i$, $a_j$, and $a_k$.

Next, we define the following quantities that also depend on the system sizes $L_i, L_j, L_k$:
\begin{align}
    d_i &\equiv \gcd(a_i^{L_i} - 1, N), \\
    d_{ij} &\equiv \gcd(a_i^{L_i} - 1, a_j^{L_j} - 1, N), \\
    d_{ijk} &\equiv \gcd(a_i^{L_i} - 1, a_j^{L_j} - 1, a_k^{L_k} - 1, N).
\end{align}
These functions will be useful when discussing global operators that act across the entire system.

Since $a_i^{L_i} - 1$ is obviously coprime to $a_i^{L_i}$, it is also coprime to $a_i$ itself. Therefore, $d_i = \gcd(a_i^{L_i} - 1, N)$ is not only a divisor of $N$ but also coprime to $a_i$. From the definition of $N_i$, it follows that $d_i$ must divide $N_i$. Similarly, one can show that $d_{ij}$ divides $N_{ij}$, and $d_{ijk}$ divides $N_{ijk}$.

\subsubsection{B\'ezout's Identity} \label{Bezout_identity}

B\'ezout's identity~\cite{bezout1779theorie} states that for any pair of integers $a$ and $b$, there always exist integers $x$ and $y$ such that
\begin{equation}
    ax + by = \gcd(a, b).
\end{equation}
In the special case where $a$ and $b$ are coprime, the identity reduces to $ax + by = 1$, implying that $x$ is a modular multiplicative inverse of $a$ modulo $b$, i.e., $ax \equiv 1 \pmod{b}$.

B\'ezout's identity can be extended to more than two integers. Given a set of integers $(a_1, a_2, \ldots, a_n)$, there always exists a corresponding set $(x_1, x_2, \ldots, x_n)$ such that
\begin{equation}
    a_1 x_1 + a_2 x_2 + \cdots + a_n x_n = \gcd(a_1, a_2, \ldots, a_n).
\end{equation}
For example, consider the case $n = 3$:
\begin{align}
    a_1 x_1 + a_2 x_2 + a_3 x_3 &= \gcd(a_1, a_2) X + a_3 x_3 \nonumber \\
    &= \gcd(\gcd(a_1, a_2), a_3) = \gcd(a_1, a_2, a_3), \nonumber
\end{align}
where we used B\'ezout's identity for two integers in the first step and the associativity of the $\gcd$ function in the second.

This identity can be generalized to arbitrary $n$ via mathematical induction. The extended version will be useful in identifying the unit charge of certain excitations in later sections. (The definition of excitation charge appears in Sec.~\ref{Appendix : gTC_particle}.) Since we are interested in $\mathbb{Z}_N$ models, the modulus $N$ should be included in the integer set $(a_1, a_2, \ldots, a_n)$ when applying the identity.

\subsection{\texorpdfstring{$2\mathrm{D}$}{$2d$} \texorpdfstring{$\mathbb{Z}_N$}{ZN} Stabilizer Models} \label{sec : 2dgTC}

\begin{figure*}
\includegraphics[width=2\columnwidth]{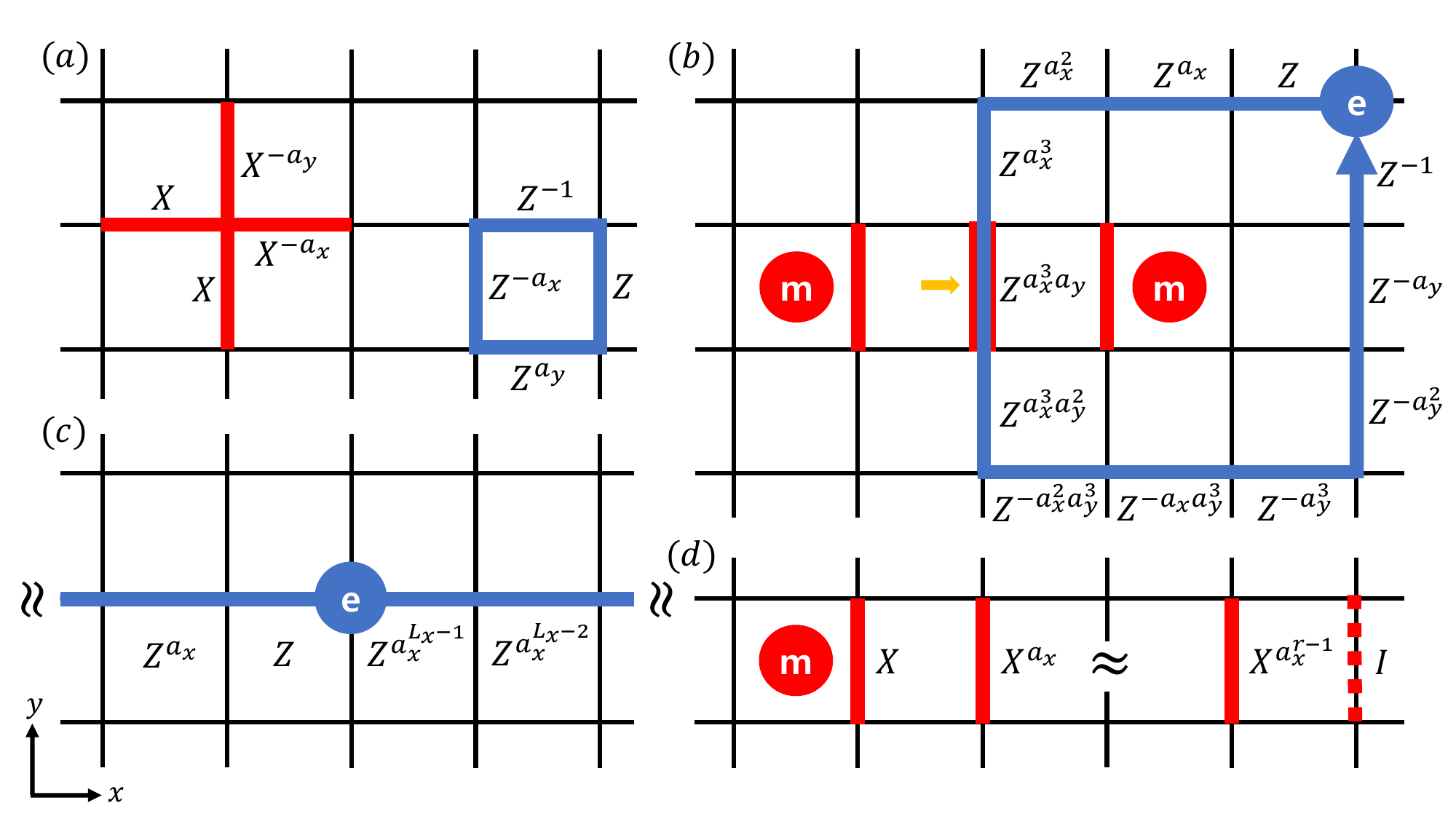}
\caption{\label{fig_2d_gTC_new} 
\textbf{$\mathbb{Z}_N$ generalization of the $2d$ toric code.} 
Each link hosts an $N$-dimensional spin. 
(a) Vertex and plaquette stabilizer terms. 
(b) A Wilson loop operator for a $\mathbb{Z}_N$ electric charge (blue circle labeled “e”) encircling a magnetic excitation (red circle labeled “m”). The Wilson loop consists of Pauli $Z$ operators with non-uniform exponents $1, a_x, a_x^2, \dots$. The yellow arrow marks the intersection of the electric and magnetic Wilson lines, where their braiding phase is generated. 
(c) A single electric charge excitation can be created by an electric Wilson loop operator winding around the system in the $\hat{x}$-direction, when $a_x^{L_x} \not\equiv 1 \pmod{N}$. 
(d) A single magnetic excitation can be created by a magnetic Wilson line operator in the $\hat{x}$-direction when $a_x^r \equiv 0 \pmod{N}$.}
\end{figure*}

Here we summarize the key results of our previous work~\cite{watanabe2023ground}. In Ref.\cite{watanabe2023ground}, a $\mathbb{Z}_N$ generalization of the $2d$ toric code was introduced and analyzed. The Hamiltonian retains the standard vertex and plaquette stabilizers as in the original $\mathbb{Z}_2$ model, but with two additional tunable integer parameters, $a_x$ and $a_y$, satisfying $1 \leq a_x, a_y \leq N$. See Fig.\ref{fig_2d_gTC_new}(a) for a pictorial representation. Despite the modifications, all stabilizer terms commute, preserving exact solvability.

Remarkably, under periodic boundary conditions on a two-torus $T^2$, the number of independent stabilizer relations, and hence the ground state degeneracy depends nontrivially on the system size and the parameters $a_x$, $a_y$. For instance, when the on-site Hilbert space dimension $N$ is prime, the GSD equals $N^2$ if and only if $L_x \equiv 0 \pmod{M_N(a_x)}$ and $L_y \equiv 0 \pmod{M_N(a_y)}$, where $L_i$ denotes the system size in direction $i$. Otherwise, the ground state is unique. As we will see, similar behavior also emerges in the $3d$ generalizations.

In addition to affecting the GSD, the parameters $a_x$ and $a_y$ also influence the nature of elementary excitations. In the conventional $\mathbb{Z}_2$ toric code, electric and magnetic excitations always occur in pairs, and their mutual braiding yields a topological phase of $(-1)$, determined solely by the linking number of their spacetime worldlines [Fig.~\ref{fig_2d_gTC_new}(b)].

Both of these features can be significantly modified in the $\mathbb{Z}_N$ generalizations. For example, if $L_x$ is not divisible by $M_N(a_x)$, then a single $\mathbb{Z}_N$ electric (magnetic) excitation can be created by a nontrivial $Z$ ($X$) Wilson loop wrapping around the $\hat{x}$ direction [Fig.\ref{fig_2d_gTC_new}(c)]. Similarly, if $a_x$ or $a_y$ is divisible by $\mathrm{rad}(N)$, then a single magnetic (electric) excitation can be created by a local $X$ ($Z$) string operator [Fig.\ref{fig_2d_gTC_new}(d)]. In such cases, the mutual braiding phase between two excitations may become trivial.

In the remainder of this paper, we explore how these $\mathbb{Z}_N$ generalizations influence the excitation structure in $3d$, where richer possibilities arise, including point particles, extended string excitations, and fractons. 

\section{X-cube Model} \label{sec : X_cube_model}

In this section, we briefly review the $\mathbb{Z}_2$ X-cube model defined on a three-torus $T^3$ (i.e., under periodic boundary conditions), and then investigate its $\mathbb{Z}_{N>2}$ generalizations and properties in detail.

\subsection{\texorpdfstring{$\mathbb{Z}_2$}{Z2} Model} \label{sec : Z2_XC}

\begin{figure*}
\includegraphics[width=2\columnwidth]{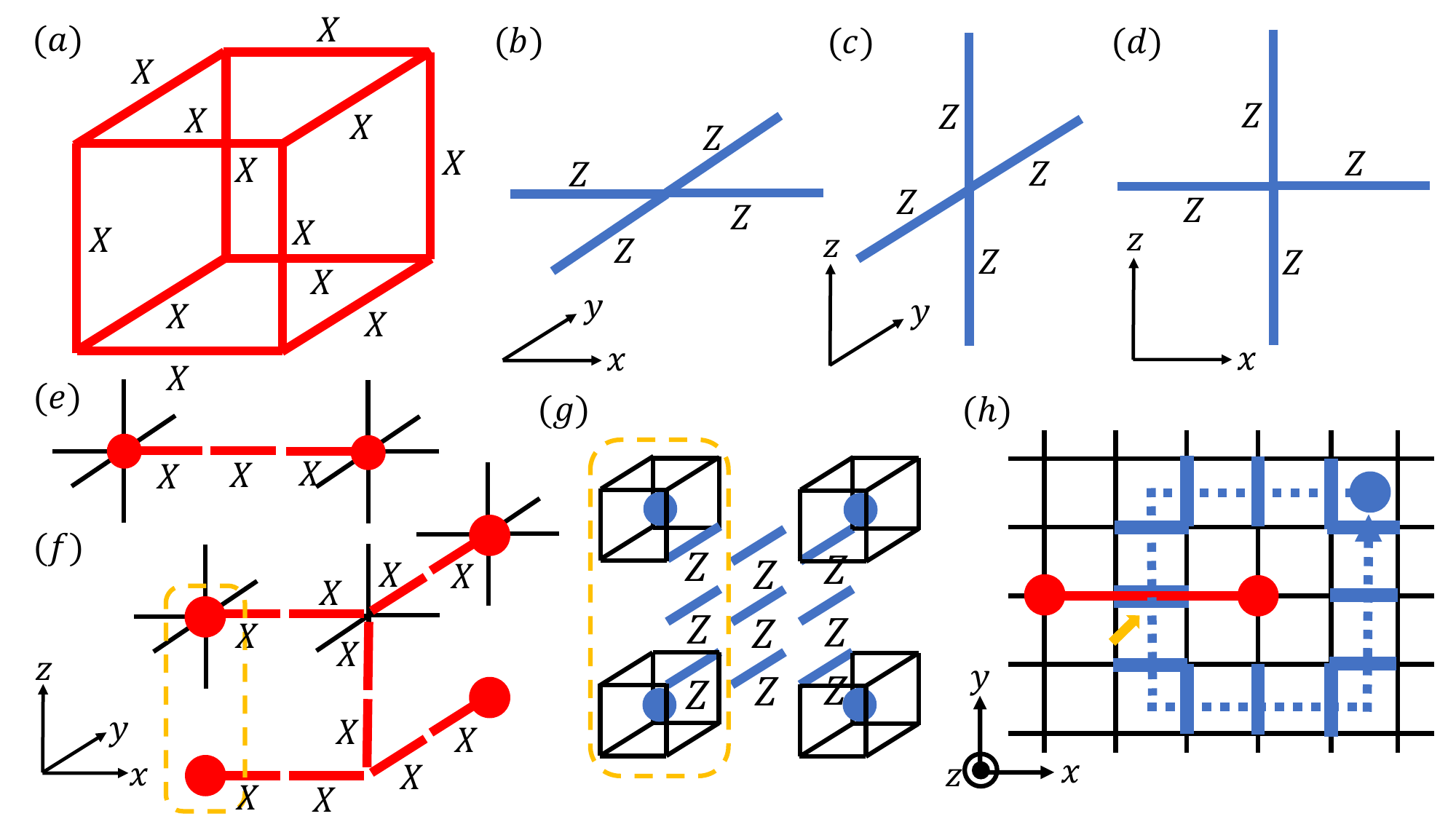}
\caption{\label{fig:Z2_XC_new} 
\textbf{$\mathbb{Z}_2$ X-cube model.} A two-dimensional spin lives on each link.
(a) Cube stabilizer $S_c$. 
(b–d) Vertex stabilizers $V_v^{xy}$, $V_v^{yz}$, and $V_v^{zx}$, defined in the corresponding planes. 
(e) A pair of $\hat{x}$-directional lineons is created by a straight line operator $W_{\mathcal{C}}$, which flips $V_v^{zx} = 1$ and $V_v^{xy} = 1$ at the endpoints of the string $\mathcal{C}$. 
(f) A lineon dipole along the $\hat{z}$ direction becomes a planon that moves freely in the $xy$-plane. 
(g) A fracton quadrupole created by a membrane operator $O_{\mathcal{M}}$ in the $zx$-plane. A fracton dipole along $\hat{z}$ becomes a planon in the $xy$-plane when acted upon by membrane operators in $yz$ and $zx$ planes. 
(h) Top view of braiding a $\hat{z}$-oriented fracton dipole around a $\hat{x}$-oriented lineon. The braiding phase $(-1)$ arises from the intersection of a $Z$ membrane along the dipole's path and the line operator $W_{\mathcal{C}}$, as indicated by the arrow.}
\end{figure*}

The X-cube model~\cite{vijay2016fracton} is defined on a cubic lattice, with two-dimensional spins placed on the links. We consider the system on a three-torus $T^3$ of size $L_x \times L_y \times L_z$. The Hamiltonian consists of mutually commuting stabilizers constructed from Pauli $X$ and $Z$ operators:
\begin{align}\label{eqn : H_Z2_XC}
H = - \sum_{c \in \mathcal{C}} S_c  - \sum_{v \in \mathcal{V}} \left(V_{v}^{xy} + V_{v}^{yz} + V_{v}^{zx} \right),
\end{align}
where $\mathcal{C}$ denotes the set of unit cubes and $\mathcal{V}$ the set of lattice vertices. Figures~\ref{fig:Z2_XC_new}(a–d) show the corresponding stabilizers.

Each cube term $S_c$ is defined on the dual site $c = \vec{r} + (1/2, 1/2, 1/2)$ with $\vec{r} \in \mathbb{Z}^3$:
\begin{align}
S_c = \prod_{\text{edges of cube } c} X_{\text{link}}, \nonumber
\end{align}
which involves twelve $X$ operators surrounding the cube. At each vertex $v = \vec{r}$, we define three types of vertex terms. For example:
\begin{align}
V_v^{xy} = Z_{v - \frac{1}{2} \hat{x}} Z_{v + \frac{1}{2} \hat{x}} Z_{v - \frac{1}{2} \hat{y}} Z_{v + \frac{1}{2} \hat{y}}. \nonumber
\end{align}
The other two, $V_v^{yz}$ and $V_v^{zx}$, are defined analogously. The ground states satisfy $S_c = V_v^{xy} = V_v^{yz} = V_v^{zx} = 1$.

Excitations are created by flipping the eigenvalue of one or more stabilizers. Those associated with $V_v^{ab}$ (with $ab \in \{xy, yz, zx\}$) are called lineons, which are restricted to move along one-dimensional lines. A straight $X$-string operator creates a pair of lineons [Fig.~\ref{fig:Z2_XC_new}(e)], with one lineon at each endpoint. Attempting to bend the string creates additional excitations, which restricts mobility of the lineons. When two lineons form a dipole (e.g., separated along $\hat{z}$), they can move freely within the plane orthogonal to the dipole moment, becoming planons [Fig.~\ref{fig:Z2_XC_new}(f)].

Excitations associated with $S_c$ include fractons and fracton dipoles. These are created at the corners of membrane operators composed of $Z$ operators [Fig.~\ref{fig:Z2_XC_new}(g)]. A single fracton appears at a corner of the membrane where $S_c$ flips sign. Isolated fractons cannot move, but dipoles (i.e., pairs of fractons) become planons constrained to two-dimensional planes. Despite their restricted mobility, nontrivial braiding statistics exist~\cite{pai2019fracton}. For instance, a fracton dipole moving around a lineon results in a $(-1)$ braiding phase, as illustrated in Fig.~\ref{fig:Z2_XC_new}(h).

Finally, the GSD on the three-torus $T^3$ is given by:
\begin{align}\label{eqn : Z2_XC_GSD}
\log_2 \mathrm{GSD} = 2(L_x + L_y + L_z) - 3.
\end{align}

\subsection{\texorpdfstring{$\mathbb{Z}_{N>2}$}{ZN} Generalization} \label{sec : gXC}

We now generalize the $\mathbb{Z}_2$ X-cube model [Eq.~\eqref{eqn : H_Z2_XC}] to its $\mathbb{Z}_N$ counterpart. As before, we consider the system defined on the three-torus $T^3$ with $L_x \times L_y \times L_z$ unit cubes. The Hamiltonian retains the same structure as in the $\mathbb{Z}_2$ case:
\begin{align}\label{eqn : H_gXC}
H = - \frac{1}{2} \sum_{c \in \mathcal{C}} S_c  - \frac{1}{2} \sum_{v \in \mathcal{V}} \left(V_{v}^{xy} + V_{v}^{yz} + V_{v}^{zx} \right) + \text{h.c.}
\end{align}

\begin{figure}
\includegraphics[width=\columnwidth]{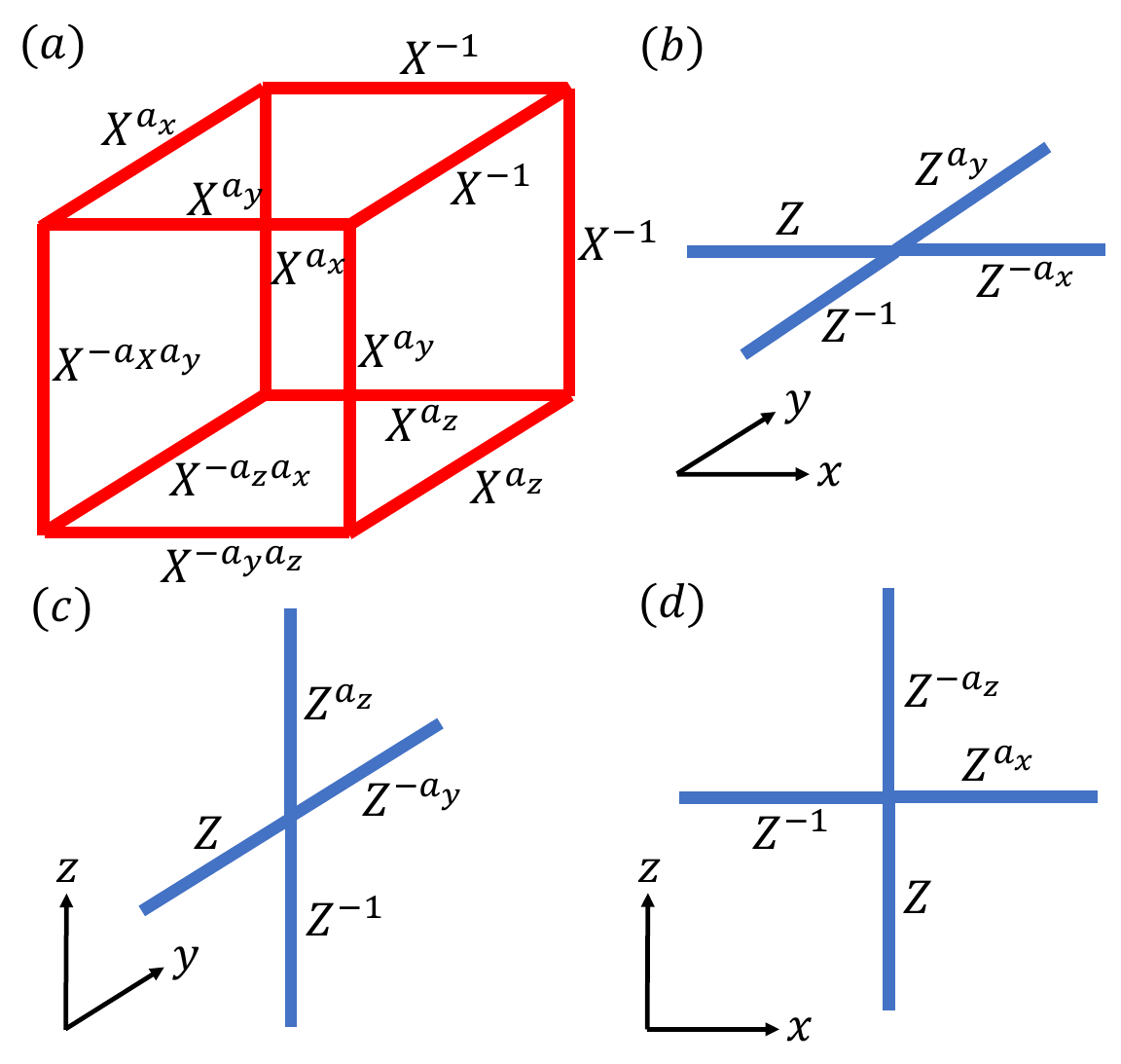}
\caption{\label{fig: gXC_stabilizer} 
\textbf{Pictorial representations of the $\mathbb{Z}_N$ generalization of the X-cube model [Eq.~(\ref{eqn : H_gXC})].} 
(a) Cube term $S_c$. 
(b–d) Vertex terms $V_v^{xy}$, $V_v^{yz}$, and $V_v^{zx}$, defined in their respective planes.}
\end{figure}

Here, the cube term $S_c$ [Fig.~\ref{fig: gXC_stabilizer}(a)] is defined at the dual lattice site $c = \vec{r} + (1/2,1/2,1/2)$ with $\vec{r} \in \mathbb{Z}^3$ and consists of $X$ operators with non-uniform integer exponents determined by $(a_x, a_y, a_z)$:
\begin{align}
S_c = \, & X_{c - \frac{1}{2} \hat{y} - \frac{1}{2} \hat{z}}^{-a_y a_z}
X_{c - \frac{1}{2} \hat{z} - \frac{1}{2} \hat{x}}^{-a_x a_z}
X_{c + \frac{1}{2} \hat{y} - \frac{1}{2} \hat{z}}^{a_z}
X_{c - \frac{1}{2} \hat{z} + \frac{1}{2} \hat{x}}^{a_z} \nonumber \\
& \times
X_{c - \frac{1}{2} \hat{x} - \frac{1}{2} \hat{y}}^{-a_x a_y}
X_{c + \frac{1}{2} \hat{x} - \frac{1}{2} \hat{y}}^{a_y}
X_{c - \frac{1}{2} \hat{x} + \frac{1}{2} \hat{y}}^{a_x}
X_{c + \frac{1}{2} \hat{x} + \frac{1}{2} \hat{y}}^{-1} \nonumber \\
& \times
X_{c - \frac{1}{2} \hat{y} + \frac{1}{2} \hat{z}}^{a_y}
X_{c + \frac{1}{2} \hat{z} - \frac{1}{2} \hat{x}}^{a_x}
X_{c + \frac{1}{2} \hat{y} + \frac{1}{2} \hat{z}}^{-1}
X_{c + \frac{1}{2} \hat{z} + \frac{1}{2} \hat{x}}^{-1}.
\end{align}

As in the $\mathbb{Z}_2$ model, each vertex $v = \vec{r}$ is associated with three vertex terms $V_v^{xy}$, $V_v^{yz}$, and $V_v^{zx}$, each constructed from $Z$ operators. For instance, the $xy$-plane vertex term [Fig.~\ref{fig: gXC_stabilizer}(b–d)] is given by:
\begin{align}
V_v^{xy} = Z_{v - \frac{1}{2} \hat{x}} 
Z_{v + \frac{1}{2} \hat{x}}^{-a_x} 
Z_{v - \frac{1}{2} \hat{y}}^{-1} 
Z_{v + \frac{1}{2} \hat{y}}^{a_y}.
\end{align}

The definitions of $V_v^{yz}$ and $V_v^{zx}$ follow analogously. The exponents of all stabilizers are determined by the integers $(a_x, a_y, a_z)$, which serve as tuning parameters for the model. These have been carefully arranged to ensure that all stabilizer terms in the Hamiltonian mutually commute. Furthermore, each stabilizer satisfies $(S_c)^N = (V_v^{xy})^N = (V_v^{yz})^N = (V_v^{zx})^N = 1$, and thus the ground state condition is $S_c = V_v^{xy} = V_v^{yz} = V_v^{zx} = 1$ for all $c \in \mathcal{C}$ and $v \in \mathcal{V}$.

In the following sections, we analyze the ground state degeneracy of the $\mathbb{Z}_N$ X-cube model, classify the two distinct types of excitations that arise, and investigate the possible topological phases this system can exhibit.

\subsection{ground state Degeneracy}

The ground state degeneracy of the $\mathbb{Z}_N$ X-cube model on a three-torus $T^3$ of size $L_x \times L_y \times L_z$ is given by
\begin{equation} \label{eqn : gXC_GSD}
    \text{GSD} = \frac{d_{yz}^{2L_x} \, d_{zx}^{2L_y} \, d_{xy}^{2L_z}}{d_{xyz}^{3}},
\end{equation}
where
\[
d_{ij} := \gcd(a_i^{L_i} - 1, \, a_j^{L_j} - 1, \, N), \quad 
d_{xyz} := \gcd(a_x^{L_x} - 1, \, a_y^{L_y} - 1, \, a_z^{L_z} - 1, \, N),
\]
for $i,j \in \{x,y,z\}$. The GSD exhibits oscillatory behavior as a function of system size, while its upper bound, $N^{2(L_x + L_y + L_z) - 3}$, grows sub-extensively with the total system size. In the special case $N=2$, Eq.~\eqref{eqn : gXC_GSD} correctly reproduces the $\mathbb{Z}_2$ result given in Eq.~\eqref{eqn : Z2_XC_GSD}.

By definition, each $d_{ij}$ is divisible by $d_{xyz}$, ensuring that the GSD is always an integer. When all $d_{ij} = 1$, the ground state is unique. If $d_{xyz} = 1$ but some $d_{ij} > 1$, the system behaves as a direct product of $(L_x + L_y + L_z)$ copies of the $2d$ toric code, yielding a topologically ordered phase without fracton characteristics. Since both $d_{ij}$ and $d_{xyz}$ depend on the system size, these phases only occur at specific values of $(L_x, L_y, L_z)$.

A particularly interesting case arises when $a_i a_j \equiv 0 \pmod{\mathrm{rad}(N)}$ for any pair $i \neq j \in \{x, y, z\}$, implying $N_{ij} = 1$ as defined previously. In this scenario, all $d_{ij} = 1$ for any system size, and the system has a unique ground state, corresponding to a trivial phase with no topological order.

Conversely, if $a_x a_y a_z \equiv 0 \pmod{\mathrm{rad}(N)}$, i.e., $N_{xyz} = 1$, then although the GSD continues to oscillate with system size, it can always reach values consistent with a direct product of $2d$ toric codes. If we ignore the fluctuations in $d_{ij}$, the upper bound of the GSD appears sub-extensive, as expected for a fracton phase. However, with $d_{xyz} = 1$ in the denominator, this scaling becomes extensive. While the system remains topologically ordered, such scaling suggests a deviation from the strict fracton phase regime.

The full derivation of Eq.~\eqref{eqn : gXC_GSD} and a detailed discussion of its upper bound are presented in Appendix~\ref{Appendix : gXC_GSD}.

\subsection{Local Vertex Excitations} \label{gXC_vertex_loacl}

The Hamiltonian of the X-cube model contains two types of stabilizers: vertex terms and cube terms. Their excitations exhibit distinct topological properties, most notably in their mobility, and thus must be analyzed separately.

We begin by discussing vertex excitations, which typically consist of a triplet of sub-excitations oriented along three directions, all localized at a common vertex.

Each vertex $v$ in the lattice has three associated stabilizers: $V_v^{xy}$, $V_v^{yz}$, and $V_v^{zx}$. We define a charge vector to characterize the excitation content at $v$:
\begin{align}
    \mathbb{V}_v \equiv (\log_\omega V_v^{yz}, \log_\omega V_v^{zx}, \log_\omega V_v^{xy}).
\end{align}
These obey the local constraint
\begin{align}
    \log_\omega V_v^{xy} + \log_\omega V_v^{yz} + \log_\omega V_v^{zx} = 0,
\end{align}
which reflects the stabilizer relation $V_v^{xy} V_v^{yz} V_v^{zx} = 1$.

In the $\mathbb{Z}_2$ case, vertex excitations are lineons—excitations that must be created in pairs and can only move along a one-dimensional subspace. A bound pair of lineons forms a planon, which can move within a two-dimensional plane. In the $\mathbb{Z}_N$ model, similar behavior may persist, but as we will show, this depends sensitively on the excitation charge.

Without loss of generality, we focus on the $(0, -1, 1)$-type excitation, created by an $X$-string operator along the $\hat{x}$-direction. A general form of this operator with length $\ell$ is
\begin{align} \label{eqn : app_gXC_line}
W_{(\vec{r}_0 - \ell \hat{x}, \vec{r}_0)} = \prod_{m=0}^{\ell-1} X_{\vec{r}_0 - (m + \frac{1}{2})\hat{x}}^{a_x^m b},
\end{align}
which creates a pair of vertex excitations at $\vec{r}_0$ and $\vec{r}_0 - \ell \hat{x}$ with charge vectors $b(0, -1, 1)$ and $a_x^\ell b(0, 1, -1)$, respectively.

\begin{figure}
    \centering\includegraphics[width=\columnwidth]{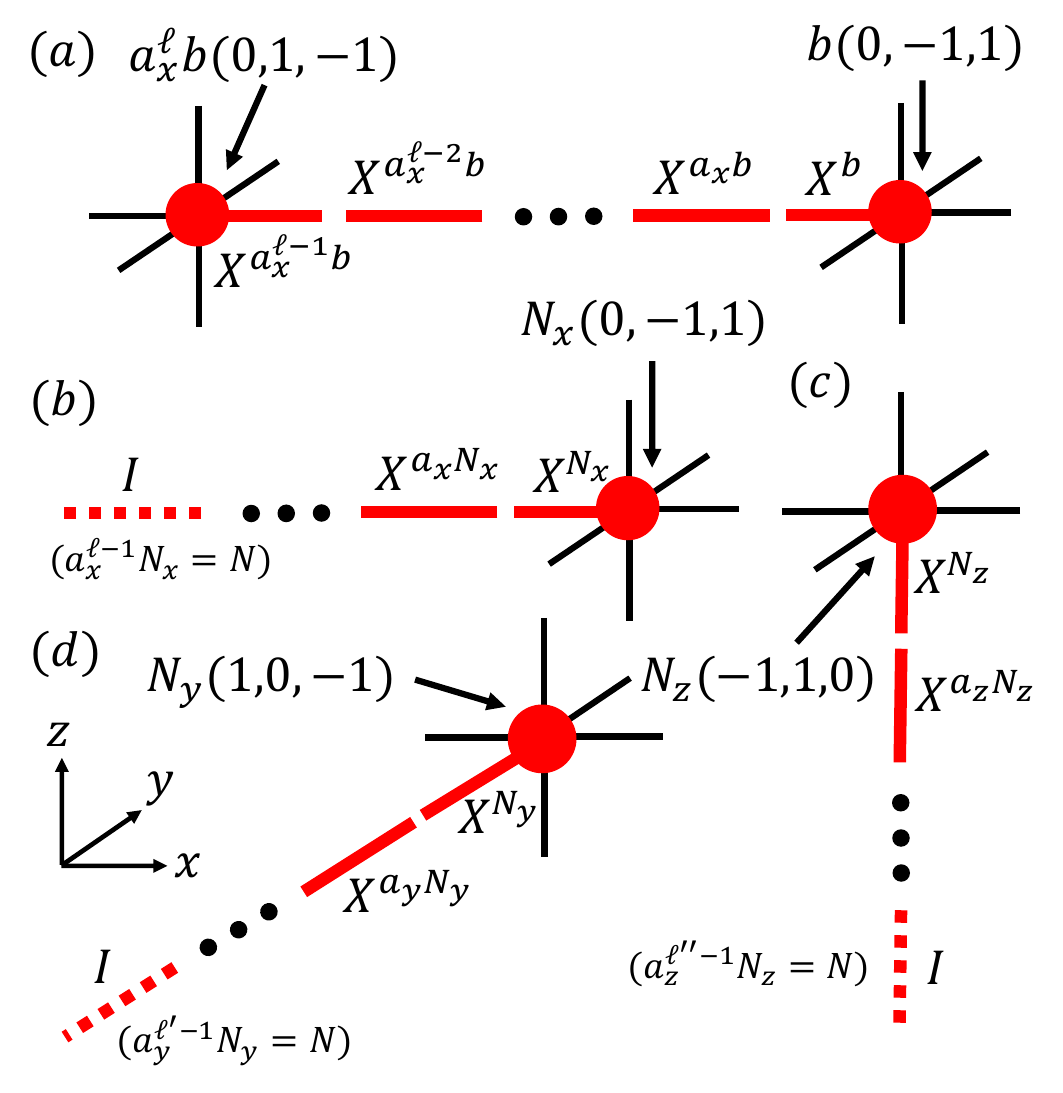}
    \caption{\label{fig_AB_gXC_single_local} 
    {\textbf{Vertex excitations of $\mathbb{Z}_N$ generalized X-cube model via local string operator $W_\mathcal{C}$ Eq.~(\ref{eqn : app_gXC_line}).}} The charge vectors of each excitation are indicated by arrows. (a) A pair of $x$-directional lineons can be created by the straight line operator $W_\mathcal{C}$ with the exponents $\{a_x^{\ell-1}b, a_x^{\ell-2}b,\cdots,a_x b, b\}$. (b) A single vertex excitation can be created by the $x$-directional local line operator $W_\mathcal{C}$ when $b=0$ mod $N_x$. (c) A single vertex excitation can be created by the $z$-directional local line operator $W_\mathcal{C}$ when $b=0$ mod $N_z$. (d) A single vertex excitation can be created by the $y$-directional local line operator $W_\mathcal{C}$ when $b=0$ mod $N_y$.}
\end{figure}

If we set $b = N_x$, then, for sufficiently large $\ell$, the excitation at $\vec{r}_0 - \ell \hat{x}$ becomes trivial, since $a_x^{\ell} N_x \equiv 0 \pmod{N}$. Thus, a single vertex excitation with charge $N_x(0, -1, 1)$ is created [Fig.~\ref{fig_AB_gXC_single_local}(b)], which can be moved freely by local operators in all directions. This excitation is therefore not a lineon, but rather a free boson.

\begin{figure*}
    \includegraphics[width=2\columnwidth]{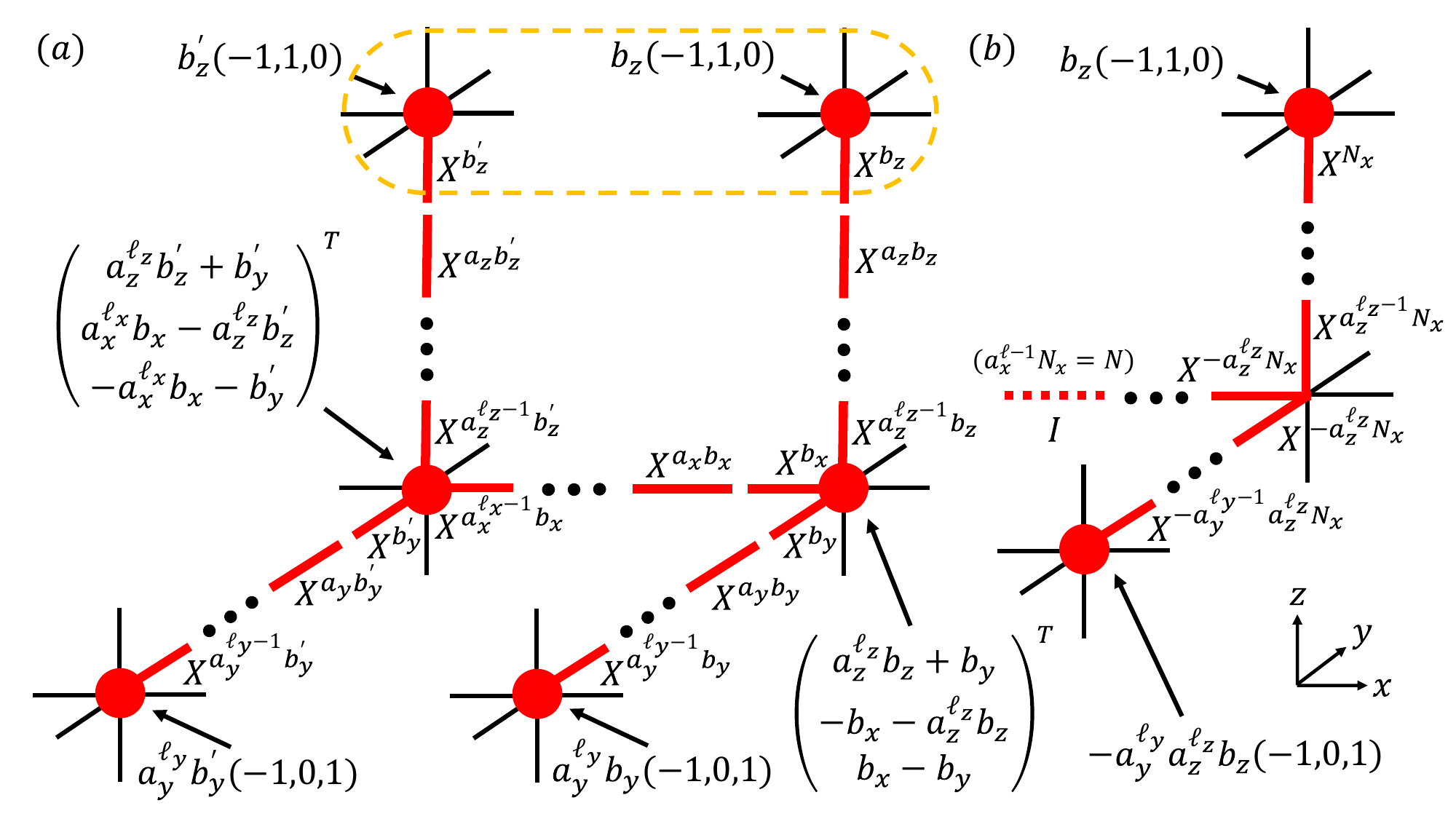}
    \caption{\label{fig_AB_gXC_lineon2planon} 
    \textbf{Promotion of vertex excitations from lineon to planon.} The charge vectors of each excitation are indicated by arrows. (a) The pictorial representation of bending the path of lineon dipole along $x$-axis indicated by dashed loop , from $\hat{z}$-direction to $\hat{y}$-direction. When $b_x=b_y=-a_z^{\ell_z}$ and $b_y'=-a_x^{\ell_x}b_x=-a_z^{\ell_z}b_z'$, lineon pair at bending points vanishes, then the lineon dipole becomes planon which is mobile on $yz$-plane. (b) A single vertex exciation can be looked as a planon instead of along $yz$-plane a lineon when $b_z=0$ mod $N_x$ in Eq.~\eqref{eqn : app_gXC_lineon2Planon}.}
\end{figure*}

Similarly, single vertex excitations of type $N_y(1, 0, -1)$ and $N_z(-1, 1, 0)$ can be created by local $X$-string operators along the $\hat{y}$- and $\hat{z}$-directions with $b = N_y$ and $b = N_z$, respectively [Figs.~\ref{fig_AB_gXC_single_local}(c–d)]. Superpositions of these excitations yield charge vectors of the form $p N_y(1, 0, -1) + q N_z(-1, 1, 0)$, which become proportional to $(0, -1, 1)$ if $p N_y \equiv q N_z \equiv \mathrm{lcm}(N_y, N_z) \pmod{N}$. Therefore, the unit charge of such $(0, -1, 1)$ excitations is $\gcd(N_x, \mathrm{lcm}(N_y, N_z))$, due to the extended Bézout identity [Appendix~\ref{Bezout_identity}]. These excitations are also free bosons.

Next, we examine how a lineon excitation can be promoted to a planon. Consider bending a lineon dipole moving along the $\hat{z}$-direction into the $\hat{y}$-direction. This is realized by composing two $X$-string operators:
\begin{align} \label{eqn : app_gXC_lineon2Planon}
W_{(\vec{r}_0 - \ell_z \hat{z} - \ell_y \hat{y}, \vec{r}_0 - \ell_z \hat{z})}
W_{(\vec{r}_0 - \ell_z \hat{z}, \vec{r}_0)} &= \prod_{m_2=0}^{\ell_y - 1} X_{\vec{r}_0 - \ell_z \hat{z} - (m_2 + \frac{1}{2})\hat{y}}^{a_y^{m_2} b_y} \nonumber \\
&\quad \times \prod_{m_3=0}^{\ell_z - 1} X_{\vec{r}_0 - (m_3 + \frac{1}{2})\hat{z}}^{a_z^{m_3} b_z}.
\end{align}
This operator creates three excitations: at $\vec{r}_0$, $\vec{r}_0 - \ell_z \hat{z}$, and $\vec{r}_0 - \ell_z \hat{z} - \ell_y \hat{y}$, with charge vectors $b_z(-1, 1, 0)$, $(a_z^{\ell_z}b_z + b_y, -a_z^{\ell_z}b_z, -b_y)$, and $a_y^{\ell_y}b_y(-1, 0, 1)$, respectively. Setting $b_y = -a_z^{\ell_z} b_z$ eliminates the charge at the bending point, leaving two well-separated excitations that can be regarded as a planon in the $yz$-plane [Fig.~\ref{fig_AB_gXC_lineon2planon}(a)].

To remove the residual excitation entirely, we introduce a third $X$-string operator along the $\hat{x}$-direction:
\begin{align} \label{eqn : app_gXC_dangling}
W_{(\vec{r}_0 - \ell_z \hat{z} - \ell_x \hat{x}, \vec{r}_0 - \ell_z \hat{z})} = \prod_{m_1 = 0}^{\ell_x - 1} X_{\vec{r}_0 - \ell_z \hat{z} - (m_1 + \frac{1}{2})\hat{x}}^{a_x^{m_1} b_y},
\end{align}
which creates a pair of vertex excitations at its endpoints with charges $b_y(0, -1, 1)$ and $a_x^{\ell_x} b_y (0, 1, -1)$. If we introduce another composite line operator like Eq.~\eqref{eqn : app_gXC_lineon2Planon} at $\vec{r}_0 - \ell_z \hat{z} - \ell_x \hat{x}$ to cancel this excitation, then the lineon dipole can bend freely, confirming its planon nature.

In the special case where $b_x = b_y = -b_z a_z^{\ell_z} \equiv 0 \pmod{N_x}$, the $\hat{x}$-string operator [Eq.~\eqref{eqn : app_gXC_dangling}] generates no excitation at its endpoint. Thus, a single vertex excitation can smoothly change its direction from $\hat{z}$ to $\hat{y}$ without energy cost. In this sense, the corresponding $yz$-planon has unit charge $N_x$, and this result holds regardless of the values of $a_z$, $a_y$, or $\ell_z$. Similarly, if the charge of a $(0, -1, 1)$ excitation is divisible by $N_y$ or $N_z$, it becomes a planon in the $zx$- or $xy$-plane, respectively.

By Bézout's identity, if a $(0, -1, 1)$ excitation has a charge that is a multiple of
\begin{align}
\text{Unit charge} &= \gcd(\gcd(N_x, \mathrm{lcm}(N_y, N_z)), N_y, N_z) \nonumber \\
&= \gcd(N_x, N_y, N_z, \mathrm{lcm}(N_y, N_z)) \nonumber \\
&= \gcd(N_x, N_y, N_z) = N_{xyz},
\end{align}
then it is equivalent to a superposition of a free boson and two types of planons, and is therefore no longer a lineon. Consequently, the allowed charges for a $(0, -1, 1)$ lineon are restricted to integers $1, 2, \dots, N_{xyz} - 1$. The same restriction applies to $(1, 0, -1)$ and $(-1, 1, 0)$ lineons, due to symmetry under coordinate rotations.

In particular, when $a_x a_y a_z \equiv 0 \pmod{\mathrm{rad}(N)}$, we have $N_{xyz} = 1$, and thus no lineon-type vertex excitations exist in the system, independent of the system sizes $L_x$, $L_y$, or $L_z$.

\subsection{Quasi-Lineons} \label{gXC_vertex_global}

In addition to the local vertex excitations discussed above, another type of vertex excitation emerges in the generalized $\mathbb{Z}_N$ X-cube model. In the original $\mathbb{Z}_2$ model with OBCs, true lineons do not exist because non-local string operators traversing the boundary can relocate or eliminate any vertex excitation. These are boundary effects and are absent under PBCs. In contrast, in the generalized $\mathbb{Z}_N$ model, certain vertex excitations remain strictly one-dimensional under local operations even with PBCs, but acquire extended mobility under non-local operations. We refer to these as quasi-lineons.

\begin{figure}
\includegraphics[width=\columnwidth]{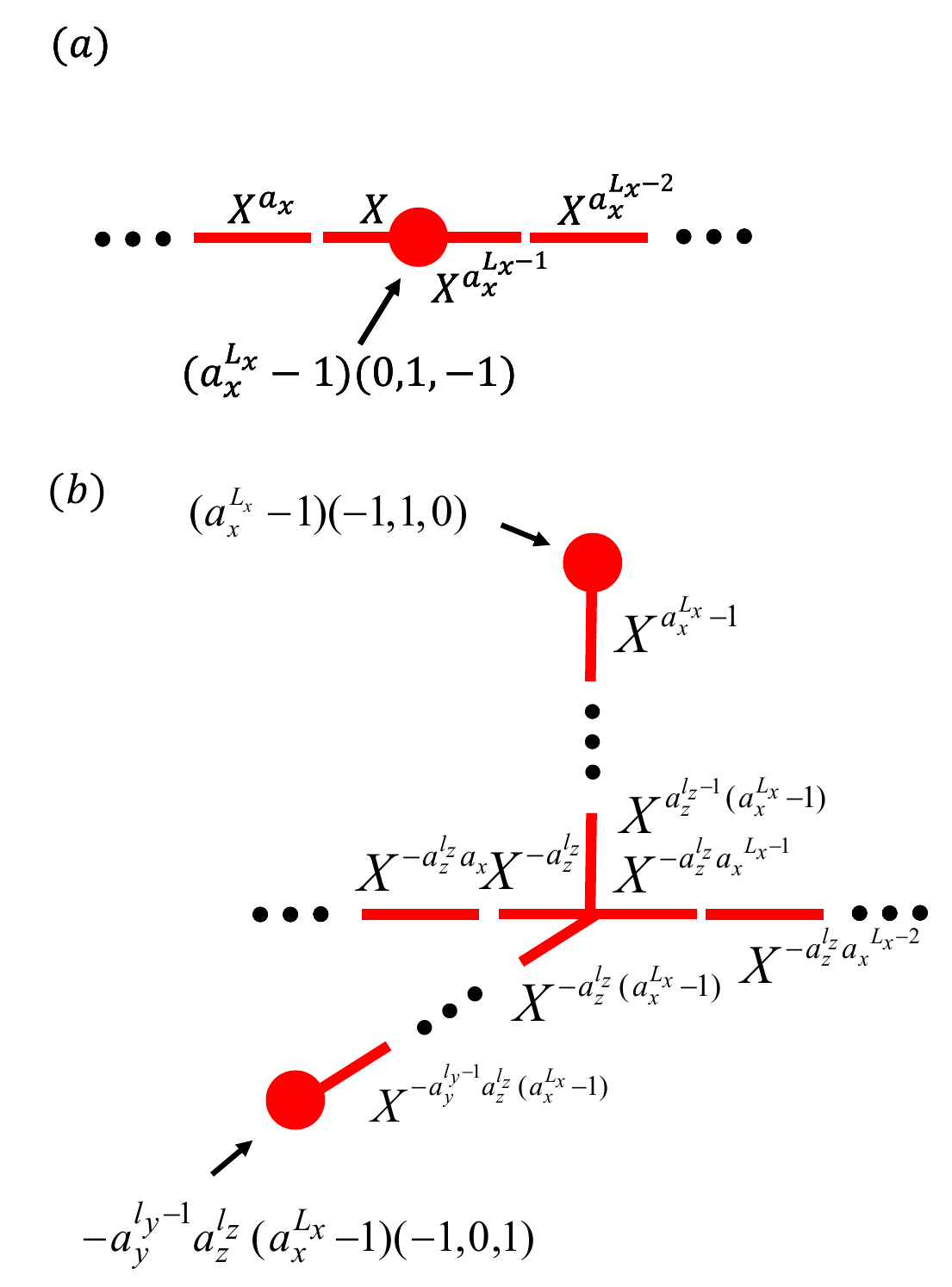}
\caption{\label{fig: gXC_non-local_lineon} \textbf{Quasi-lineons of $\mathbb{Z}_N$ generalized X-cube model via non-local loop operator $W_\mathcal{C}$ Eq.~(\ref{eqn : app_gXC_line}).} The charge vectors of each excitation are indicated by arrows. (a) A single $x$-directional quasi-lineon can be created by the straight line operator $\overline{W}_\mathcal{C}$ wrapping the system along $\hat{x}$-axis. Although it can only move along the $x$-direction under the action of local operators, it can be created and annihilated at any position like a free particle under the action of non-local operators. (b) After a pair of vertex excitations, whose charges are integer multiples of $d_x=\gcd(a_x^{L_x-1},N)$, are created by an ordinary local string operator, a non-local string operator wrapping around the $x$-direction can be used to change their direction of motion. Therefore, such pairs of excitations can effectively move within the $yz$-plane, exhibiting planon-like behavior under non-local operators. However, local operators can only move them along fixed directions, making them a distinct type of quasi-lineon, different from (a).
}
\end{figure}

A single vertex excitation can be created by a non-local $X$-string operator [Eq.~(\ref{eqn : app_gXC_line})] wrapping the system along the $i$-th direction by choosing $\ell = L_i$, for $i \in \{x, y, z\}$. For instance, when $L_x \not\equiv 0 \pmod{M_N(a_x)}$, a non-local operator wrapping the system along the $\hat{x}$-direction generates a single excitation:
\begin{align}
\overline{W}_{\vec{r}_0, \hat{x}} = \prod_{m=0}^{L_x - 1} X_{\vec{r}_0 - (m + \frac{1}{2})\hat{x}}^{a_x^m}, 
\end{align}
carrying charge $(a_x^{L_x} - 1)(0,1,-1)$ [Fig.~\ref{fig: gXC_non-local_lineon}(a)]. Since charges are defined modulo $N$, the true unit charge of such excitations is $d_x = \gcd(a_x^{L_x} - 1, N)$.

Although this excitation resembles a free boson created by local operators, its mobility differs: it can be annihilated and recreated at arbitrary locations using non-local operators, effectively making it mobile in all directions globally. However, under local operations, it only moves along the $\hat{x}$-direction using string operators of finite length. This directional asymmetry leads us to classify it as a quasi-lineon—it behaves like a lineon under local operators, but exhibits enhanced mobility under global operations.

By symmetry, quasi-lineons of types $(1,0,-1)$ and $(-1,1,0)$ can similarly be generated via non-local loops along $\hat{y}$ and $\hat{z}$, respectively, with unit charges $d_y = \gcd(a_y^{L_y} - 1, N)$ and $d_z = \gcd(a_z^{L_z} - 1, N)$.

\begin{figure*}[hbt!]
    \includegraphics[width=2\columnwidth]{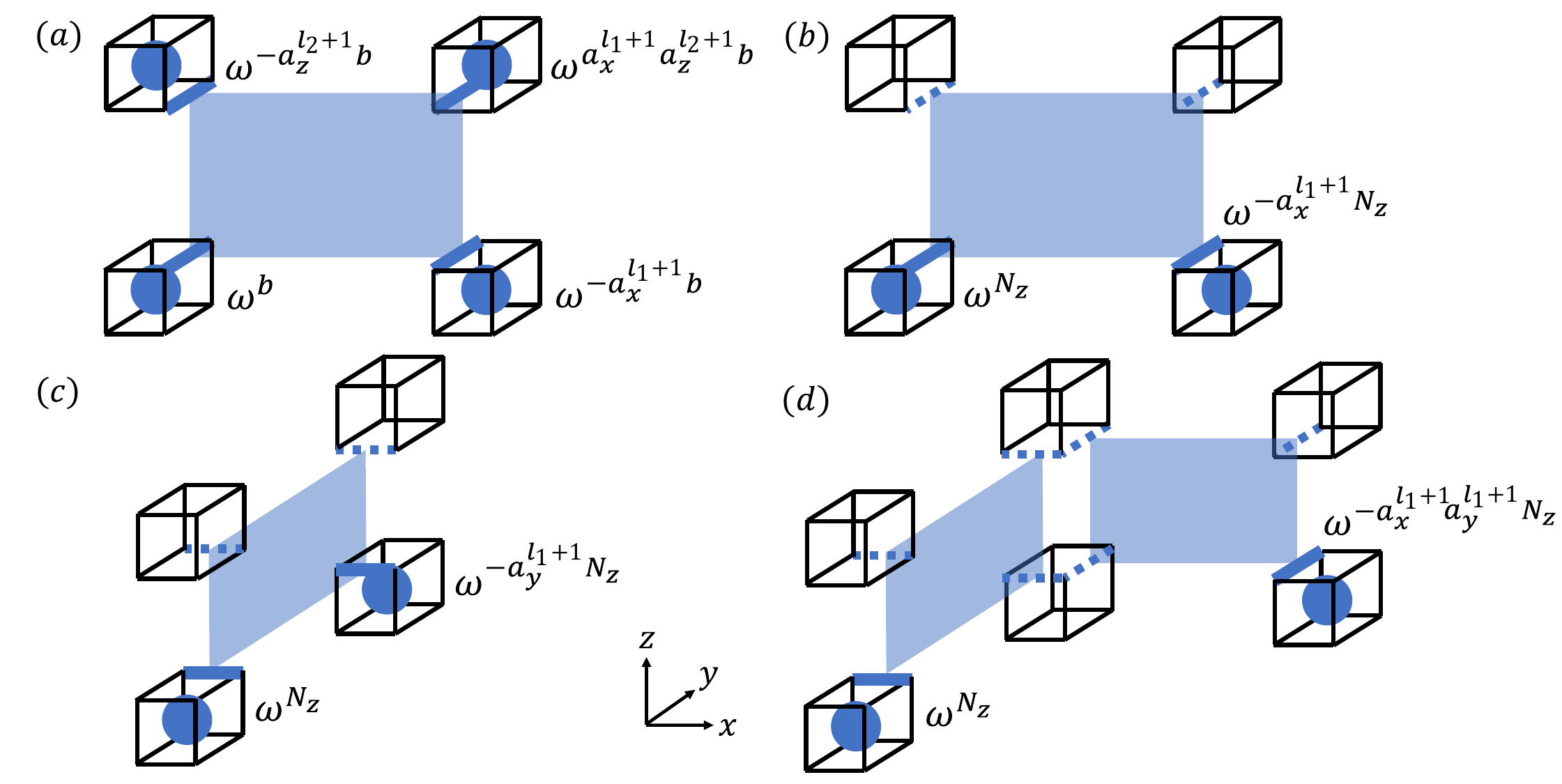}
    \caption{\label{fig_app_gXC_cube_local} 
    \textbf{Cube excitations of $\mathbb{Z}_N$ generalized X-cube model via local membrane operator $O_\mathcal{M}$ Eq.~(\ref{eqn : app_gXC_membrane})}. Each fracton's eigenvalue of $S_c$ is denoted. Translucent rectangles represents $Z$-membrane operator $O_\mathcal{M}$, and dotted edge of cubes indicates identity operator.
    (a) A fracton quadrupole can be created by the rectangular membrane operator $O_\mathcal{M}$ along $zx$-plane.
    (b) A cube excitation pair along $\hat{x}$-direction can be created by the local Membrane operators $O_\mathcal{M}$ along $zx$-plane when $b=N_z$. 
    (c) A cube excitation pair along $\hat{y}$-direction can be created by the local Membrane operators $O_\mathcal{M}$ along $yz$-plane when $b=N_z$.
    (d) Using the two membrane operators in Figures (b) and (c), a cube excitation with charge $N_z$ can move within the $xy$-plane. Therefore, it can be regarded as a planon.
    }
\end{figure*}

Furthermore, composite excitations can be formed by combining two non-local loop operators along orthogonal directions. For instance, when $L_x \not\equiv 0 \pmod{M_N(a_x)}$ and $L_y \not\equiv 0 \pmod{M_N(a_y)}$, the operator:
\begin{align}
\overline{W}_{\vec{r}_0, \hat{x}} \, \overline{W}_{\vec{r}_0, \hat{y}}^{b_y} 
= \prod_{m=0}^{L_x - 1} X_{\vec{r}_0 - (m + \frac{1}{2})\hat{x}}^{a_x^m}
  \prod_{n=0}^{L_y - 1} X_{\vec{r}_0 - (n + \frac{1}{2})\hat{y}}^{a_y^n b_y}
\end{align}
creates a composite excitation at $\vec{r}_0$ with charge vector:
\[
(-b_y(a_y^{L_y} - 1), \, a_x^{L_x} - 1, \, -a_x^{L_x} + 1 + b_y(a_y^{L_y} - 1)).
\]
This object is generically immobile, being a bound state of orthogonal lineons. However, if $a_x^{L_x} - 1 - b_y(a_y^{L_y} - 1) \equiv 0 \pmod{N}$, the total charge reduces to $(a_x^{L_x} - 1)(-1,1,0)$, violating only two stabilizers $V_v^{zx}$ and $V_v^{yz}$, and becomes a quasi-lineon along $\hat{z}$.

As with the free boson case, superpositions of quasi-lineons of types $(1,0,-1)$ and $(-1,1,0)$ allow for mobility transformations. The minimal unit charge for a $(0,-1,1)$ quasi-lineon becomes $\gcd(d_x, \mathrm{lcm}(d_y, d_z))$.

Now consider a vertex excitation with charge $d_y(0, -1, 1)$, which can move along $\hat{x}$ via local operators but not along $\hat{y}$. A global $X$-loop along $\hat{y}$:
\begin{align}
\overline{W}_{\vec{r}_0, \hat{y}}^b 
= \prod_{m=0}^{L_y - 1} X_{\vec{r}_0 - (m + \frac{1}{2})\hat{y}}^{a_y^m b}
\end{align}
with $b(a_y^{L_y} - 1) \equiv d_y \pmod{N}$ can convert the excitation to $d_y(-1,1,0)$ via Bézout's identity, thereby altering its local mobility direction from $\hat{x}$ to $\hat{z}$.

Thus, this excitation behaves as a planon within the $xz$-plane under global operators, while retaining lineon-like mobility under local operators. We refer to this as a planon-like quasi-lineon distinct from the free-particle-like quasi-lineon discussed earlier. These represent two qualitatively different classes of quasi-lineons.

We define quasi-lineons as vertex excitations that are lineons under local operations but gain additional mobility through global operators. They can be further classified based on their global mobility patterns.

By Bézout's theorem, the minimal unit charge from the superposition of the three types of quasi-lineons is:
\begin{align}
\text{Unit Charge} &= \gcd(\gcd(d_x, \mathrm{lcm}(d_y, d_z)), d_y, d_z) \nonumber \\
&= \gcd(d_x, d_y, d_z, \mathrm{lcm}(d_y, d_z)) \nonumber \\
&= \gcd(d_x, d_y, d_z) \nonumber \\
&= \gcd(a_x^{L_x} - 1, \, a_y^{L_y} - 1, \, a_z^{L_z} - 1, \, N) = d_{xyz}.
\end{align}

Since each $a_i^{L_i} - 1$ is coprime with $a_i$, it follows that $d_{xyz}$ must be coprime with $a_x a_y a_z$ and divides $N$. The definition of $N_{xyz}$—the largest divisor of $N$ coprime with $a_x a_y a_z$—implies $d_{xyz} \leq N_{xyz} \leq N$.

Therefore, the previously stated upper bound $N_{xyz} - 1$ encompasses both true lineons and quasi-lineons. If we restrict to strictly immobile (i.e., lineon-type) excitations, their possible charges are limited to $1, 2, \dots, d_{xyz} - 1$. Notably, this upper bound depends sensitively on the system size.

\subsection{Local Cube Excitations}\label{gXC_cube_local}

We now focus on the excitations of the cube stabilizers, i.e., violations of the condition $S_c = 1$ in Eq.~(\ref{eqn : H_gHC}). A rectangular $Z$-membrane operator along the $zx$-plane can be written as
\begin{align}\label{eqn : app_gXC_membrane}
O_{\{\vec{r}_0 + \tfrac{1}{2}\hat{y}, \, l_1\hat{x} + l_2\hat{z}\}} 
= \prod_{m_1 = 0}^{l_1} \prod_{m_2 = 0}^{l_2} 
Z_{\vec{r}_0 + \tfrac{1}{2}\hat{y} + m_1\hat{x} + m_2\hat{z}}^{a_x^{m_1} a_z^{m_2} b}, 
\end{align}
which creates four cube excitations at its corners, with stabilizer eigenvalues $\omega^b$, $\omega^{-a_x^{l_1+1}b}$, $\omega^{-a_z^{l_2+1}b}$, and $\omega^{a_x^{l_1+1} a_z^{l_2+1} b}$ [Fig.~\ref{fig_app_gXC_cube_local}(a)].

In the original $\mathbb{Z}_2$ X-cube model, these excitations are immobile fractons that cannot move independently. In the generalized $\mathbb{Z}_N$ model, however, their mobility depends on the system parameters and the membrane operator indices. In some cases, they can even become mobile or trivial.

For instance, setting $b = N_z$ in Eq.~\eqref{eqn : app_gXC_membrane} ensures that the membrane edge at $z - z_0 = l_2$ acts trivially when $a_z^{l_2} N_z \equiv 0 \pmod{N}$. This produces a pair of cube excitations forming a dipole along the $\hat{x}$-direction [Fig.~\ref{fig_app_gXC_cube_local}(b)]. Similarly, a membrane operator in the $yz$-plane with $b = N_z$ creates a dipole along the $\hat{y}$-direction [Fig.~\ref{fig_app_gXC_cube_local}(c)]. Since both types of excitations can be locally created and annihilated, cube excitations whose charges are multiples of $N_z$ become planons within the $xy$-plane.

By symmetry, the unit charge of cube planons along the $zx$- and $yz$-planes is given by $N_y$ and $N_x$, respectively.

According to Bézout's identity, the unit charge of a superposition of these three types of planons is the greatest common divisor:
\[
N_{xyz} = \gcd(N_x, N_y, N_z).
\]
Therefore, cube excitations are considered true fractons only if their charges lie in the range $1, 2, \dots, N_{xyz} - 1$.

In particular, when $a_x a_y a_z \equiv 0 \pmod{\mathrm{rad}(N)}$, then $N_{xyz} = 1$, and thus no fracton-type cube excitations exist in the system regardless of the system size $(L_x, L_y, L_z)$.

\begin{figure}
\includegraphics[width=\columnwidth]{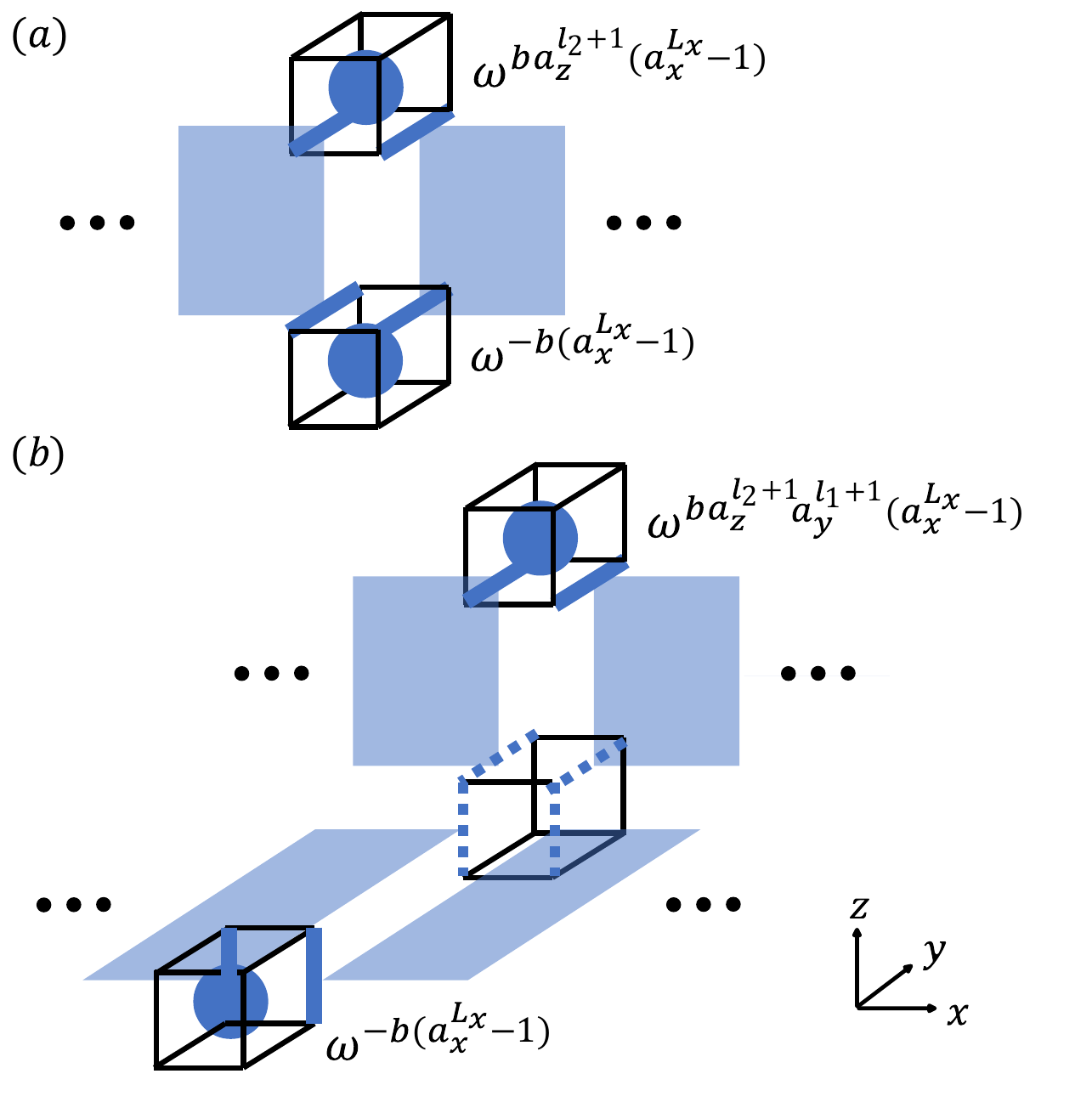}
\caption{\label{fig_app_gXC_cube_non-local} \textbf{Quasi-Fractons of $\mathbb{Z}_N$ generalized X-cube model via non-local membrane operator $O_\mathcal{M}$ Eq.~(\ref{eqn : app_gXC_membrane}).} Each excitation's eigenvalue of $S_c$ is denoted. (a) A cube excitation pair along $\hat{z}$-direction can be created by the non-local membrane operator $O_\mathcal{M}$ along $zx$-plane with $l_1=L_x-1$ when $a_x^{L_x}-1\ne 0$ mod $N$. (b) By using different types of non-local membrane operators that wrap around the system along the \( x \)-direction, such excitations in (a) can move in a two-dimensional manner, like planons, within the \( yz \)-plane. However, no local operator can move them without causing other effects. Such excitations are referred to as quasi-fractons.
}
\end{figure}

\subsection{Quasi-Fractons}

Analogous to vertex excitations, cube excitations in the original $\mathbb{Z}_2$ model with OBCs are not true fractons, as membrane operators that wrap around the boundaries allow them to move freely in all directions. Since this is a boundary effect, it does not occur under PBCs. However, in the generalized $\mathbb{Z}_N$ model, the situation differs: even with PBCs, certain cube excitations remain immobile under local operations but gain mobility through non-local ones. We refer to these as quasi-fractons.

For example, when $L_x \not\equiv 0 \pmod{M_N(a_x)}$, a pair of cube excitations along the $\hat{z}$-direction can be created by a non-local membrane operator on the $zx$-plane wrapping around the system in the $\hat{x}$-direction:
\begin{align}\label{eqn : app_gXC_membrane_global}
O_{\{\vec{r}_0 + \tfrac{1}{2}\hat{y}, \, (L_x - 1)\hat{x} + l_2 \hat{z} \}} 
= \prod_{m_1 = 0}^{L_x - 1} \prod_{m_2 = 0}^{l_2} 
Z_{\vec{r}_0 + \tfrac{1}{2}\hat{y} + m_1 \hat{x} + m_2 \hat{z}}^{a_x^{m_1} a_z^{m_2} b},
\end{align}
which generates charges $-b(a_x^{L_x} - 1)$ and $b a_z^{l_2+1} (a_x^{L_x} - 1)$ [Fig.~\ref{fig_app_gXC_cube_non-local}(a)]. By Bézout's identity, the unit charge of such excitations is $d_x = \gcd(a_x^{L_x} - 1, N)$.

Unlike standard fractons, these excitations do not require fourfold creation; instead, they can be generated as dipoles. While a single cube excitation remains immobile, a dipole created by this membrane operator can move along the $\hat{z}$-direction. Additionally, another membrane operator wrapping around the $\hat{x}$-direction in the $xy$-plane can produce a pair of cube excitations along the $\hat{y}$-direction, also with unit charge $d_x$. This implies that cube excitations with charge $d_x$ can effectively move in both the $\hat{y}$- and $\hat{z}$-directions, exhibiting planon-like behavior within the $yz$-plane.

However, these excitations are distinct from the planons discussed earlier, which are created and moved by local operators with unit charge $N_x$. The quasi-fractons discussed here must be created in pairs by non-local membrane operators, and their motion also requires global operations. Thus, their existence and behavior are strongly dependent on the system size and boundary conditions. Under local operations, they are completely immobile, behaving like standard fractons. Hence, similar to our definition of quasi-lineons, we refer to these as planon-like quasi-fractons.

By symmetry, quasi-fractons in the $xy$- and $zx$-planes have unit charges $d_z = \gcd(a_z^{L_z} - 1, N)$ and $d_y = \gcd(a_y^{L_y} - 1, N)$, respectively. By Bézout’s identity, the minimal unit charge arising from the superposition of the three types of quasi-fractons is
\[
d_{xyz} = \gcd(d_x, d_y, d_z),
\]
which must be a divisor of $N_{xyz} = \gcd(N_x, N_y, N_z)$. Therefore, if we restrict our attention to true fractons—i.e., cube excitations that cannot be decomposed into globally movable quasi-fractons—their allowed charges must lie in the range $1, 2, \dots, d_{xyz} - 1$. Notably, the upper bound $d_{xyz} - 1$ depends sensitively on the system size $(L_x, L_y, L_z)$.

\subsection{Phase Classification}

In the previous sections, we have shown that the ground state degeneracy, the nature of vertex and cube excitations in the $\mathbb{Z}_N$ X-cube model all depend on the parameters $\{N, a_x, a_y, a_z\}$ in the Hamiltonian, and in some cases also on the system sizes $\{L_x, L_y, L_z\}$. Therefore, by tuning these parameters, the model may realize distinct topological phases. This section provides a detailed classification of such possibilities.

\subsubsection{Trivial Phase}

When $N_{xyz} = 1$, the model enters a trivial phase. Here, the term "trivial" does not imply the absence of topological order altogether. In fact, the system still exhibits features akin to topological order, such as a ground state degeneracy and anyon-like excitations, similar to multiple copies of the $2d$ toric code. However, since the X-cube model is typically viewed as a prototype of the fracton phase, a mere collection of $2d$ toric codes is considered trivial in this context.

Because $d_{xyz}$ is a divisor of $N_{xyz}$, it must also equal 1 regardless of system size. In this case, the ground state degeneracy becomes
\begin{align}
    \text{GSD} = d_{yz}^{2L_x} d_{zx}^{2L_y} d_{xy}^{2L_z},
\end{align}
which resembles a stack of decoupled $2d$ toric codes. While this GSD may still oscillate with system size, and in some cases even reduce to 1, such behaviors merely reflect the behavior of $2L_x + 2L_y + 2L_z$ independent toric codes and do not indicate a genuine fracton phase.

This perspective is further supported by examining the excitations. According to previous discussions, the maximal nontrivial charge of any excitation is $N_{xyz} - 1 = 0$, implying that all vertex and cube excitations can be regarded as trivial. They either behave as free bosons or as combinations of planons. In this phase, planons are considered trivial excitations because they resemble anyons in the $2d$ toric code and do not contribute to the defining features of a fracton phase.

\subsubsection{Fracton Phase}

Conversely, when $N_{xyz} \neq 1$, the system may realize a nontrivial fracton phase. In this case, the denominator in the GSD formula,
\[
d_{xyz} = \gcd(a_x^{L_x} - 1, a_y^{L_y} - 1, a_z^{L_z} - 1, N)^3,
\]
oscillates with system size. Even in the thermodynamic limit, there exist large but specific system sizes where $d_{xyz} = 1$, and the GSD reduces to that of decoupled $2d$ toric codes. However, topological phases are generally expected to be robust against changes in system size. Therefore, even when $d_{xyz} = 1$, the system should still be regarded as being in a fracton phase.

This conclusion is reinforced by analyzing the excitations. As previously established, $d_{xyz} - 1$ is the upper bound for the charge of conventional fractons and lineons. When $d_{xyz} = 1$, these traditional excitations are absent. Nonetheless, the condition $N_{xyz} \neq 1$ ensures the existence of quasi-fractons and quasi-lineons. All cube and vertex excitations in this case can be viewed as composites of such quasi-excitations. Although their restricted mobility is partially relaxed, allowing planon-like or even particle-like behavior, their movement still requires non-local operators, granting them protection from local perturbations. In this sense, they play the same effective role as conventional fractons and lineons and should be considered characteristic of the same fracton phase.

Moreover, quasi-fractons and quasi-lineons resemble boundary-induced effects in the original $\mathbb{Z}_2$ model under OBCs. However, if not all of the $d_{ij}$ are equal to 1, the generalized $\mathbb{Z}_N$ model with PBCs is fundamentally different from its OBC counterpart. In OBCs, even planons can be reduced to free bosons due to boundary effects, and the system lacks any topological degeneracy of the form $d_{yz}^{2L_x} d_{zx}^{2L_y} d_{xy}^{2L_z}$. Hence, the existence of quasi-excitations with system-size-dependent GSD under PBCs reflects genuine fracton behavior rather than boundary-induced artifacts under OBCs.

\section{\texorpdfstring{$3d$}{3d} Toric Code}\label{sec : 3d_toric_code}

In this section, we perform the same analysis for the $3d$ toric code as we did for the X-cube model in Section~\ref{sec : X_cube_model}.

\subsection{\texorpdfstring{$\mathbb{Z}_2$}{Z\_2} Model}

\begin{figure*}[hbt!]
\includegraphics[width=2\columnwidth]{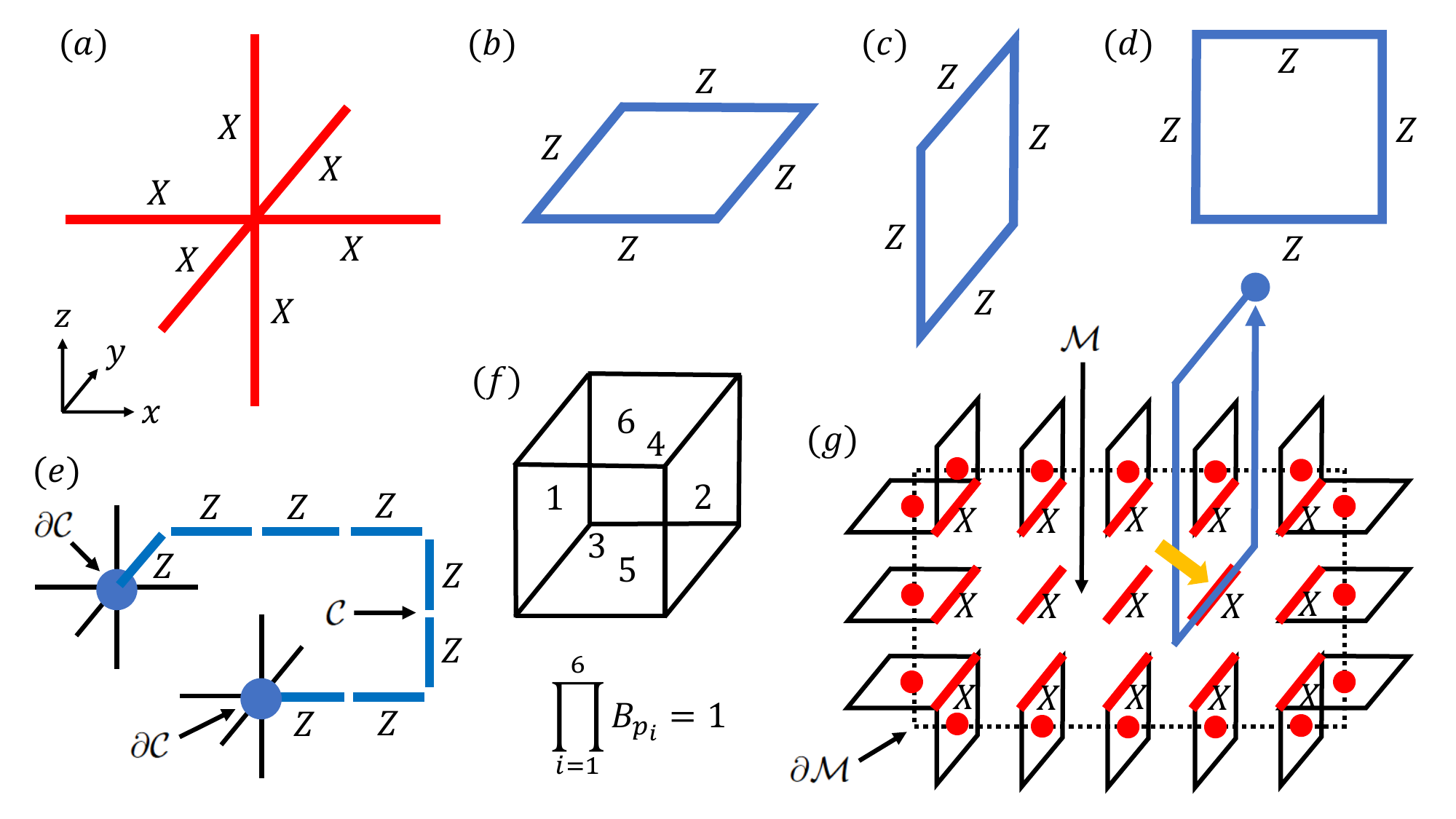}
\caption{\label{fig_Z2_TC_new} \textbf{$3d$ $\mathbb{Z}_2$ toric code.} A two-dimensional spin lives on each link. (a) A vertex term $S_v$. (b--d) Plaquette terms $B_p$ in the $xy$, $yz$, and $zx$ planes. (e) A pair of particle-like excitations can be created by a string operator $W_{\mathcal{C}}$, which violates $S_v = 1$ at the endpoints of the curve $\mathcal{C}$. (f) The product of six plaquette terms around a cube is equal to 1, enforcing magnetic flux conservation. (g) A loop-like flux excitation is created by a membrane operator $O_{\mathcal{M}}$ whose boundary $\partial \mathcal{M}$ defines the loop. Braiding between the particle and loop excitations leads to a nontrivial phase factor $e^{i\pi} = -1$, arising from their intersection, indicated by a yellow arrow.}
\end{figure*}

The toric code~\cite{PhysRevB.72.035307,Kong2020} is the simplest three-dimensional model considered in this work. Although many of its properties are well-known, we begin with a brief review of the $\mathbb{Z}_2$ toric code. Its local Hilbert space is two-dimensional and defined on the links of a cubic lattice, with conventional Pauli operators acting on each spin. The Hamiltonian consists of two types of mutually commuting stabilizers: vertex and plaquette terms [Fig.~\ref{fig_Z2_TC_new} (a--d)].
\begin{align}\label{eqn : H_Z2_TC}
H = -\sum_{v \in \mathcal{V}} S_v -\sum_{p \in \mathcal{P}} B_p.
\end{align}
Here, $\mathcal{V}$ denotes the set of all vertices, and $\mathcal{P}$ the set of all plaquettes. Each vertex $v$ has six surrounding spins [Fig.~\ref{fig_Z2_TC_new} (a)], and the vertex term is given by
\begin{align}
S_{v} =  X_{v-\frac{1}{2}\hat{x}} X_{v+\frac{1}{2}\hat{x}} X_{v-\frac{1}{2}\hat{y}} X_{v+\frac{1}{2}\hat{y}} X_{v-\frac{1}{2}\hat{z}} X_{v+\frac{1}{2}\hat{z}}. \nonumber
\end{align}
Plaquette terms are defined on square faces in the $ij$-plane ($ij \in \{xy, yz, zx\}$), where a plaquette $p = \vec{r} + \frac{1}{2}(\hat{i} + \hat{j})$ has four edges:
\begin{align}
B_{p} =  Z_{p-\frac{1}{2}\hat{j}} Z_{p+\frac{1}{2}\hat{j}} Z_{p-\frac{1}{2}\hat{i}} Z_{p+\frac{1}{2}\hat{i}}. \nonumber
\end{align}
Since $X$ and $Z$ anticommute only on the same qubit, all $S_v$ and $B_p$ terms commute with each other. In addition, $S_v^2 = B_p^2 = 1$ because they are composed of $\mathbb{Z}_2$ Pauli operators. Therefore, the ground states satisfy $S_v = B_p = 1$ for all $v$ and $p$.

There are two types of excitations in this model. The first is the particle-like $\mathbb{Z}_2$ electric charge [Fig.~\ref{fig_Z2_TC_new} (e)], created by a $Z$-string operator $W_{\mathcal{C}}$ along a curve $\mathcal{C}$. This operator flips the eigenvalue of $S_v$ at the endpoints $\partial \mathcal{C}$, resulting in a pair of electric charges. These excitations are always created in pairs. The second is the loop-like $\mathbb{Z}_2$ magnetic flux excitation [Fig.~\ref{fig_Z2_TC_new} (g)], created by an $X$-membrane operator $O_{\mathcal{M}}$ acting over a surface $\mathcal{M}$. The boundary $\partial \mathcal{M}$ of this membrane marks the location of the magnetic flux loop, as plaquette terms $B_p$ are violated along it. This loop must be closed due to the constraint
\begin{align}
\prod_{p \in \partial C} B_p = 1, \nonumber
\end{align}
where $C$ is any cube in the lattice [Fig.~\ref{fig_Z2_TC_new} (f)]. This ensures conservation of magnetic flux and prohibits open-loop excitations.

Unlike in fracton models, both types of excitations are fully mobile: electric charges can move freely in three dimensions, and magnetic flux loops can deform or shrink without creating additional excitations. Notably, when an electric charge winds around a magnetic flux loop, a mutual braiding phase $e^{i\pi} = -1$ is acquired, arising from the intersection of the string and membrane operators [Fig.~\ref{fig_Z2_TC_new} (g)].

Finally, the ground state degeneracy on the three-torus $T^3$ is $\text{GSD} = 2^3 = 8$, independent of system size.

\subsection{\texorpdfstring{$\mathbb{Z}_{N>2}$}{Z\_N} Generalization}\label{sec : gTC}

\begin{figure}
\includegraphics[width=\columnwidth]{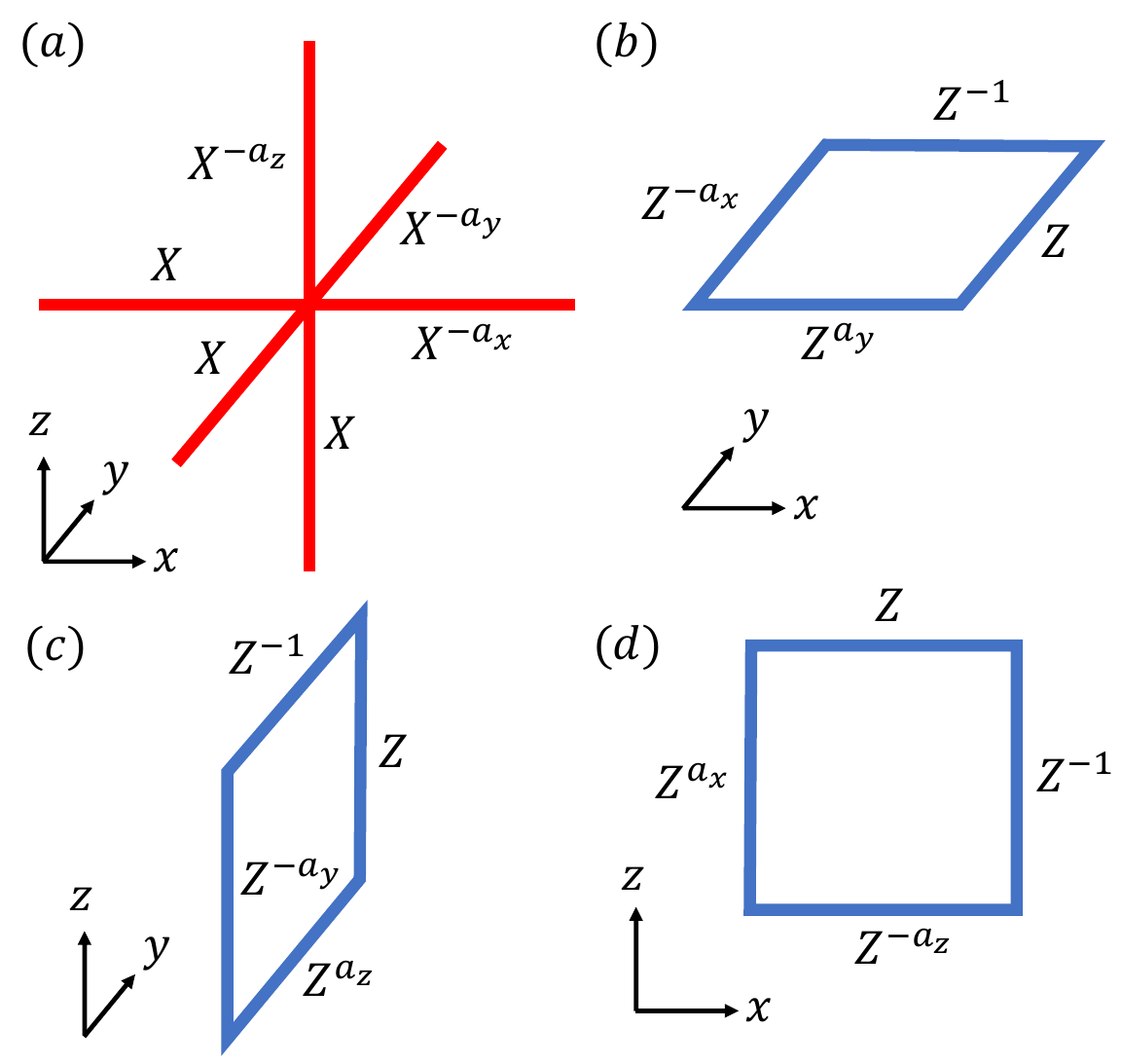}
\caption{\label{fig_gTC_stabilizer_new}\textbf{Stabilizer terms of the generalized $3d$ $\mathbb{Z}_N$ toric code (Eq.~\eqref{eqn : H_gTC}).} (a) Vertex term $S_v$. (b--d) Plaquette terms $B_p$ in the $xy$, $yz$, and $zx$ planes.}
\end{figure}

We now consider the $\mathbb{Z}_N$ generalization of the $3d$ toric code for $N > 2$. The Hamiltonian retains the same form as in the $\mathbb{Z}_2$ case:
\begin{align}{\label{eqn : H_gTC}}
H = -\frac{1}{2}\sum_{v \in \mathcal{V}} S_v -\frac{1}{2}\sum_{p \in \mathcal{P}} B_p + \text{h.c.}
\end{align}
The vertex terms $S_v$ and plaquette terms $B_p$ are now constructed using the generalized $\mathbb{Z}_N$ Pauli operators acting on $N$-dimensional spins on the links of the cubic lattice. Their definitions depend on three integer parameters $(a_x, a_y, a_z)$.

The vertex term at site $v = \vec{r} \in \mathbb{Z}^3$ is defined as [Fig.~\ref{fig_gTC_stabilizer_new} (a)]:
\begin{align}
S_{v} =  X_{v-\frac{1}{2}\hat{x}} X_{v+\frac{1}{2}\hat{x}}^{-a_x} X_{v-\frac{1}{2}\hat{y}} X_{v+\frac{1}{2}\hat{y}}^{-a_y} X_{v-\frac{1}{2}\hat{z}} X_{v+\frac{1}{2}\hat{z}}^{-a_z}. \nonumber 
\end{align}
Each plaquette term is defined on a square face at $p = \vec{r} + \frac{1}{2}\hat{i} + \frac{1}{2}\hat{j}$ for $(i,j) \in \{(x,y), (y,z), (z,x)\}$, as follows [Fig.~\ref{fig_gTC_stabilizer_new} (b--d)]:
\begin{align}
B_{p} =  Z^{-a_i}_{p-\frac{1}{2}\hat{i}} Z_{p+\frac{1}{2}\hat{i}} Z^{a_j}_{p-\frac{1}{2}\hat{j}} Z^{-1}_{p+\frac{1}{2}\hat{j}}. \nonumber
\end{align}
The exponents $(a_x, a_y, a_z)$ are chosen to ensure that all stabilizer terms commute. Because these operators are of order $N$, we have $S_v^N = B_p^N = 1$. When $N = 2$ and $(a_x, a_y, a_z) = (1, 1, 1)$, the model reduces to the standard $\mathbb{Z}_2$ toric code Hamiltonian in Eq.~\eqref{eqn : H_Z2_TC}. The ground states satisfy $S_v = B_p = 1$ for all $v$ and $p$, as in the $\mathbb{Z}_2$ case.

Next, we will show that the generalized model Eq.~\eqref{eqn : H_gTC} exhibits behavior beyond that of the continuum $\mathbb{Z}_N$ gauge theory~\cite{horowitz1989exactly,wang2019topological,chen2016bulk}, whose Lagrangian is:
\begin{align}
\mathcal{L} = \frac{N}{2\pi} \epsilon^{\mu\nu\lambda\rho} B_{\mu\nu}\partial_\lambda A_\rho - J_\mu A^{\mu} - \frac{1}{2} B_{\mu\nu}\Sigma^{\mu\nu}. \nonumber
\end{align}
In this continuum theory, the ground state degeneracy on the three-torus is always $N^3$, independent of the system size. The point-like electric excitation carries charge $q \bmod N$, and the loop-like magnetic excitation carries flux $2\pi n / N$ (or equivalently, integer label $n \bmod N$). Both excitations are fully mobile and fuse modulo $N$. The mutual statistics between an electric and magnetic excitation linked in space gives a Berry phase $\exp(2\pi i nq/N)$.

However, in our lattice generalization, depending on the values of $(a_x, a_y, a_z)$, we find deviations from this naive expectation.

\begin{figure*}
\includegraphics[width=2\columnwidth]{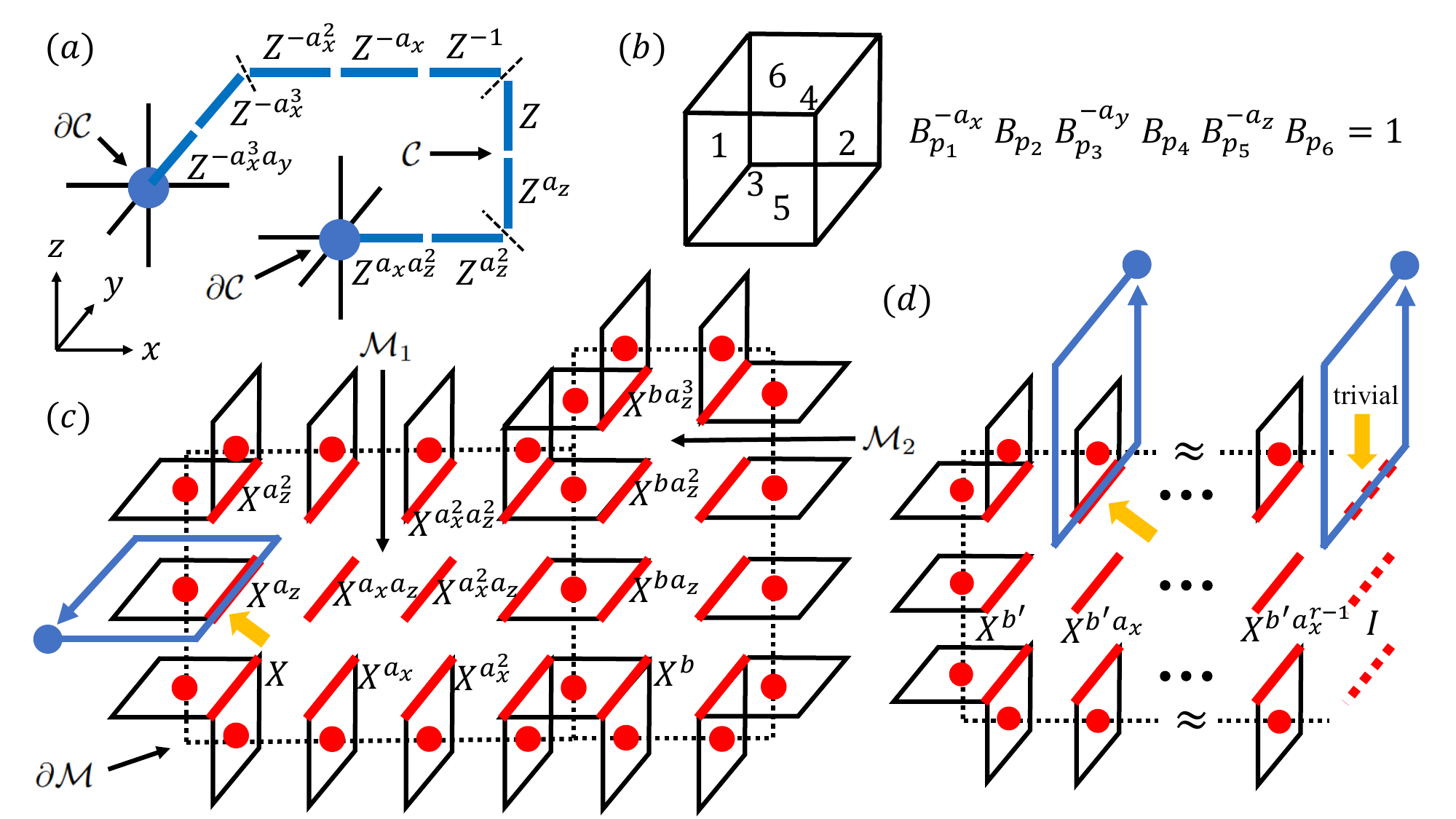}
\caption{\label{fig: gTC_excitation_new} \textbf{Excitations in the generalized $3d$ $\mathbb{Z}_N$ toric code.} (a) A pair of particle-like excitations is created by a string operator $W_\mathcal{C}$ with position-dependent exponents $\{ a_x a_z^2, a_z^2, \cdots \}$. (b) The product of plaquette terms around a cube, with nonuniform exponents, equals 1. (c) A magnetic flux loop is created by a membrane operator $O_\mathcal{M}$ defined on $\mathcal{M} = \mathcal{M}_1 \cup \mathcal{M}_2$. The exponents of the $X$ operators vary across the membrane. When the membranes merge, excitations may appear at the boundary. Braiding between the particle and the loop yields a statistical angle dependent on the particle's path. (d) Unlike in the $\mathbb{Z}_2$ case, open-loop flux excitations are possible when $a_i \equiv 0 \mod \text{rad}(N)$. An example is shown in the $zx$-plane when $a_x \equiv 0 \mod \text{rad}(N)$. In this case, the braiding phase between particle and loop may be path-dependent or even trivial.}
\end{figure*}

\subsection{Ground State Degeneracy}

The ground state degeneracy on a 3-torus $T^3$ with system sizes $(L_x, L_y, L_z)$ is given by
\begin{equation}\label{eqn : gTC_GSD}
     \text{GSD}=d_{xyz}^3 = \gcd(a_{x}^{L_x} - 1, a_{y}^{L_y} - 1, a_{z}^{L_z} - 1, N)^3,
\end{equation}
which strongly oscillates between $1$ and $N^3$, depending on the system sizes $(L_x, L_y, L_z)$ and the parameters $(a_x, a_y, a_z)$. This expression resembles a natural extension of the $2d$ toric code to three dimensions. Even in the thermodynamic limit, where the system size becomes arbitrarily large, there are still specific system sizes that yield a unique ground state.

We note that $d_{xyz}$ is always a divisor of $N_{xyz}$. In particular, when $N_{xyz} = 1$, the GSD is necessarily $1$, independent of system size. For $N = 2$ and $a_x = a_y = a_z = 1$, this formula reproduces the familiar $2^3$-fold degeneracy of the standard $\mathbb{Z}_2$ toric code on a three-torus.

A detailed derivation of the GSD formula is provided in Appendix~\ref{Appendix : gTC_GSD}.

\subsection{Local Vertex Excitations}\label{Appendix : gTC_particle}

The Hamiltonian of the $3d$ toric code contains two types of stabilizers: vertex terms and plaquette terms. The excitations corresponding to these stabilizers have distinct topological features. Vertex excitations appear at the endpoints of string operators, while plaquette excitations occur on the boundaries of membrane operators. We will treat these cases separately.

We begin with the particle-like excitations that violate the vertex terms, $S_v = 1$. A generic $Z$-string operator along the $\hat{i}$-direction of length $l$ can be written as
\begin{align}\label{eqn : app_gTC_string}
W_{(\vec{r}_0 - l\hat{i}, \vec{r}_0)} = \prod_{m=0}^{l-1} Z_{\vec{r}_0 - (m + \frac{1}{2})\hat{i}}^{a_i^m b},
\end{align}
where $\hat{i} \in \{\hat{x}, \hat{y}, \hat{z} \}$, $b \in \mathbb{Z}$ is arbitrary, and $\vec{r}_0 = (x_0, y_0, z_0)$ denotes the starting position of the string.

One can readily verify that the operator $W_{(\vec{r}_0 - l\hat{i}, \vec{r}_0)}$ commutes with all stabilizers in the Hamiltonian Eq.~\eqref{eqn : H_gTC}, except for the two vertex terms at the endpoints of the string, namely $S_{\vec{r}_0}$ and $S_{\vec{r}_0 - l\hat{i}}$. Specifically:

\begin{align}
& S_{\vec{r}_0 - d\hat{i}} W = W S_{\vec{r}_0 - d\hat{i}} \quad \text{for } 1 \le d \le l-1, \nonumber\\
& S_{\vec{r}_0} W = \omega^{-b} W S_{\vec{r}_0}, \nonumber\\
& S_{\vec{r}_0 - l\hat{i}} W = \omega^{a_i^l b} W S_{\vec{r}_0 - l\hat{i}}, \nonumber
\end{align}
where $\omega = e^{2\pi i / N}$ is the $N$-th root of unity.

Thus, this operator creates a pair of point-like excitations at the two endpoints of the string, just like in the $\mathbb{Z}_2$ case.

In the $\mathbb{Z}_2$ toric code, such vertex excitations must always be created in pairs and exhibit nontrivial braiding with magnetic flux loops. In contrast, the $\mathbb{Z}_N$ generalization admits a richer spectrum of excitations. Some vertex excitations can even be created individually. This can occur via two fundamentally distinct mechanisms, each corresponding to a different type of topological structure and statistics. We now explain them one by one.

First, consider setting $b = N_i$ in Eq.~\eqref{eqn : app_gTC_string}. Then, for sufficiently large $l$ such that $a_i^{l-1} N_i = N$, the operator becomes trivial at one end of the string, effectively producing a single excitation with vertex charge $\omega^{-N_i}$ [Fig.~\ref{fig_AA_gTC_particle}(b)]. We refer to the exponent of the eigenvalue as the charge of the excitation.

Similarly, one can generate excitations with charges $\omega^{-N_j}$ and $\omega^{-N_k}$ by string operators along the $\hat{j}$ and $\hat{k}$ directions using $b = N_j$ and $b = N_k$, respectively. Superposing such excitations gives a net charge of $\omega^{-pN_x - qN_y - rN_z}$. By the extended Bézout identity, this is always a multiple of $\gcd(N_x, N_y, N_z) = N_{xyz}$. Hence, the unit charge of such single-excitation states is $N_{xyz}$.

Importantly, string operators cannot locally create single excitations with smaller charges $1, 2, \ldots, N_{xyz} - 1$—those can only appear as pairs. Therefore, single vertex excitations with unit charge $N_{xyz}$ are locally creatable, do not braid non-trivially with fluxes, and obey trivial (bosonic) statistics. They are topologically trivial and do not contribute to the long-range entanglement structure of the ground state.

\subsection{Single Topological Point-like Excitations}

\begin{figure}
\includegraphics[width=\columnwidth]{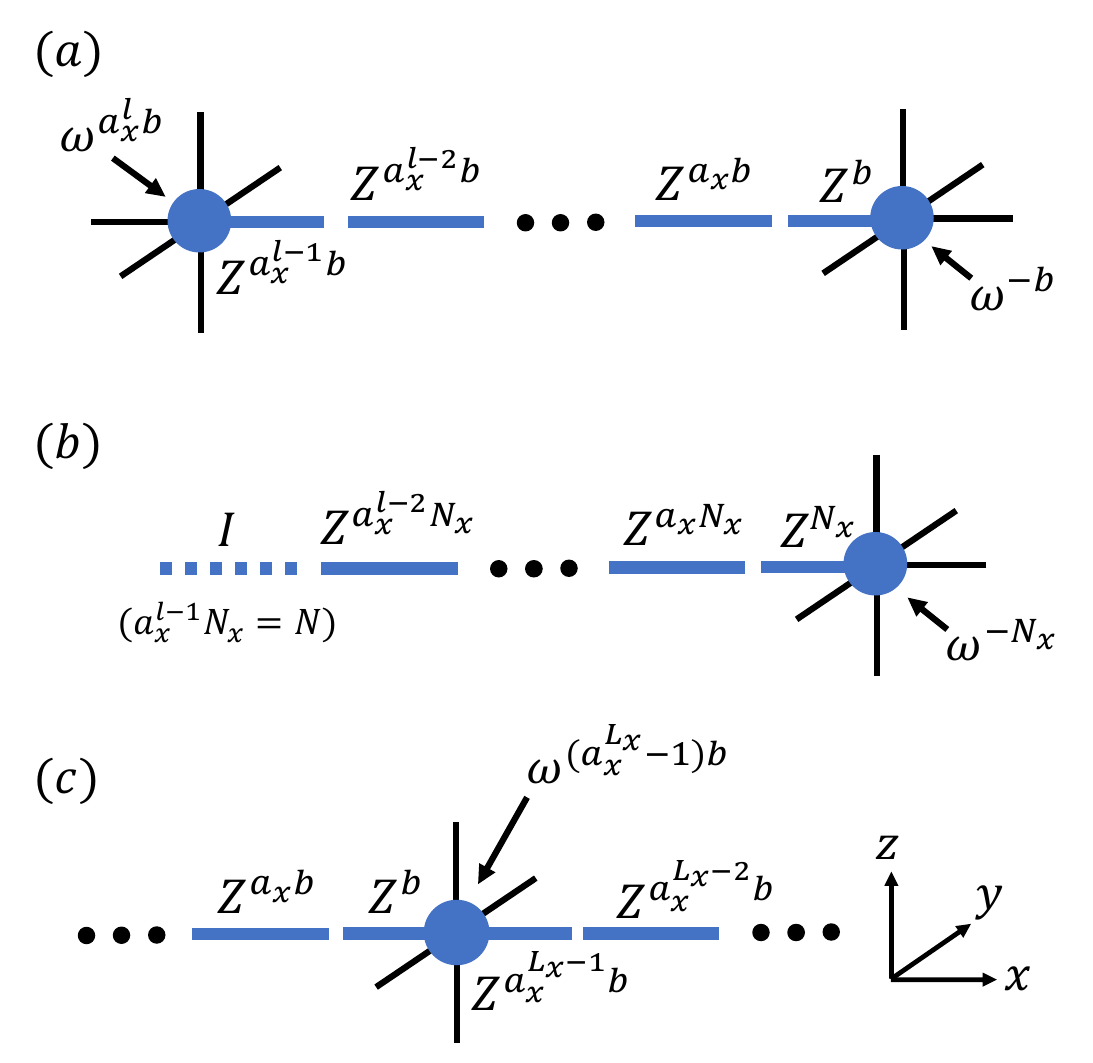}
\caption{\label{fig_AA_gTC_particle}\textbf{Point-like vertex excitations of $\mathbb{Z}_N$ generalized $3d$ toric code.} Charges of each particle are denoted as power of $\omega=e^{2\pi i/N}$. (a) A pair of particles can be created by the straight line operator $W_\mathcal{C}$ Eq.~(\ref{eqn : app_gTC_string}) along $x$-axis. (b) A single particle can be created by the $x$-directional local line operator $W_\mathcal{C}$ when $b=N_i$. (c) A single particle can be created by the $x$-directional non-local line operator $W_\mathcal{C}$ wrapping around the system, when $L_x\ne 0$ mod $M_N(a_x)$.}
\end{figure}

We now consider the case where $l = L_i$, so that the string operator in Eq.~(\ref{eqn : app_gTC_string}) winds around the entire system along the $\hat{i}$-direction. In this setup, the line operator commutes with all terms in the Hamiltonian Eq.~(\ref{eqn : H_gTC}) except for a single vertex term $S_{\vec{r}_0}$, whose eigenvalue is shifted to $\omega^{(a_i^{L_i} - 1) b}$ [Fig.~\ref{fig_AA_gTC_particle} (c)]. Since all charges—i.e., the exponents of $\omega$—are defined modulo $N$, the unit charge of the excitation created by this non-local string operator is not $a_i^{L_i} - 1$, but rather $d_i = \gcd(a_i^{L_i} - 1, N)$. 

By wrapping the non-local string operator around the three independent directions of the torus, one can generate three such types of single vertex excitations. Taking their linear combinations, the minimal unit charge becomes
\begin{equation}
d_{xyz} = \gcd(a_x^{L_x} - 1, a_y^{L_y} - 1, a_z^{L_z} - 1, N),
\end{equation}
implying that it is impossible to create any single vertex excitation with a smaller charge using non-local string operators.

These excitations appear to be individually creatable and freely mobile, resembling the trivial bosons discussed earlier. However, this resemblance is only superficial. Unlike bosons with unit charge $N_{xyz}$, these excitations cannot be created or annihilated by local operators; only global, non-local operations allow for their individual manipulation. On the other hand, their movement can still be achieved via local string operators. 

This asymmetric behavior-globally creatable but locally mobile—implies that these particles are capable of forming nontrivial braiding with loop-like magnetic excitations, and therefore they exhibit fractional statistics. As such, they are topologically nontrivial and should be regarded as the true analogs of anyons in $2d$ systems, albeit in a $3d$ setting.

Interestingly, in the $\mathbb{Z}_2$ toric code with OBCs, it is also possible to create isolated vertex excitations by absorbing one end of the string at the boundary. These excitations are also known to exhibit topological properties. However, in the $\mathbb{Z}_N$ model, such topological single-particle excitations can exist even under PBCs, which is not possible in the $\mathbb{Z}_2$ case.

\subsection{Plaquette Excitations}\label{Appendix : gTC_loop}

\begin{figure*}
\includegraphics[width=2\columnwidth]{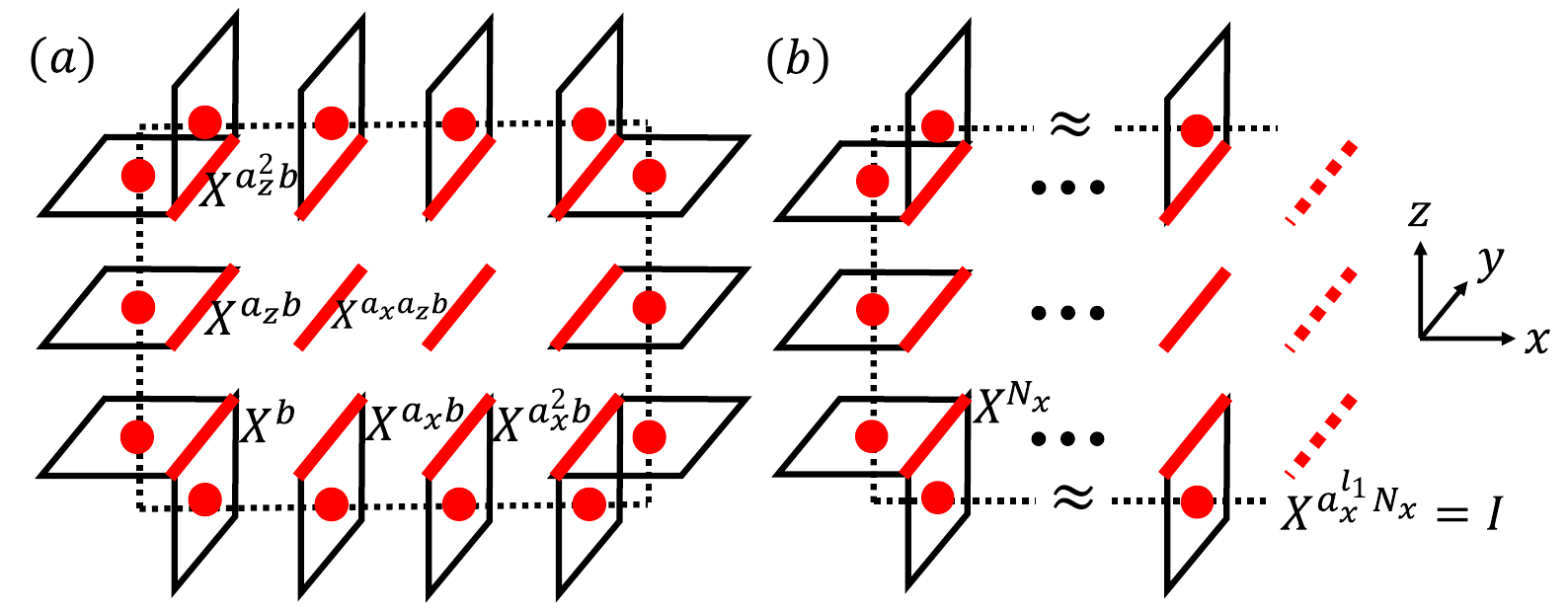}
\caption{\label{fig_AA_gTC_loop}\textbf{Loop-like excitations in the $\mathbb{Z}_N$ generalized $3d$ toric code.} (a) A closed loop excitation created by the rectangular membrane operator $O_\mathcal{M}$ in Eq.~(\ref{eqn : app_gTC_membrane}) along the $xz$-plane. (b) An open-string excitation that avoids plaquette violations along the line $x-x_0=l_1$, created by the rectangular membrane operator $O_\mathcal{M}$ when $b=N_x$.}
\end{figure*}

We now turn to the loop-like excitations generated by violations of $B_p = 1$. Consider the rectangular $X$-membrane operator defined on the $ij$-plane:
\begin{align}\label{eqn : app_gTC_membrane}
O_{\{\Vec{r}_0+\frac{1}{2}\hat{k},\ l_1\hat{i}+l_2\hat{j}\}} = \prod_{m_1=0}^{l_1}\prod_{m_2=0}^{l_2} X_{\Vec{r}_0+\frac{1}{2}\hat{k}+m_1\hat{i}+m_2\hat{j}}^{a_i^{m_1}a_j^{m_2}b}, 
\end{align}
where $\hat{i}, \hat{j}, \hat{k} \in \{\hat{x}, \hat{y}, \hat{z}\}$ are distinct unit vectors, and $b \in \mathbb{Z}$ is an arbitrary integer. The vector $\Vec{r}_0 + \frac{1}{2} \hat{k}$ denotes the corner of the rectangular membrane, while $l_1 \hat{i} + l_2 \hat{j}$ spans its diagonal. This operator commutes with all Hamiltonian terms except the plaquette stabilizers $B_p$ along its boundary, thus creating a closed loop excitation—verified by a direct computation as in~\ref{Appendix : gTC_particle}.

In the original $\mathbb{Z}_2$ model with PBCs, all plaquette excitations form closed loops. These loops can be deformed freely but cannot be broken into open segments. In the $\mathbb{Z}_N$ model, however, counterexamples exist. For instance, setting $b=N_i$ in Eq.~(\ref{eqn : app_gTC_membrane}) renders the membrane edge at $m_1 = l_1$ trivial when $l_1$ is large enough to satisfy $a_i^{l_1} N_i = N$. This results in an open-string excitation with no plaquette violation along $\Vec{r}_0 + \frac{1}{2} \hat{k} + l_1 \hat{i}$ [Fig.~\ref{fig_AA_gTC_loop} (b)].

We define the charge of a string operator as the greatest common divisor of the charges of the plaquette excitations it contains. The existence of open strings implies that if a closed loop excitation has charge divisible by $N_i$, its length along the $\hat{i}$-direction is bounded above. It cannot be stretched arbitrarily and will be truncated once $N_i a_i^{l_i} = N$ holds.

To ensure topological invariance under local perturbations, closed loop-like excitations must remain deformable without degrading into open strings—except via global effects (e.g., through PBCs). Thus, a closed string excitation with charge divisible by any of $N_x$, $N_y$, or $N_z$ can be truncated into an open string under local operations. According to Bézout’s identity, the minimal nontrivial charge for a closed loop-like excitation is bounded by $1, \dots, N_{xyz} - 1$.

This distinction is crucial: closed loop excitations support nontrivial braiding with point-like excitations, while open-string excitations do not. In $3+1$ dimensions, the worldline of a point particle cannot form nontrivial braiding with that of an open string. Even if specific paths of string and membrane operators fail to commute, such paths can be continuously deformed into commuting ones, making them topologically trivial.

One may wonder if global membrane operators—like those wrapping around $T^3$—can truncate closed loops, further restricting their charge to $d_{xyz} = \gcd(a_x^{L_x} - 1, a_y^{L_y} - 1, a_z^{L_z} - 1, N)$. Unfortunately, this is not viable. Unlike vertex excitations (localized at string ends), plaquette excitations appear not only at corners but along the entire boundary of the membrane. Extending a membrane to system size necessarily creates redundant excitations, possibly forming new closed loops. These global effects cannot be eliminated, making it impossible to truncate a loop without introducing additional artifacts.

\subsection{Phase Classification}

In the previous sections, we have established that the ground state degeneracy, the nature of vertex excitations, and the nature of plaquette excitations in the $\mathbb{Z}_N$ $3d$ toric code all depend on the parameters $\{N, a_x, a_y, a_z\}$ in the Hamiltonian and, in some cases, on the system sizes $\{L_x, L_y, L_z\}$. Therefore, this model can realize different topological phases depending on these choices. This section provides a detailed classification of these possible phases.

\subsubsection{Trivial Phase}

When $N_{xyz} = 1$, the model lies in a trivial phase without topological order. Since $d_{xyz}$ divides $N_{xyz}$, we also have $d_{xyz} = 1$ for any system size. In this case, the ground state degeneracy becomes
\begin{align}
    \text{GSD} = d_{xyz}^3 = 1, \quad \forall \ L_x, L_y, L_z \in \mathbb{N}_+,
\end{align}
which resembles a trivial generalization of the $2d$ toric code. As the ground state remains unique at all scales, one does not expect any topological order in such a system.

This conclusion is also supported by the behavior of excitations. As discussed previously, the maximal charge for nontrivial excitations is $N_{xyz} - 1 = 0$, which implies that no nontrivial excitations exist. All vertex excitations reduce to trivial bosons, and any closed loop of plaquette excitations can be truncated into open strings. Therefore, such a system possesses neither topologically degenerate ground states nor nontrivial excitations, confirming its trivial nature.

\subsubsection{Topologically Ordered Phase}

Conversely, when $N_{xyz} \neq 1$, the ground state degeneracy
\[
d_{xyz}^3 = \gcd(a_x^{L_x}-1, a_y^{L_y}-1, a_z^{L_z}-1, N)^3
\]
fluctuates with the system size. Even in the thermodynamic limit, for certain values of $(L_x, L_y, L_z)$, $d_{xyz}$ may still be $1$, leading to a unique ground state. However, the topological phase should not depend sensitively on the exact system size. Therefore, we argue that the system remains in a topologically ordered phase even when $d_{xyz} = 1$.

This can be further understood from the excitation perspective. As previously shown, closed strings formed by plaquette excitations cannot be truncated via non-local mechanisms involving $d_{xyz}$. Hence, all such excitations form genuine closed loops capable of participating in nontrivial topological braiding with vertex excitations.

Let us now consider vertex excitations. The quantity $d_{xyz} - 1$ bounds the charge of vertex excitations that must be created in pairs. When $d_{xyz} = 1$, all vertex excitations can, in principle, be created individually. However, due to $N_{xyz} \neq 1$, many such excitations can only be realized through non-local string operators wrapping around the $3$-torus. When such a non-local string operator intersects a closed plaquette excitation, the corresponding operators fail to commute, and this non-commutativity cannot be removed by continuous deformation. As a result, a nontrivial topological braiding phase and hence fractional statistics appear. Therefore, these individually creatable vertex excitations are indeed topological.

A similar phenomenon arises in the original $\mathbb{Z}_2$ model with OBCs. The boundary can absorb one of the vertex excitations, allowing the other to remain isolated, and it can also truncate closed string excitations into open ones. However, the open boundary itself has the effect of obstructing the smooth deformation of worldlines, so its truncation and absorption of topological excitations do not affect the nontrivial topological braiding between the two types of excitations. Thus, the $\mathbb{Z}_2$ model with OBCs, despite lacking topological ground state degeneracy, remains in a topologically ordered phase. The generalized $\mathbb{Z}_N$ model realizes this mechanism intrinsically under PBCs, for the first time.

\section{Haah's code} 
In this section, we perform the same analysis for Haah's code as we did for the X-cube model and the $3d$ toric code in Sections~\ref{sec : X_cube_model} and \ref{sec : 3d_toric_code}, respectively.

\subsection{\texorpdfstring{$\mathbb{Z}_2$}{Z\_2} Model}\label{sec : Z2_HC}

In the literature~\cite{PhysRevB.100.155137,pretko2020fracton,nandkishore2019fractons}, fracton models are categorized into type-I and type-II phases based on the mobility of their topological excitations. The $\mathbb{Z}_2$ Haah's code belongs to the type-II class, where all topological excitations and their composites are strictly immobile. This sharply contrasts with type-I phases, such as the $\mathbb{Z}_2$ X-cube model, which support not only immobile fractons but also lineons and planons, which exhibit restricted but nonzero mobility. We begin with a brief review of the original $\mathbb{Z}_2$ Haah's code~\cite{PhysRevA.83.042330}.

\begin{figure*}
\includegraphics[width=2\columnwidth]{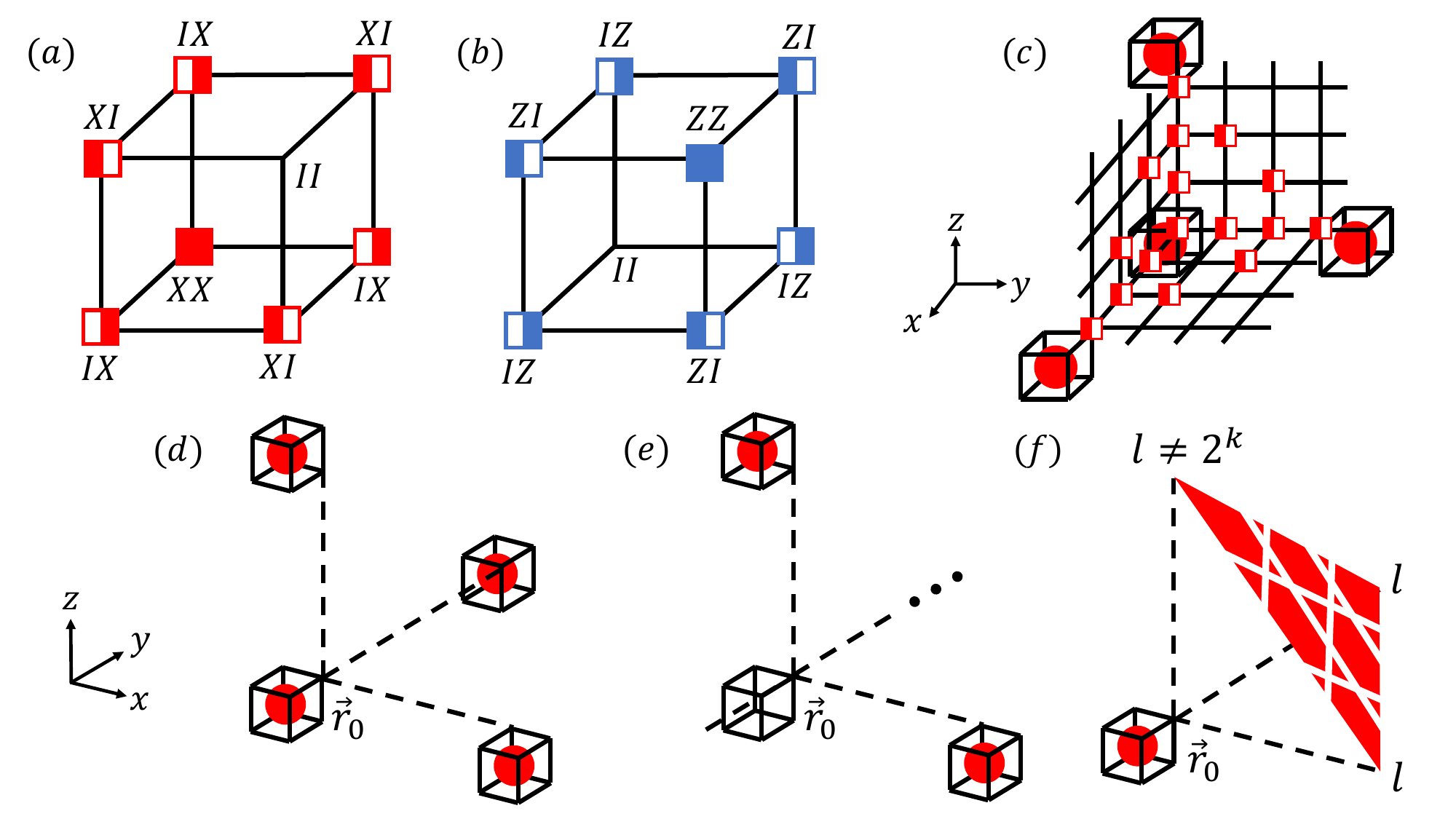}% Here is how to import EPS art
\caption{\label{fig : Z2_HC_new}
\textbf{$\mathbb{Z}_2$ Haah code.} 
A pair of two-dimensional spins lives on each vertex. (a) Pictorial representation of the cube term $S_c^X$. (b) Pictorial representation of the cube term $S_c^Z$. (c) A fracton quadrupole can be created by a tetrahedron operator in Eq.~\eqref{Z2HaahO} with $l=2^k$. In this figure, we have chosen $k=2$ as an example. (d) A side view of the fracton quadrupole in (c) is shown for clarity. The locations of the excitations are represented by red spheres. (e) A dipole of the fractons can be created by a tetrahedron operator in Eq.~\eqref{Z2HaahO} with $l=2^k=L_y$. (f) When $l\neq 2^k$, the surface $(x-x_0) + (y-y_0)+(z-z_0)=l-1$ of the tetrahedron is filled with the fractons except along the stripes $x+y=2^r$, $y+z=2^r$, and $z+x=2^r$ for $2^r<l$. }
\end{figure*}

The $\mathbb{Z}_2$ Haah's code is defined on a cubic lattice with a pair of qubits at each vertex $\vec{r} \in \mathbb{Z}^3$, unlike the toric code and X-cube models, which place spins on links. The Hamiltonian comprises commuting stabilizers formed from Pauli $X$ and $Z$ operators [Fig.~\ref{fig : Z2_HC_new} (a--b)]:
\begin{align}\label{eqn : Z2_HC_Hamiltonian}
H = - \sum_{c \in \mathcal{C}} (S_c^X+S_c^Z),
\end{align}
where $\mathcal{C}$ denotes the set of all unit cubes in the lattice. Each term $S_c^{X}$ or $S_c^{Z}$ is defined for each dual site $c = \vec{r}+(1/2, 1/2, 1/2)$ with $\vec{r} \in \mathbb{Z}^3$ as follows:
\begin{align}
S_{c}^X &=  (X\otimes X)_{c-\frac{1}{2}\hat{x}-\frac{1}{2}\hat{y}-\frac{1}{2}\hat{z}} 
(I\otimes X)_{c+\frac{1}{2}\hat{x}-\frac{1}{2}\hat{y}-\frac{1}{2}\hat{z}} \nonumber \\
&\quad (I\otimes X)_{c-\frac{1}{2}\hat{x}+\frac{1}{2}\hat{y}-\frac{1}{2}\hat{z}} 
(I\otimes X)_{c-\frac{1}{2}\hat{x}-\frac{1}{2}\hat{y}+\frac{1}{2}\hat{z}} \nonumber \\
&\quad (X\otimes I)_{c+\frac{1}{2}\hat{x}+\frac{1}{2}\hat{y}-\frac{1}{2}\hat{z}} 
(X\otimes I)_{c-\frac{1}{2}\hat{x}+\frac{1}{2}\hat{y}+\frac{1}{2}\hat{z}} \nonumber \\
&\quad (X\otimes I)_{c+\frac{1}{2}\hat{x}-\frac{1}{2}\hat{y}+\frac{1}{2}\hat{z}},  \nonumber  
\end{align}
\begin{align}
S_{c}^Z &=  (I\otimes Z)_{c+\frac{1}{2}\hat{x}-\frac{1}{2}\hat{y}-\frac{1}{2}\hat{z}} 
(I\otimes Z)_{c-\frac{1}{2}\hat{x}+\frac{1}{2}\hat{y}-\frac{1}{2}\hat{z}} \nonumber \\
&\quad (I\otimes Z)_{c-\frac{1}{2}\hat{x}-\frac{1}{2}\hat{y}+\frac{1}{2}\hat{z}} 
(Z\otimes I)_{c+\frac{1}{2}\hat{x}+\frac{1}{2}\hat{y}-\frac{1}{2}\hat{z}} \nonumber \\
&\quad (Z\otimes I)_{c-\frac{1}{2}\hat{x}+\frac{1}{2}\hat{y}+\frac{1}{2}\hat{z}} 
(Z\otimes I)_{c+\frac{1}{2}\hat{x}-\frac{1}{2}\hat{y}-\frac{1}{2}\hat{z}} \nonumber \\
&\quad (Z\otimes Z)_{c+\frac{1}{2}\hat{x}+\frac{1}{2}\hat{y}+\frac{1}{2}\hat{z}}. \nonumber    
\end{align}

The ground state satisfies $S_c^{Z}=S_c^{X}=1$ for all $c$.

\begin{widetext}

Fractons arise from violations of these conditions. A hallmark of this model is the absence of local string operators for creating or moving these excitations. Instead, they are generated by a Sierpinski tetrahedron operator $O_{\{\vec{r}_0,l\}}^{XI}$ composed of $X \otimes I$ operators [Fig.~\ref{fig : Z2_HC_new} (c)], defined as:
\begin{align}\label{Z2HaahO}
O_{\{\vec{r}_0,l\}}^{XI} = \prod_{m_1=0}^{l-1} \prod_{m_2=0}^{l-1-m_1} \prod_{m_3=0}^{l-1-m_1-m_2} 
(X\otimes I)_{(x_0+m_1, y_0+m_2, z_0+m_3)}^{\binom{m_1+m_2+m_3}{m_1+m_2}\binom{m_1+m_2}{m_1}},
\end{align}
where $\vec{r}_0=(x_0,y_0,z_0)$ specifies the corner of the tetrahedron, and $l$ is its edge length.

\end{widetext}

This operator commutes with the Hamiltonian within the tetrahedron, but anti-commutes with certain $S_c^Z$ stabilizers on the tetrahedral surface [Fig.~\ref{fig : Z2_HC_new} (f)], specifically along the plane:
\begin{align}
(x-x_0) + (y-y_0) + (z-z_0) = l-1. \nonumber
\end{align}
Interestingly, there exist stripe-like regions where excitations vanish, namely at the intersections of the form $x+y = 2^r$, $y+z = 2^r$, or $z+x = 2^r$ for $2^r < l$, due to the trinomial coefficients vanishing modulo 2. See Appendix~\ref{Appendix : gHC_tetrahedron} for more details.

When $l=2^k$ for integer $k$, the excitations localize to the four corners of the tetrahedron [Fig.~\ref{fig : Z2_HC_new} (d)], forming a quadrupole of fractons~\cite{PhysRevA.83.042330,PhysRevB.109.205125}. Furthermore, if $l=2^k=L_i$ matches the system size in direction $i$, then overlapping excitations at the origin cancel, leaving a dipole [Fig.~\ref{fig : Z2_HC_new} (e)].

Other tetrahedron operators constructed from $I \otimes X$, $Z \otimes I$, and $I \otimes Z$ produce distinct types of excitations, all of which are immobile.

The ground state degeneracy for a system of size $L \times L \times L$ lies within the bounds $2^2 \leq \text{GSD} \leq 2^{4L}$ and fluctuates irregularly with $L$~\cite{PhysRevA.83.042330}. The lower bound arises from global stabilizer identities, such as the product of all $S_c^X$ or all $S_c^Z$ being equal to identity. The upper bound involves more advanced arguments, which we refer readers to in Ref.~\cite{PhysRevA.83.042330}.

\subsection{\texorpdfstring{$\mathbb{Z}_{N>2}$}{} Generalization}\label{sec : gHC}

\begin{figure}
\includegraphics[width=\columnwidth]{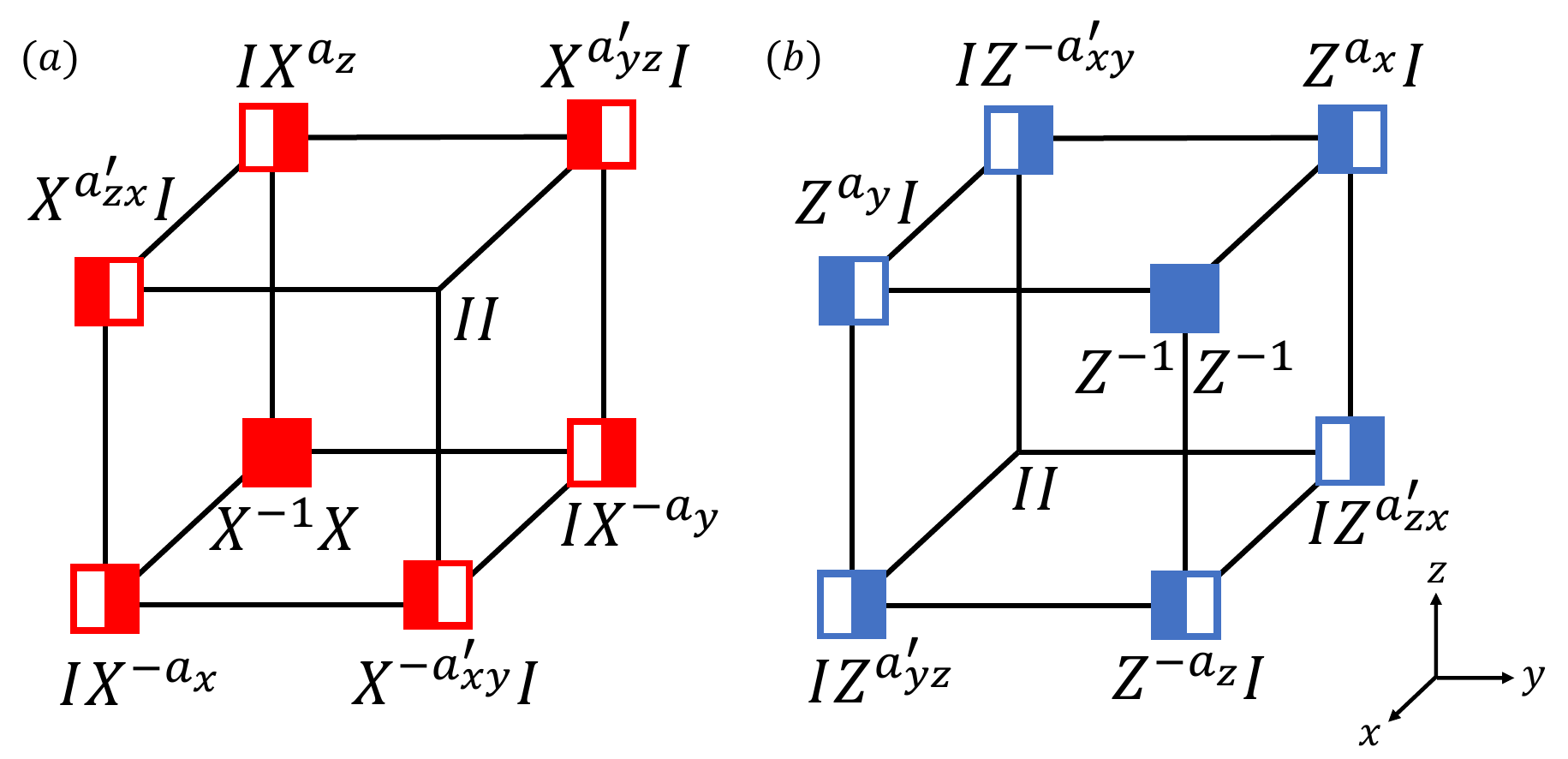}
\caption{\label{fig : gHC_stabilizer_new}
\textbf{Pictorial representations of the $\mathbb{Z}_N$ generalization Eq.~(\ref{eqn : H_gHC}) of Haah's code.} (a) The cube term $S_c^X$. (b) The cube term $S_c^Z$.}
\end{figure}

We now generalize the $\mathbb{Z}_2$ Haah's code in Eq.~\eqref{eqn : Z2_HC_Hamiltonian} to a $\mathbb{Z}_N$ model. For technical simplicity, we consider only the case where $N$ is a prime number, leaving the general non-prime case for future investigation. The Hamiltonian takes the same form as in the $\mathbb{Z}_2$ case:
\begin{align} \label{eqn : H_gHC}
H = - \frac{1}{2}\sum\limits_{c \in \mathcal{C}} (S_c^X + S_c^Z) + \text{h.c.},
\end{align}
where the cube stabilizers are modified with additional $\mathbb{Z}_N$ exponents:

\begin{align}
S_{c}^X &= (X^{-1}\otimes X)_{c-\frac{1}{2}\hat{x}-\frac{1}{2}\hat{y}-\frac{1}{2}\hat{z}} (I\otimes X^{-a_x})_{c+\frac{1}{2}\hat{x}-\frac{1}{2}\hat{y}-\frac{1}{2}\hat{z}} \nonumber \\
&\quad (I\otimes X^{-a_y})_{c-\frac{1}{2}\hat{x}+\frac{1}{2}\hat{y}-\frac{1}{2}\hat{z}} (I\otimes X^{a_z})_{c-\frac{1}{2}\hat{x}-\frac{1}{2}\hat{y}+\frac{1}{2}\hat{z}} \nonumber \\
&\quad (X^{-a_{xy}'}\otimes I)_{c+\frac{1}{2}\hat{x}+\frac{1}{2}\hat{y}-\frac{1}{2}\hat{z}} (X^{a_{yz}'}\otimes I)_{c-\frac{1}{2}\hat{x}+\frac{1}{2}\hat{y}+\frac{1}{2}\hat{z}} \nonumber \\
&\quad (X^{a_{zx}'}\otimes I)_{c+\frac{1}{2}\hat{x}-\frac{1}{2}\hat{y}+\frac{1}{2}\hat{z}}, \nonumber
\end{align}

\begin{align}
S_{c}^Z &= (I\otimes Z^{a_{yz}'})_{c+\frac{1}{2}\hat{x}-\frac{1}{2}\hat{y}-\frac{1}{2}\hat{z}} (I\otimes Z^{a_{zx}'})_{c-\frac{1}{2}\hat{x}+\frac{1}{2}\hat{y}-\frac{1}{2}\hat{z}} \nonumber \\
&\quad (I\otimes Z^{-a_{xy}'})_{c-\frac{1}{2}\hat{x}-\frac{1}{2}\hat{y}+\frac{1}{2}\hat{z}} (Z^{-a_z}\otimes I)_{c+\frac{1}{2}\hat{x}+\frac{1}{2}\hat{y}-\frac{1}{2}\hat{z}} \nonumber \\
&\quad (Z^{a_x}\otimes I)_{c-\frac{1}{2}\hat{x}+\frac{1}{2}\hat{y}+\frac{1}{2}\hat{z}} (Z^{a_y}\otimes I)_{c+\frac{1}{2}\hat{x}-\frac{1}{2}\hat{y}+\frac{1}{2}\hat{z}} \nonumber \\
&\quad (Z^{-1}\otimes Z^{-1})_{c+\frac{1}{2}\hat{x}+\frac{1}{2}\hat{y}+\frac{1}{2}\hat{z}}. \nonumber
\end{align}

Figure~\ref{fig : gHC_stabilizer_new} illustrates the stabilizer structure. Each stabilizer satisfies $(S_c^X)^N = (S_c^Z)^N = 1$, so the ground state satisfies $S_c^X = S_c^Z = 1$ for all $c \in \mathcal{C}$.

\subsection{ground state Degeneracy}\label{sec : gHC_GSD}

\begin{figure}
\includegraphics[width=\columnwidth]{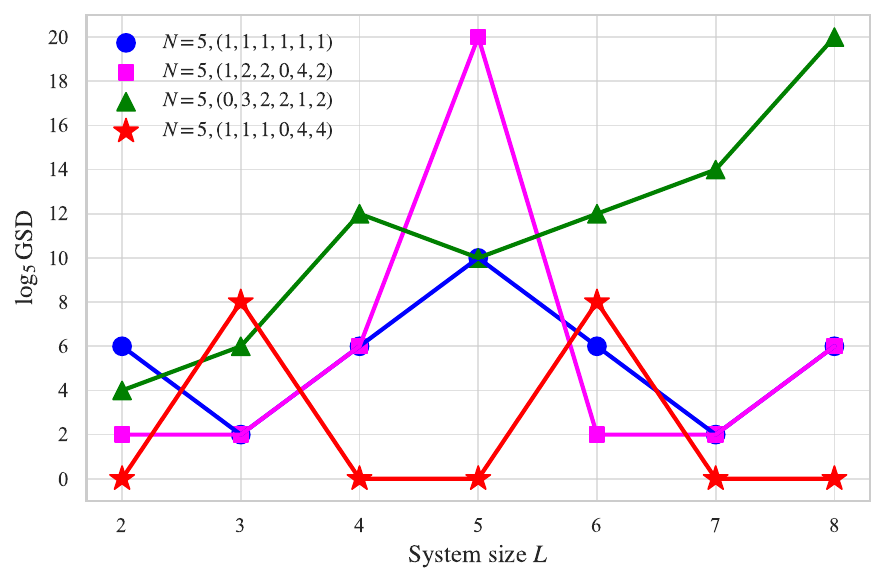}
\caption{\label{fig : gHC_GSD_N5}\textbf{$\log_5\text{GSD}$ plot of the $\mathbb{Z}_5$ generalized Haah's code.} The plot shows $\log_5\text{GSD}$ for several different sets of exponent parameters in Eq.~\eqref{eqn : H_gHC} for system sizes $L\times L\times L$, with $N=5$ and $(a_x,a_y,a_z,a_{xy}',a_{yz}',a_{zx}')$. Blue circles represent $(1,1,1,1,1,1)$; magenta squares: $(1,2,2,0,4,2)$; green triangles: $(0,3,2,2,1,2)$; and red stars: $(1,1,1,0,4,4)$. The largest values of $\log_5\text{GSD}$ at each $L$ occur for those saturating the maximum. In particular, the red stars correspond to a case where the GSD oscillates but peaks at $L=3$.}
\end{figure}

As in the $\mathbb{Z}_2$ case, the ground state degeneracy of Eq.~\eqref{eqn : H_gHC} fluctuates depending on the system size and the exponent parameters appearing in the Hamiltonian. We numerically compute the GSD for the $\mathbb{Z}_5$ model with several choices of exponents and system sizes, following the method of Ref.~\cite{PhysRevA.83.042330} [Fig.~\ref{fig : gHC_GSD_N5}]. As discussed in Section~\ref{sec : gHC_local_cube_excitation}, one can set certain parameters to zero while still preserving fracton order, which is characterized by immobile excitations and a GSD greater than one that depends on the system size.

When the exponent parameters satisfy both $a_x + a_y - a_z - 1 = 0$ and $-a_{xy}' + a_{yz}' + a_{zx}' - 1 = 0$, the model has a guaranteed GSD lower bound of $N^2$, due to the existence of two global stabilizer relations—products over all $S_c^X$ and $S_c^Z$ equal identity—regardless of the system size. We say that any $\mathbb{Z}_N$ Haah's code satisfying these conditions inherits the stabilizer structure of the original $\mathbb{Z}_2$ model described in Section~\ref{sec : Z2_HC}. Similarly, the $\mathbb{Z}_N$ X-cube model and $3d$ toric code with $a_x = a_y = a_z = 1$ have GSDs of $N^{2(L_x + L_y + L_z) - 3}$ and $N^3$, respectively, and can be seen as direct generalizations of their $\mathbb{Z}_2$ counterparts. Notably, setting all $a_i = 1$ maximizes the GSD across all system sizes, which aligns with the general observation that nontrivial $\mathbb{Z}_N$ exponents reduce the number of independent stabilizer constraints.

Interestingly, for Haah's code, we observe that certain parameter sets not satisfying the above conditions can still yield larger GSD at specific system sizes than those that do. By exhaustively scanning all exponent combinations in $\mathbb{Z}_5$, we find that $(1,1,1,0,4,4)$—marked by red stars in Fig.~\ref{fig : gHC_GSD_N5}—is the only such configuration (up to permutations of $a_x, a_y, -a_z$) that achieves maximal GSD at $L=3$. This suggests that certain generalized $\mathbb{Z}_N$ exponents can introduce additional stabilizer relations or symmetries not present in the original model. A full analytic understanding of the GSD behavior and upper bounds in the generalized $\mathbb{Z}_N$ case is left to future work.

\begin{figure}
\includegraphics[width=\columnwidth]{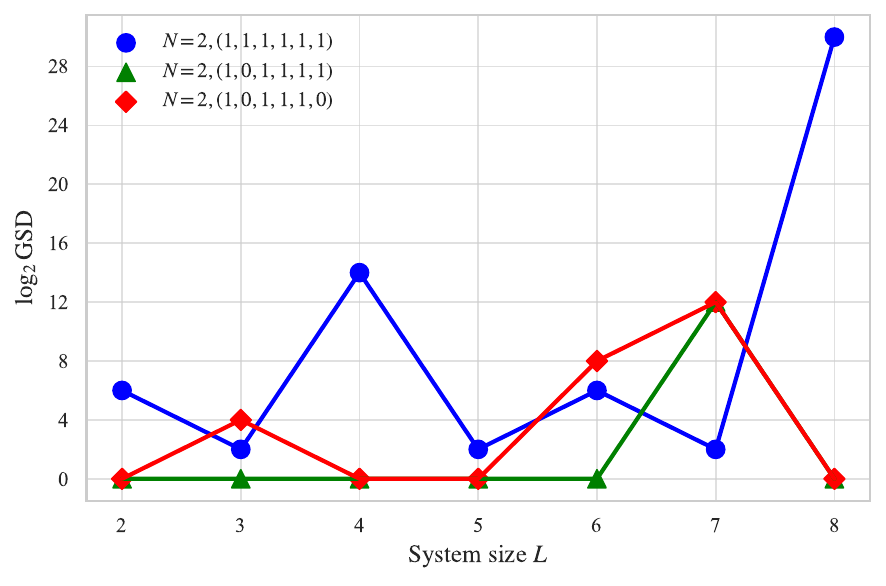}
\caption{\label{fig : gHC_GSD_N2}\textbf{$\log_2\text{GSD}$ plot of the $\mathbb{Z}_2$ generalized Haah's code.} The plot shows $\log_2\text{GSD}$ for several different exponent sets in Eq.~\eqref{eqn : H_gHC} on $L\times L\times L$ lattices with $N=2$ and $(a_x,a_y,a_z,a_{xy}',a_{yz}',a_{zx}')$. Blue circles: original Haah code $(1,1,1,1,1,1)$; green triangles: $(1,0,1,1,1,1)$; red diamonds: $(1,0,1,1,1,0)$. Note that models with zero parameters can exhibit oscillating GSD between 1 and higher values.}
\end{figure}

Moreover, for $N=2$, we observe that setting certain parameters to zero causes the GSD to oscillate between 1 and higher values [Fig.~\ref{fig : gHC_GSD_N2}]. This behavior, first seen in the $\mathbb{Z}_N$ generalized $2d$ toric code~\cite{watanabe2023ground}, now appears in the Haah model as well, showing it is not unique to $N > 2$. The original study~\cite{PhysRevA.83.042330} dismissed such $\mathbb{Z}_2$ models with vanishing parameters as candidates for quantum memory due to the absence of a guaranteed $N^2$ lower bound in their GSD. However, under the broader definition of topological order adopted in~\cite{watanabe2023ground}, these models still exhibit fracton order. A detailed comparison of the zero-parameter $\mathbb{Z}_2$ Haah model with other previously known models will be presented in Appendix~\ref{appendix : Z2_HC_zeros}.

\begin{figure*}
\includegraphics[width=2\columnwidth]{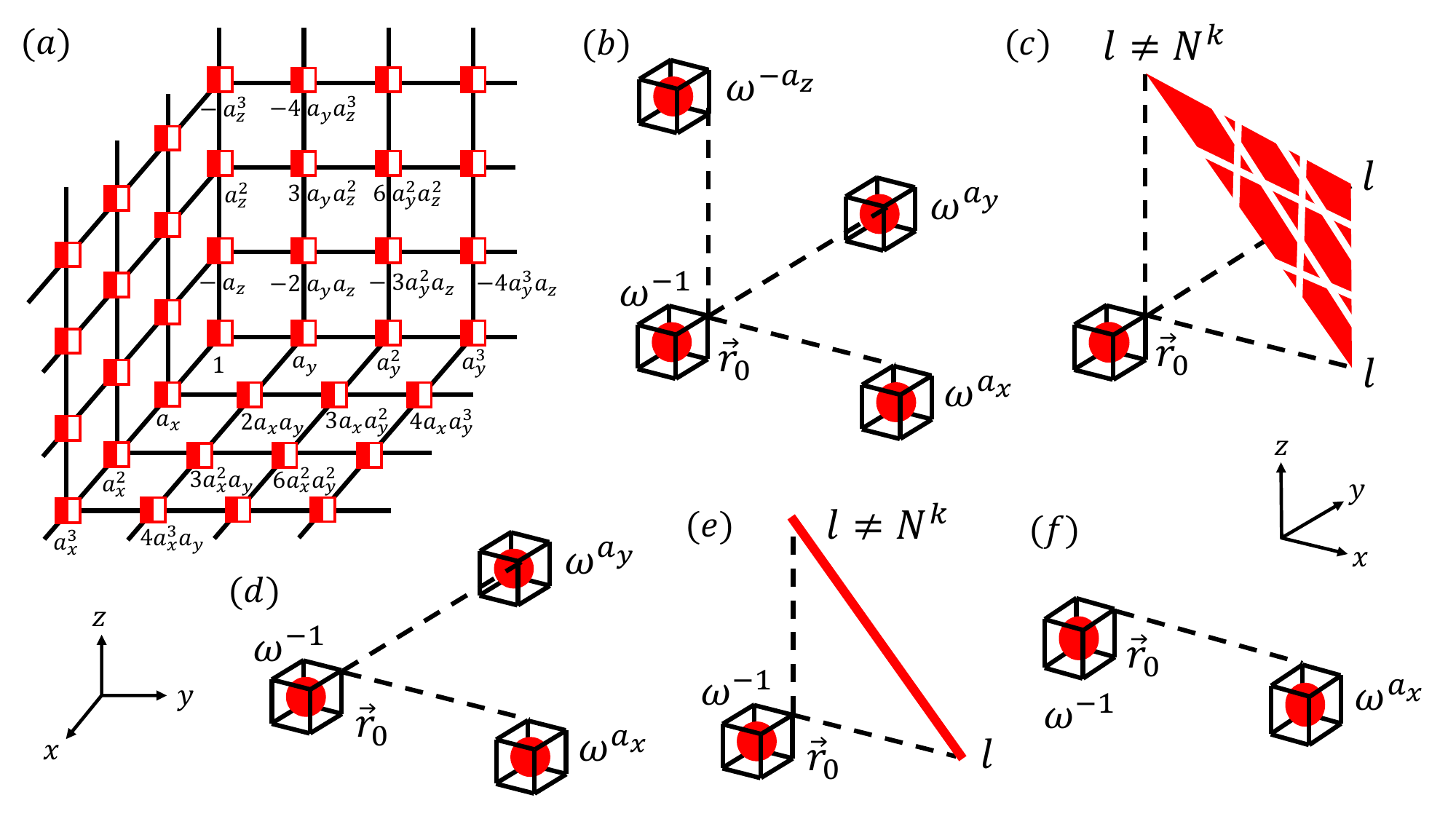}
\caption{\label{fig : gHC_local_excitation}
\textbf{Local cube excitations in the $\mathbb{Z}_N$ generalized Haah's code.} Each vertex hosts a pair of $N$-dimensional spins. (a) A schematic of the tetrahedron operator $O$ [Eq.~\eqref{eqn : gHC_tetrahedron_bulk}], with exponents shown on two selected surfaces for clarity. (b) A fracton quadrupole created by the tetrahedron operator $O$ with $l = N^k$. The eigenvalues of $S_c^Z$ at the locations of the fractons are denoted, where $\omega = e^{2\pi i/N}$. (c) When $l \neq N^k$, fractons are generated across the surface $(x - x_0) + (y - y_0) + (z - z_0) = l - 1$ of the tetrahedron, except along the stripes where $(x - x_0) + (y - y_0) = N^r$, $(y - y_0) + (z - z_0) = N^r$, and $(z - z_0) + (x - x_0) = N^r$ for all $N^r < l$. (d) A fracton tripole created by the triangle operator [Eq.~\eqref{eqn : gHC_zero_triangle}] with $l = N^k$, when $a_z = 0$ and $L_i \neq N^k$ for all $i = x, y$. (e) When $a_y = 0$ and $l \neq N^k$, fractons appear along the line $(x - x_0) + (z - z_0) = l - 1$ on the triangle. (f) When $a_y = a_z = 0$, a lineon pair is created along the $\hat{x}$-direction via the line operator in Eq.~\eqref{eqn : gHC_zero_line}.}
\end{figure*}

\begin{widetext}
    
\subsection{Local Cube Excitations}\label{sec : gHC_local_cube_excitation}

The excitations in the $\mathbb{Z}_N$ Haah's code are generated by a Sierpinski tetrahedron operator, similar to Eq.~\eqref{Z2HaahO}, which violates either $S_c^X = 1$ or $S_c^Z = 1$, as in the $\mathbb{Z}_2$ version. The appropriate generalization is given as follows [Fig.~\ref{fig : gHC_local_excitation} (a)]:
\begin{align}\label{eqn : gHC_tetrahedron_bulk}
O_{\{\vec{r}_0,l\}}^{XI} = \prod_{m_1=0}^{l-1} \prod_{m_2=0}^{l-1-m_1} \prod_{m_3=0}^{l-1-m_1-m_2} (X\otimes I)_{(x_0+m_1,y_0+m_2,z_0+m_3)}^{\binom{m_1+m_2+m_3}{m_1+m_2} \binom{m_1+m_2}{m_1} a_x^{m_1} a_y^{m_2} (-a_z)^{m_3}},
\end{align}
Here, $\vec{r}_0$ denotes the corner of the tetrahedron, and $l$ is its side length. The exponents of $X\otimes I$ in Eq.~\eqref{eqn : gHC_tetrahedron_bulk} are determined by the generalized Pascal’s tetrahedron sum rule for the $\mathbb{Z}_N$ case:
\begin{align}\label{eqn : generalized Pascal's sum rule}
a_x s_{\vec{r}+(0,1,1)} + a_y s_{\vec{r}+(1,0,1)} - a_z s_{\vec{r}+(1,1,0)} = s_{\vec{r}+(1,1,1)} \mod N,
\end{align}
where $s_{\vec{r}}$ is the exponent of $X\otimes I$ at position $\vec{r}$ in Eq.~\eqref{eqn : gHC_tetrahedron_bulk}. See Appendix~\ref{App_sum_rule} for the derivation. Other tetrahedron operators composed of $I\otimes X$, $Z\otimes I$, and $I\otimes Z$ can be similarly constructed, each generating independent excitations. For instance, a tetrahedron operator made of $I\otimes X$ can be derived by requiring a similar sum rule:
\begin{align}
a'_{yz} s_{\vec{r}+(1,0,0)} + a'_{zx} s_{\vec{r}+(0,1,0)} - a'_{xy} s_{\vec{r}+(0,0,1)} = s_{\vec{r}+(1,1,1)} \mod N. \nonumber
\end{align}
In the following, we focus on the $X\otimes I$ tetrahedron operator in Eq.~\eqref{eqn : gHC_tetrahedron_bulk} and analyze the behavior of its excitations.  
\end{widetext} 

We first consider a local tetrahedron operator $O_{\{\vec{r}_0,l\}}^{XI}$ from Eq.~\eqref{eqn : gHC_tetrahedron_bulk}, with size $l \times l \times l$ and $l < L_i$ for all $i = x, y, z$. When $l = N^k$, the operator creates a quadrupole of fractons at the four corners of the tetrahedron [Fig.~\ref{fig : gHC_local_excitation} (b)]. No excitations appear on the surface due to the identity
\begin{align}\label{eqn : gHC_combinatorics}
\binom{p^k}{x} = 0 \mod p
\end{align}
for any integer $1 \leq x \leq p^k - 1$ and prime $p$. A detailed proof is provided in Appendix~\ref{Appendix : gHC_prove}.

In the more general case where $l \neq N^k$ for any integer $k$, the tetrahedron operator produces cube excitations at both the corner $\vec{r}_0$ and along its surface [Fig.~\ref{fig : gHC_local_excitation} (c)], satisfying
\begin{align}
(x - x_0) + (y - y_0) + (z - z_0) = l - 1. \nonumber
\end{align}
Almost every point on the surface violates $S_c^Z = 1$, resulting in fracton excitations. However, some stripe regions remain unaffected when the following conditions are satisfied: $(x - x_0) + (y - y_0) = N^r$, $(y - y_0) + (z - z_0) = N^r$, and $(z - z_0) + (x - x_0) = N^r$ for some $N^r < l$ [Fig.~\ref{fig : gHC_local_excitation} (c)]. Similar features are observed in the original $\mathbb{Z}_2$ Haah's code, as discussed in Appendix~\ref{Appendix : gHC_tetrahedron}.

We also investigate cases where some exponents are set to zero, which do not appear in the original $\mathbb{Z}_2$ model. For example, consider the case where a single parameter is set to zero, such as $a_y = 0$. Then, the generalized exponent sum rule in Eq.~(\ref{eqn : generalized Pascal's sum rule}) reduces to
\begin{equation}
    -a_z s_{\vec{r}+(1,1,0)} + a_x s_{\vec{r}+(0,1,1)} = s_{\vec{r}+(1,1,1)} \mod N. \nonumber
\end{equation}
In this case, the $X \otimes I$ operator with a commuting bulk becomes
\begin{equation}\label{eqn : gHC_zero_triangle}
\prod_{m_1=0}^{l-1} \prod_{m_3=0}^{l-1} (X \otimes I)_{(x_0 + m_1, y_0, z_0 + m_3)}^{\binom{m_1 + m_3}{m_1} a_x^{m_1} (-a_z)^{m_3}},
\end{equation}
where the previous trinomial factors in Eq.~\eqref{eqn : gHC_tetrahedron_bulk} are replaced by binomial coefficients.

Notably, similar forms of the operator Eq.~\eqref{eqn : gHC_zero_triangle} appear for other zero parameter cases, such as $a_x = 0$ or $a_z = 0$, with appropriate spatial rotations. In these cases, fracton excitations appear along the boundary of the Sierpinski triangle operator. For a local triangle of size $l = N^k < L_i$ for all $i = x, y, z$, a fracton tripole is created at the three corners of the triangle [Fig.~\ref{fig : gHC_local_excitation} (d)] where $a_z = 0$. For the more general case where $l \neq N^k$, excitations appear at $\vec{r}_0$ and along the line
\begin{align}
(x - x_0) + (z - z_0) = l - 1, \nonumber
\end{align}
as shown in [Fig.~\ref{fig : gHC_local_excitation} (e)], where $a_y = 0$.

When two parameters are set to zero, such as $a_y = a_z = 0$, the sum rule reduces to
\begin{equation}
    a_x s_{\vec{r}+(0,1,1)} = s_{\vec{r}+(1,1,1)} \mod N. \nonumber
\end{equation}
This yields a one-dimensional $X \otimes I$ string operator along the $x$-axis:
\begin{equation}\label{eqn : gHC_zero_line}
\prod_{m_1=0}^{l-1} (X \otimes I)_{(x_0 + m_1, y_0, z_0)}^{a_x^{m_1}}.
\end{equation}
Since $l$ is arbitrary, this string operator creates a pair of cube excitations that behave as lineons moving along the $\hat{x}$-direction [Fig.~\ref{fig : gHC_local_excitation} (f)]. This is reminiscent of the string operators in the X-cube model Eq.~\eqref{eqn : app_gXC_line}, naturally giving rise to lineon behavior. Moreover, if two of the parameters in $(a_{xy}', a_{yz}', a_{zx}')$ are also set to zero, the excitation becomes a planon, which is mobile in two directions via both $X \otimes I$ and $I \otimes X$ operators.

Finally, when all three parameters $a_x = a_y = a_z = 0$, a single cube excitation can be created or annihilated by a single $X \otimes I$ or $I \otimes Z$ operator. Likewise, when $a_{xy}' = a_{yz}' = a_{zx}' = 0$, a cube excitation can be created or annihilated by $I \otimes X$ or $Z \otimes I$, thus behaving as a free particle.

\subsection{Fracton Multipoles and Quasi-Fractons} \label{sec : multoipoles_quasi_fracton}

We now consider the case where $l = L_i$, such that the tetrahedron operator in Eq.~\eqref{eqn : gHC_tetrahedron_bulk} wraps around the system along the $\hat{i}$-direction. The resulting behavior is similar to that of the $\mathbb{Z}_2$ model, provided at least one of the following conditions is satisfied: (i) $L_i \neq N^{k}$ for all $i = x,y,z$; (ii) $L_x = N^k$ for some integer $k$ and $a_x \equiv 1 \mod N$; (iii) $L_y = N^k$ for some integer $k$ and $a_y \equiv 1 \mod N$; or (iv) $L_z = N^k$ for some integer $k$ and $-a_z \equiv 1 \mod N$. Under condition (i), a tetrahedron operator with $l = N^k$ generates a fracton quadrupole. If one of the conditions (ii)--(iv) holds, then a tetrahedron with $l = L_i = N^k$, where $i \in \{x,y,z\}$, generates a fracton dipole, resembling the behavior observed in the $\mathbb{Z}_2$ model discussed in Section~\ref{sec : Z2_HC}.

\begin{figure}
\includegraphics[width=\columnwidth]{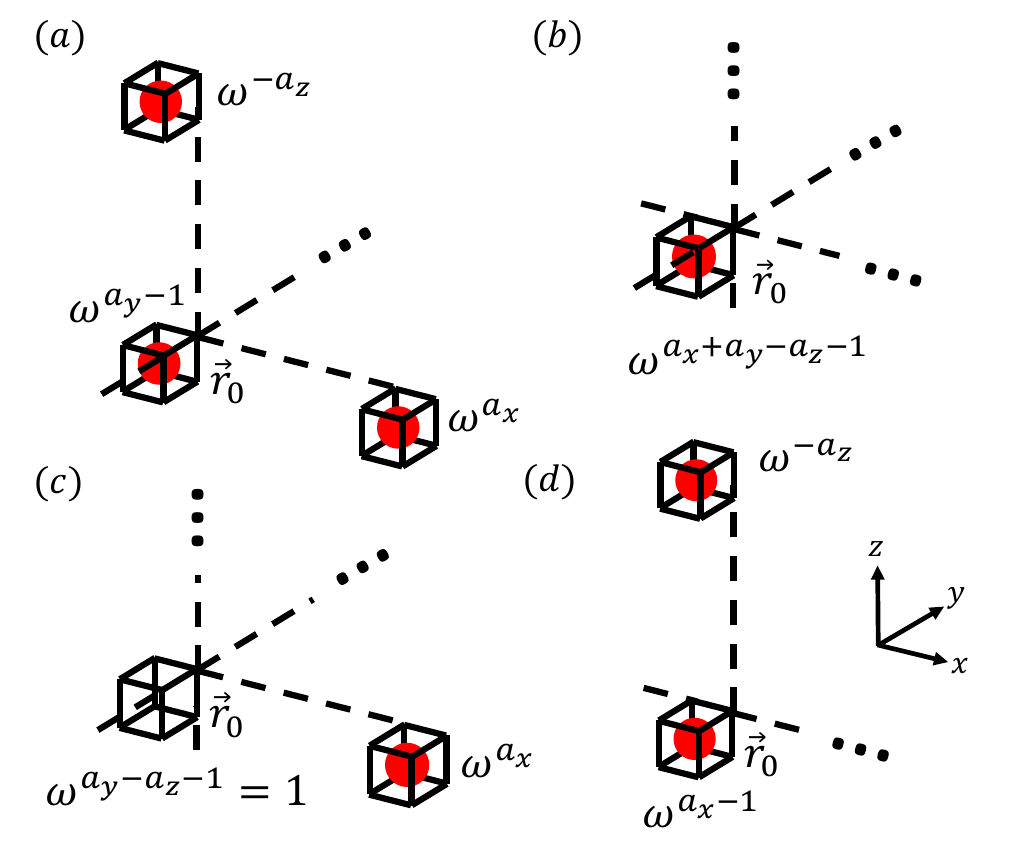}
\caption{\label{fig : gHC_global_excitation} 
\textbf{Fracton multipoles and quasi-fractons in the $\mathbb{Z}_N$ generalized Haah's code via non-local tetrahedron operators $O$ [Eq.~\eqref{eqn : gHC_tetrahedron_bulk}].} The eigenvalue of $S_c^Z$ at each fracton is indicated, where $\omega = e^{2\pi i/N}$. (a) A fracton tripole is generated by a tetrahedron operator $O$ with $l = L_y = N^k$ and $a_y \not\equiv 1 \mod N$. If $a_y \equiv 1 \mod N$, the same operator yields a dipole, as in the $\mathbb{Z}_2$ model. (b) A fracton monopole is created by $O$ with $l = L_y = L_z = N^k$ and $a_y - a_z \not\equiv 1 \mod N$. If $a_y - a_z \equiv 1 \mod N$, the same operator creates a dipole instead. (c) Another fracton monopole arises when $l = L_x = L_y = L_z = N^k$ and $a_x + a_y - a_z \not\equiv 1 \mod N$. If $a_x + a_y - a_z \equiv 1 \mod N$, the operator does not generate any excitation. (d) A fracton dipole is produced by the triangle operator [Eq.~\eqref{eqn : gHC_zero_triangle}] with $l = L_x = N^k$, $L_x \neq L_z$, and $c \not\equiv 1 \mod N$. If $c \equiv 1 \mod N$, a monopole is instead generated.}
\end{figure}

If all of the above conditions are violated, the tetrahedron operator can create novel types of excitations exclusive to the $\mathbb{Z}_N$ generalization in Eq.~\eqref{eqn : H_gHC}, such as fracton tripoles or monopoles, which do not arise in the $\mathbb{Z}_2$ model Eq.~\eqref{eqn : Z2_HC_Hamiltonian}. Consider, for example, the case where $a_y \not\equiv 1 \mod N$ and $L_y = N^k$ for some integer $k$, with $L_y \neq L_x$ and $L_y \neq L_z$. In this case, a tetrahedron operator with $l = L_y$ creates a fracton tripole [Fig.~\ref{fig : gHC_global_excitation} (a)]. Likewise, if $L_x = L_y = L_z = L = N^k$ and $a_x + a_y - a_z \not\equiv 1 \mod N$, a fracton monopole can be created [Fig.~\ref{fig : gHC_global_excitation} (b)]. More generally, fracton monopoles can be generated when exactly one of the following conditions is satisfied: (i) $L_x = L_y = N^k$ and $a_x + a_y \equiv 1 \mod N$; (ii) $L_y = L_z = N^k$ and $a_y - a_z \equiv 1 \mod N$ [Fig.~\ref{fig : gHC_global_excitation} (c)]; or (iii) $L_z = L_x = N^k$ and $-a_z + a_x \equiv 1 \mod N$. These fracton monopoles are termed \emph{quasi-fractons}, in the sense that although they cannot move under any local operator, they can still be created or annihilated by global operators. In other words, their mobility is restricted in the same way as conventional fractons under local operations, but they possess nontrivial global dynamics. A detailed discussion of these conditions and their connection to the excitation structure is provided in Appendix~\ref{Appendix : multipole}.

A triangle operator in Eq.~\eqref{eqn : gHC_zero_triangle} with $l = L_x = N^k$ and $a_y = 0$ can also generate a fracton dipole [Fig.~\ref{fig : gHC_global_excitation} (d)]. Similarly, a quasi-fracton can be created by a triangle operator with $a_y = 0$, using a mechanism analogous to that of the tetrahedron operators shown in Figs.~\ref{fig : gHC_global_excitation} (b) and (c).

The discussion in this section, concerning fracton multipoles generated by non-local $X \otimes I$ and $I \otimes Z$ operators as a function of the parameters $(a_x, a_y, a_z)$, can be analogously extended to cover non-local $I \otimes X$ and $Z \otimes I$ operators governed by the exponents $(a_{xy}', a_{yz}', a_{zx}')$, with only minor structural differences.

\subsection{Phase classification}

In the previous sections, we have demonstrated that both the ground state degeneracy and the nature of cube excitations in the $\mathbb{Z}_N$ Haah's code depend on the parameters $N, a_x, a_y, a_z, a_{xy}', a_{yz}', a_{zx}'$ as well as the system sizes $L_x, L_y, L_z$. In this section, we provide a comprehensive classification of the distinct phases exhibited by the model, drawing analogies with the X-cube model and the $3d$ toric code analyzed earlier.

\subsubsection{Trivial phase}

We define the trivial phase, in analogy to the X-cube model, as the regime in which the system does not exhibit any form of fracton order. Specifically, if at least two of the parameters in the set $(a_x, a_y, a_z)$ are zero, or at least two in $(a_{xy}', a_{yz}', a_{zx}')$ are zero, the model resides in a trivial phase. More precisely, when $a_x = a_y = a_z = 0$ or $a_{xy}' = a_{yz}' = a_{zx}' = 0$, all excitations become trivial bosonic particles that can be created and annihilated via local operators. In such cases, it is straightforward to verify that there exist no nontrivial stabilizer relations among the cube terms in Eq.~\eqref{eqn : H_gHC}. Consequently, the model possesses a unique ground state for all system sizes.

When both sets contain only two vanishing parameters, the excitations reduce to planons, which are constrained to move within two-dimensional planes. Additionally, if two of $(a_x, a_y, a_z)$ vanish while at most one of $(a_{xy}', a_{yz}', a_{zx}')$ vanishes, the resulting excitations are lineons with one-dimensional mobility. The converse is also true: if two of $(a_{xy}', a_{yz}', a_{zx}')$ vanish while at most one of $(a_x, a_y, a_z)$ vanishes, the excitations remain lineons.

\subsubsection{Fracton phase}

The model enters a fracton phase when both $(a_x, a_y, a_z)$ and $(a_{xy}', a_{yz}', a_{zx}')$ contain at least two nonzero parameters. In this regime, the cube excitations become immobile under all local operations, i.e., they are genuine fractons. When either set contains three nonzero entries, the tetrahedron operator defined in Eq.~\eqref{eqn : gHC_tetrahedron_bulk} generates fracton multipoles and extended surface excitations. Conversely, when either set contains exactly two nonzero entries, the triangle operator in Eq.~\eqref{eqn : gHC_zero_triangle} produces fracton multipoles and line-like excitations.

Of particular interest is the emergence of quasi-fractons—effectively behaving as fracton monopoles—that can be created or annihilated via non-local operators under specific conditions, as discussed in Section~\ref{sec : multoipoles_quasi_fracton}. These excitations are immobile under local operations yet exhibit topological characteristics similar to true fractons.

Furthermore, in the fracton phase, the ground state degeneracy exhibits strong oscillatory behavior as a function of both the system size and the choice of nonzero parameters. This size-dependent GSD is a hallmark of fracton topological order.

\section{Conclusions \& Outlooks}
In this work, we have investigated the $\mathbb{Z}_N$ generalizations of several three-dimensional stabilizer models, including the toric code, X-cube model, and Haah’s code, focusing on their ground state degeneracies and topological excitations under periodic boundary conditions. Our findings reveal novel and unexpected behaviors that strongly depend on system size, offering new insights into the interplay between topology, dimensionality, and symmetry in fracton and topologically ordered phases.

In the $\mathbb{Z}_N$ X-cube model, we identified a class of quasi-fracton excitations that are strictly immobile under local operations but can move with relaxed constraints via nonlocal operators. Remarkably, such excitations persist even under PBCs. Unlike the trivial behavior seen under open boundary conditions, the system can still exhibit nontrivial ground state degeneracy in this regime. These observations point toward a broader theoretical understanding of fracton phases and their classification.

For the $3d$ toric code, we observed that even when the ground state is unique, nontrivial closed-string excitations can braid with point-like excitations to form a genuine topologically ordered phase. This behavior, which also survives under PBCs, challenges the conventional view that topological order necessarily implies ground state degeneracy on manifolds with nontrivial topology.

In the case of Haah’s code, we discovered surface and line excitations, along with fracton tripoles and monopoles (quasi-fractons), which are absent in the original $\mathbb{Z}_2$ version. The GSD profiles suggest that, at certain system sizes, $\mathbb{Z}_N$ generalizations may introduce emergent symmetries. Furthermore, we confirmed that the oscillation of the GSD between 1 and higher values is not exclusive to $N > 2$, as similar behavior occurs even in $\mathbb{Z}_2$ models with vanishing parameters.

Collectively, these results broaden our understanding of mobility constraints, excitation structure, and ground state properties in generalized fracton models. Our work opens new directions for the study of $\mathbb{Z}_N$ stabilizer codes in higher dimensions and provides a foundation for their potential realization in quantum simulation platforms. Future research could explore how these findings extend to other classes of stabilizer codes or inform practical applications in quantum information and materials science.

\begin{acknowledgments}
C. L. and G.Y.C. are financially supported by Samsung Science and Technology Foundation under Project Number SSTF-BA2002-05 and SSTF-BA2401-03, the NRF of Korea (Grants No.~RS-2023-00208291, No.~2023M3K5A1094810, No.~2023M3K5A1094813, No.~RS-2024-00410027, No.~RS-2024-00444725, No.~2023R1A2C1006542, No.~2021R1A6A1A10042944) funded by the Korean Government (MSIT), the Air Force Office of Scientific Research under Award No.~FA2386-22-1-4061 and FA23862514026, and Institute of Basic Science under project code IBS-R014-D1. The work of H.W. is supported by JSPS KAKENHI Grant No. JP24K00541. 
\end{acknowledgments}

\appendix
\section{Minimal model of the \texorpdfstring{$2d$}{} toric code}\label{Appendix : 2d gTC}

\begin{figure}
+\includegraphics[width=\columnwidth]{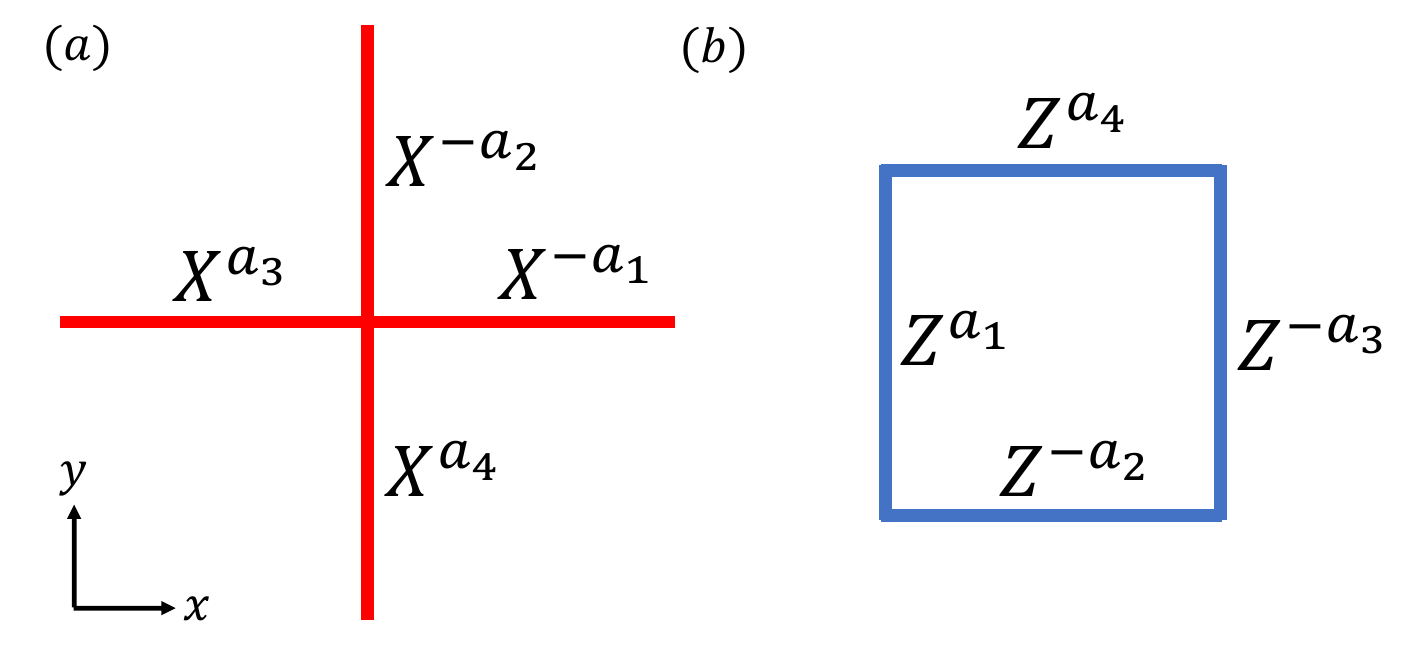}
\caption{\label{fig:4_varialbe_2d_TC}\textbf{Pictorial representation of the $2d$ toric code with four exponents.} (a)The vertex term. (b) The plaquette term}
\end{figure}

The $\mathbb{Z}_N$ generalized $2d$ toric code introduced in Ref.~\cite{watanabe2023ground} has only two exponents. However, the vertex and plaquette terms can each have at most four exponents without losing their commuting property [Fig.~\ref{fig:4_varialbe_2d_TC} (a-b)]. When examining the form of the Wilson line operator depicted in [Fig.~\ref{fig_2d_gTC_new} (b-d)], it becomes clear that the ratio of exponents between neighboring edges plays an important role in the physics, as it crucial for constructing operators that commute with each stabilizer term in the bulk. Therefore, even if we set the two exponents $a_3=a_4=1$, the remained exponents $a_1, a_2$ are sufficient to capture the essential physics, as explained in Section~\ref{sec : 2dgTC}). Thus, we can conclude that the model in Ref.~\cite{watanabe2023ground} represents a minimal model of the $\mathbb{Z}_N$ generalized $2d$ toric code. 

In the main text, we have constructed $3d$ $\mathbb{Z}_N$ generalized models as the minimal models, which consist of minimum exponents, for the same reason for the $2d$ toric code. This approach ensures that the models capture the essential physics while maintaining the simplest form of the operator structure.

\begin{widetext}

\section{ground state Degeneracy}\label{Appendix : mathematical_backgrounds}
In this section, we present the calculation details of the ground state degeneracy for the three $\mathbb{Z}_N$ generalized $3d$ models discussed in this paper.

\subsection{\texorpdfstring{$X$}{X}-cube Model}\label{Appendix : gXC_GSD}
In this section we calculate the ground state degeneracy of the generalized $\mathbb{Z}_N$ X-cube model Eq.~(\ref{eqn : H_gXC}), dependent on the parameter set of $(N,a_x,a_y,a_z,L_x,L_y,L_z)$.

In the model, a single $N$-dimensional spin lives on each link of the cubic lattice with the system size $L_x\times L_y\times L_z$, thus there are total $3L_xL_yL_z$ spins and its Hilbert space dimension is $N^{3L_xL_yL_z}$. 
This calculation can follow the standard procedure, just as we have done before in the $\mathbb{Z}_N$ generalized toric code~\ref{Appendix : gTC_GSD}. 
First, we calculate the dimension of the Hilbert space of the entire system. The system consists of $3L_xL_yL_z$ $\mathbb{Z}_N$ spins, so the dimension of its Hilbert space should be $N^{3L_xL_yL_z}$. 
Then, we select some observables that commute with the Hamiltonian and also commute with each other as candidates for the CSCO. 
It is clear that the cube terms $S_c$ and vertex terms $V_v^{xy}$, $V_v^{yz}$, $V_v^{zx}$ in the Hamiltonian Eq.~(\ref{eqn : H_gXC}) are all suitable candidates to be members of the CSCO. However, we still need to confirm whether these local operators are complete, that is, whether the common eigenvectors of these operators alone can span the entire Hilbert space. In simple terms, we just need to confirm the dimension of the spanned subspace of these operators.

In this model, we have $4L_xL_yL_z$ $N$-fold local operators(contains $3L_xL_yL_z$ vertex terms and $L_xL_yL_z$ cube terms) in the Hamiltonian Eq.~\eqref{eqn : H_gXC}. All these operators are not independent of each other. In fact, there are $L_xL_yL_z$ obvious local constraints on each vertices
\begin{align}
    V_{v}^{xy}V_{v}^{yz}V_{v}^{zx}=1 \quad\text{for}\quad v\in\mathcal{V}.\label{localconstrains}\nonumber
\end{align}
Therefore, without loss of generality, we can retain only the $V_{v}^{yz}$ and $V_{v}^{zx}$ vertex operators in CSCO, while eliminating all the $V_{v}^{xy}$ terms.
For the operators in the remaining set, there are some other global constraints that may make them not mutually independent.
As such, the dimension of the subspace they span could be less than $N^{3L_xL_yL_z}$, thus they may not constitute a complete CSCO.

Let's find the global constraints for the cube terms $S_{(m_1+\frac{1}{2},m_2+\frac{1}{2},m_3+\frac{1}{2})}$. Without loss of generality, multiplying the cube operators in the layer $m_3=m_{3,0}=constant$ together with some specific exponents:
\begin{equation}
\begin{aligned}
    &\prod_{m_1=0}^{L_x-1}\prod_{m_2=0}^{L_y-1}S_{(m_1+\frac{1}{2},m_2+\frac{1}{2},m_{3,0}+\frac{1}{2})}^{n_{xy}a_x^{L_x-1-m_1}a_y^{L_y-1-m_2}}=
    \prod_{m_1=0}^{L_x-1}\left(X_{(m_1+\frac{1}{2},0,m_{3,0})}^{-a_z} X_{(m_1+\frac{1}{2},0,m_{3,0}+1)}\right)^{n_{xy}a_x^{L_x-1-m_1}(a_y^{L_x}-1)}\\
    &\cdot\prod_{m_2=0}^{L_y-1}\left(X_{(0,m_2+\frac{1}{2},m_{3,0})}^{-a_z} X_{(0,m_2+\frac{1}{2},m_{3,0}+1)}\right)^{n_{xy}(a_x^{L_x}-1)a_y^{L_y-1-m_2}}    
    \cdot X_{(0,0,m_{3,0}+\frac{1}{2})}^{-n_{xy}(a_x^{L_x}-1)(a_y^{L_y}-1)}=1,\\
    &\text{for}\quad m_{3,0}=0,...,L_z-1.\label{productBC}
\end{aligned}
\end{equation}

Here we define that
\begin{equation}
n_{xy}\equiv \frac{N}{d_{xy}},\quad d_{xy}\equiv \gcd(a_x^{L_x}-1,a_y^{L_y}-1,N),
\label{ndefinition}\nonumber
\end{equation}
and $n_{yz}$, $n_{zx}$ are defined in the same way.
This is a global constraint in the layer $m_3=m_{3,0}$, which make these $L_xL_y$ cubic operators not independent of each other.
One can choose $S_c$, $c\neq (L_x-\frac{1}{2},L_y-\frac{1}{2},m_{3,0}+\frac{1}{2})$ still independent $N$-fold operators in the CSCO.
Then the $S_{(L_x-\frac{1}{2},L_y-\frac{1}{2},m_{3,0}+\frac{1}{2})}$ becomes $n_{xy}$-fold operators, 
because the eigenvalues of $S_{(L_x-\frac{1}{2},L_y-\frac{1}{2},m_{3,0}+\frac{1}{2})}^{n_{xy}}$ has been confirmed from other $L_xL_y-1$ operators in this layer. Precisely, one can choose each eigenvalue of $S_c$, $c\neq (L_x-\frac{1}{2},L_y-\frac{1}{2},m_{3,0}+\frac{1}{2})$ to any $\mathbb{Z}_N$ value, whereas $S_{(L_x-\frac{1}{2},L_y-\frac{1}{2},m_{3,0}+\frac{1}{2})}$ can  have only $n_{xy}$ kinds of eigenvalues : 
$\omega^{x+d_{xy}k} \  (k=0,1,\cdots ,n_{xy}-1)$, where $x=0,1,\cdots ,d_{xy}-1$ is automatically determined in the constraint Eq.~(\ref{productBC})
From the layer on other directions, we can derive global constraints by the same way,
therefore $S_{(m_{1,0}+\frac{1}{2},L_y-\frac{1}{2},L_z-\frac{1}{2})}$ are $n_{yz}$-fold and $S_{(L_x-\frac{1}{2},m_{2,0}+\frac{1}{2},L_z-\frac{1}{2})}$ are $n_{zx}$-fold, 
for $m_{1,0}=1,...,L_x-1$ and $m_{2,0}=1,...,L_y-1$.

There is a specific cube term $S_{(L_x-\frac{1}{2},L_y-\frac{1}{2},L_z-\frac{1}{2})}$ locating on the corner of this system, which is 
restricted by all three global constraints from different layers on three directions.
Therefore, $S_{(L_x-\frac{1}{2},L_y-\frac{1}{2},L_z-\frac{1}{2})}$ should be $\mathrm{gcd}(n_{xy},n_{yz},n_{zx})$-fold.
All of these cube operators can span a subspace with dimensions

\begin{align}
    N^{L_xL_yL_z-L_x-L_y-L_z+2}n_{yz}^{L_x-1}n_{zx}^{L_y-1}n_{xy}^{L_z-1}\gcd(n_{xy},n_{yz},n_{zx})\nonumber
\end{align}

Now we will find the global constraints for the vertex terms $V_v^{xy}$, $V_v^{yz}$ and $V_v^{zx}$. Because of the local constraints, we only focus on  $V_v^{yz}$ and $V_v^{zx}$, and neglect the $V_v^{xy}$. First, one can multiply all vertex operators on the plane $m_1=m_{1,0}=constant$ together
\begin{equation}
\begin{aligned}
    \prod_{m_2=0}^{L_y-1}\prod_{m_3=0}^{L_z-1}(V_{(m_{1,0},m_2,m_3)}^{yz})^{n_{yz}a_y^{m_2} a_z^{m_3}}=
    \prod_{m_2=0}^{L_y-1}Z_{(m_{1,0},m_2,-\frac{1}{2})}^{n_{yz}a_y^{m_2}(a_z^{L_z}-1)}\prod_{m_3=0}^{L_z-1}Z_{(m_{1,0},-\frac{1}{2},m_3)}^{-n_{yz}(a_y^{L_y}-1)a_z^{m_3}}
    =1, \label{vertexconstraint1}\nonumber
\end{aligned}
\end{equation}
and get a set of global constraints. 
One can choose $V_{v}^{yz}, v\neq (m_{1,0},0,0)$ still $N$-fold independent operators in the CSCO, 
then $V_{(m_{1,0},0,0)}^{yz}$ becomes $n_{yz}$-fold operators, for $m_{1,0}=0,...,L_1-1$.

Both we can multiply all vertex operators on the plane $m_2=m_{2,0}=constant$ together
\begin{equation}
    \begin{aligned}
        \prod_{m_3=0}^{L_z-1}\prod_{m_1=0}^{L_x-1}(V_{(m_1,m_{2,0},m_3)}^{zx})^{n_{zx}a_z^{m_3}a_x^{m_1}}=
        \prod_{m_3=0}^{L_z-1}Z_{(-\frac{1}{2},m_{2,0},m_3)}^{n_{zx}a_z^{m_3}(a_x^{L_x}-1)}\prod_{m_1=0}^{L_x-1}Z_{(m_1,m_{2,0},-\frac{1}{2})}^{-n_{zx}(a_z^{L_z}-1)a_x^{m_1}}
        =1, \label{vertexconstraint2}\nonumber
    \end{aligned}
\end{equation}
and also on the plane $m_3=m_{3,0}=constant$ together, with local constraints on each vertex
\begin{equation}
    \begin{aligned}
        \prod_{m_1=0}^{L_x-1}\prod_{m_2=0}^{L_y-1}(V_{(m_1,m_2,m_{3,0})}^{xy})^{n_{xy}a_x^{m_1}a_y^{m_2}}
        &=\prod_{m_1=0}^{L_x-1}\prod_{m_2=0}^{L_y-1}\left(V_{(m_1,m_2,m_{3,0})}^{yz}V_{(m_1,m_2,m_{3,0})}^{zx}\right)^{-n_{xy}a_x^{m_1}a_y^{m_2}}\\
        &=\prod_{m_1=0}^{L_x-1}Z_{(m_1,-\frac{1}{2},m_{3,0})}^{-n_{xy}a_x^{m_1}(a_y^{L_y}-1)}\prod_{m_2=0}^{L_y-1}Z_{(-\frac{1}{2},m_2,m_{3,0})}^{n_{xy}(a_x^{L_x}-1)a_y^{m_2}}
        =1, \label{vertexconstraint3}\nonumber
    \end{aligned}
\end{equation}
Therefore just choosing all $V_{(0,m_{2,0},0)}^{zx}$ as $n_{zx}$-fold operators is not enough to contain all affection from global constraints,
we should also confirm that every $V_{(0,0,m_{3,0})}^{zx}$ is $n_{xy}$-fold.
As same as what happened in cube terms, there is also a specific vertex operator $V_{(0,0,0)}^{yz}$ which is restricted by two different constraints.
It should be $\gcd(n_{zx},n_{xy})$-fold.

All these vertex operators can span a subspace with dimensions
\begin{align}
    N^{2L_xL_yL_z-L_x-L_y-L_z+1}n_{yz}^{L_x}n_{zx}^{L_y-1}n_{xy}^{L_z-1}\gcd(n_{xy},n_{zx}).\nonumber
\end{align}

In conclusion, all these local terms in the Hamiltonian can span a
\begin{equation}
\begin{aligned}
    D_{\text{local terms}}\equiv
     N^{3L_xL_yL_z-2L_x-2L_y-2L_z+3}n_{yz}^{2L_x-2}n_{zx}^{2L_y-2}n_{xy}^{2L_z-2}n_{yz}\mathrm{gcd}(n_{xy},n_{zx})\gcd(n_{xy},n_{yz},n_{zx})\label{localtermdimension}\nonumber
\end{aligned}
\end{equation}
-dimensional subspace.
Since we can choose all eigenvalues of this operator to be 1 to find the ground state of the Hamiltonian,
the dimension of degenerate ground state subspace is just
\begin{equation}
    \begin{aligned}
        N_{\text{deg}}&=\frac{N^{3L_xL_yL_z}}{D_{\text{local terms}}}
        =\frac{d_{yz}^{2L_x-2}d_{zx}^{2L_y-2}d_{13}^{2L_z-2}N^3}{n_{yz}\gcd(n_{zx},n_{xy})\gcd(n_{yz},n_{zx},n_{xy})}
        =\frac{d_{yz}^{2L_x}d_{zx}^{2L_y}d_{xy}^{2L_z}}{d_{xyz}^3}\\
        &=\frac{\gcd(a_y^{L_y}-1,a_z^{L_z}-1,N)^{2L_x}\gcd(a_x^{L_x}-1,a_z^{L_z}-1,N)^{2L_y}\gcd(a_x^{L_x}-1,a_y^{L_y}-1,N)^{2L_z}}{\gcd(a_x^{L_x}-1,a_y^{L_y}-1,a_z^{L_z}-1,N)^3}.\label{GSD}
    \end{aligned}
\end{equation}
Here $d_{xyz}$ is defined by
\begin{align}
d_{xyz}\equiv \gcd(a_x^{L_x}-1,a_y^{L_y}-1,a_z^{L_z}-1,N).\nonumber
\end{align}

\end{widetext}

Now, we are going to prove the inequality : 
\begin{align}\label{App_gXC_inequality}
    \frac{d_{yz}^{2L_x} d_{zx}^{2L_y} d_{xy}^{2L_z}}{d_{xyz}^{3}}\le N^{2(L_x+L_y+L_z)-3}
\end{align}
We first define the new numbers $d_{i}:=\text{gcd}(a_{i}^{L_i}-1,N)$ for $i=x,y,z$.

\begin{figure}
\includegraphics[width=\columnwidth]{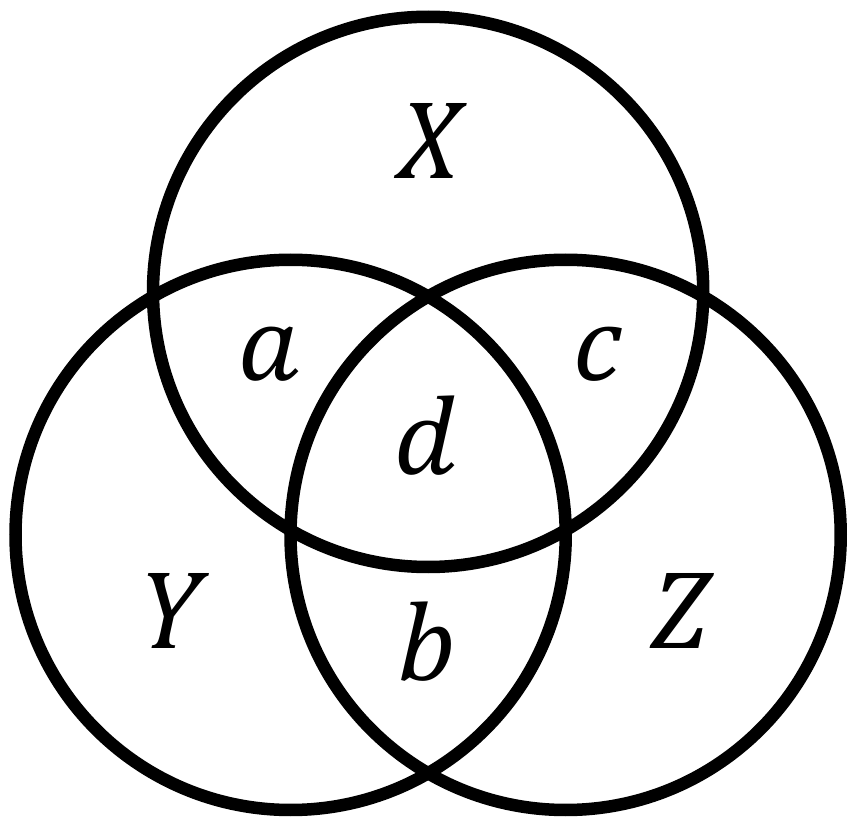}
\caption{\label{fig : gXC_diagram} \textbf{Venn diagram used to prove the inequality Eq.~\ref{App_gXC_inequality}} In the diagram, the three circles represent $d_x$, $d_y$, $d_z$. The product of numbers in the intersection of two different sets correspond to the greatest common divisor of the product all elements in each set.}
\end{figure}

To prove the inequality above, it is convenient to draw a Venn diagram representing the relationship between $d_x$, $d_y$, and $d_z$ [Fig.~\ref{fig : gXC_diagram}]. We then associate each number with the product of elements in the corresponding partial sets, such that $d_x=Xacd$, $d_y=Yabd$ and $d_z=Zbcd$. We introduce a rule stating that product of all numbers in the intersection of two different sets is the greatest common divisor (gcd) of the two numbers represented by those sets. Explicitly, this gives $d_{xy}=ad$, $d_{yz}=bd$, $d_{zx}=cd$, $d_{xyz}=d$. The smallest possible $N$ is given by least common multiple (lcm) of $d_x,d_y,d_z$, leading to $\text{lcm}(d_x,d_y,d_z)$, thus $N=XYZabcd$.
Now substituting the Venn diagram representations of each number into Eq.~(\ref{App_gXC_inequality}), the inequality takes the form :
\begin{align}
    \frac{(bd)^{2L_x}(cd)^{2L_y}(ad)^{2L_z}}{d^{3}}\le (XYZabcd)^{2(L_x+L_y+L_z)-3}.\nonumber
\end{align}
To prove the inequality, we compare the left-hand side with the lower bound of the right-hand side. By setting $X=Y=Z=1$, the expression simplifies to
\begin{align}
    b^{2L_x}c^{2L_y}a^{2L_z}\le (abc)^{2(L_x+L_y+L_z)-3}.\nonumber
\end{align}
The final inequality holds for all system sizes $(L_x,L_y,L_z)$ where each is a natural numbers. The equality condition is met when $a=b=c=1$, which implies that $d_{xyz}=d_{ij}=N$ for all $i,j$. In particular, when $a_x=a_y=a_z=1$, the equality holds, meaning that $N_{\text{deg}}(a_x=a_y=a_z=1)$ reaches the saturated upper bound of $N_{\text{deg}}$ Eq.~\eqref{eqn : gXC_GSD}. 

\begin{widetext}

\subsection{\texorpdfstring{$3d$}{} Toric Code}\label{Appendix : gTC_GSD}
In this section, we calculate the ground state degeneracy of the $\mathbb{Z}_N$ generalized $3d$ toric code Eq.~(\ref{eqn : H_gTC}), dependent on the parameter set of $(N,a_x,a_y,a_z,L_x,L_y,L_z)$. 

In the model, a single $N$-dimensional spin lives on each link of the cubic lattice with the system size $L_x\times L_y\times L_z$, thus there are total $3L_xL_yL_z$ spins and its Hilbert space dimension is $N^{3L_xL_yL_z}$. 
To calculate the degeneracy of energy levels, the general practice is to find a set of observables that commute with the Hamiltonian and with each other, such that their common eigenstates span the entire Hilbert space. At this point, we can label each common eigenstate with the eigenvalues of this set of operators. We refer to this set of operators as a CSCO (complete set of commuting observables).
Obviously, the local terms in the Hamiltonian are a suitable choice, but there are too many of them. In our Hamiltonian Eq.~(\ref{eqn : H_gTC}), there are a total of $3L_xL_yL_z$ plaquette operators $B_p$ and $L_xL_yL_z$ vertex operators $S_v$, and they commute with each other. 
This number greatly exceeds the need to span the $N^{3L_xL_yL_z}$-dimensional Hilbert space, so we should clarify some constraint equations between the operators, to restrict that they are not all independent $N$-fold operators.
When the number of constraint equations is sufficient, the subspace spanned by these local terms will have a dimension less than $N^{3L_xL_yL_z}$, and determining the vectors in it can uniquely specify the energy of the system. 
Only when all the eigenvalues of the local terms are 1, the system is specified in the ground state, so the remaining part of the Hilbert space describes the degenerate ground state subspace, and its dimension is the degeneracy of the ground state.

The $\mathbb{Z}_N$ generalized $3d$ toric code has constraints between vertex operators $S_c$ and between plaquette operators $B_p$. Let's discuss these constraints one by one.

To construct a global constraint about vertex terms, multiply all the vertex terms together with specific exponents $a_x^{m_1}a_y^{m_2}a_z^{m_3}$ to ensure that there are no residual spin operators in the body of the system.
\begin{equation}
    \begin{aligned}
        &\prod_{m_1=0}^{L_x-1}\prod_{m_2=0}^{L_y-1}\prod_{m_3=0}^{L_z-1}S_{(m_1,m_2,m_3)}^{a_x^{m_1}a_y^{m_2}a_z^{m_3}}=\\
        &\prod_{m_2=0}^{L_y-1}\prod_{m_3=0}^{L_z-1}X_{(-\frac{1}{2},m_2,m_3)}^{-(a_x^{L_x}-1)a_y^{m_2}a_z^{m_3}}\prod_{m_3=0}^{L_z-1}\prod_{m_1=0}^{L_x-1}X_{(m_1,-\frac{1}{2},m_3)}^{-(a_y^{L_y}-1)a_z^{m_3}a_x^{m_1}}\prod_{m_1=0}^{L_x-1}\prod_{m_2=0}^{L_y-1}X_{(m_1,m_2,-\frac{1}{2})}^{-(a_z^{L_z}-1)a_x^{m_1}a_y^{m_2}}
    \end{aligned}
\end{equation}

with PBCs, if we want the boundary of the system to also have no residual spin operators, it is necessary to further introduce the parameter $n_{xyz}\equiv \frac{N}{d_{xyz}}$ in the exponent. 
Here $d_{xyz}\equiv \gcd(a_x^{L_x}-1,a_y^{L_y}-1,a_z^{L_z}-1,N)$.
Due to $n_{xyz}(a_x^{L_x}-1)=n_{xyz}(a_y^{L_y}-1)=n_{xyz}(a_z^{L_z}-1)=0\mod N$, we derive the global constraint.
\begin{equation}
    \begin{aligned}
        &\prod_{m_1=0}^{L_1-1}\prod_{m_2=0}^{L_2-1}\prod_{m_3=0}^{L_3-1}S_{(m_1,m_2,m_3)}^{n_{xyz}a_x^{m_1}a_y^{m_2}a_z^{m_3}}=\\
        &\prod_{m_2=0}^{L_2-1}\prod_{m_3=0}^{L_3-1}X_{(-\frac{1}{2},m_2,m_3)}^{-n_{xyz}(a_x^{L_1}-1)a_y^{m_2}a_z^{m_3}}\prod_{m_3=0}^{L_3-1}\prod_{m_1=0}^{L_1-1}X_{(m_1,-\frac{1}{2},m_3)}^{-n_{xyz}(a_y^{L_2}-1)a_z^{m_3}a_x^{m_1}}\prod_{m_1=0}^{L_1-1}\prod_{m_2=0}^{L_2-1}X_{(m_1,m_2,-\frac{1}{2})}^{-n_{xyz}(a_z^{L_3}-1)a_x^{m_1}a_y^{m_2}}=1.
    \end{aligned}\label{vertexProduct}
\end{equation}

One can choose $S_v$, $v\neq (0,0,0)$ still independent $N$-fold operators in the CSCO.
Then the $S_{(0,0,0)}$ becomes $n_{xyz}$-fold operators, because the value of $S_{(0,0,0)}^{n_{xyz}}$ has been confirmed from other $L_xL_yL_z-1$ operators. Precisely, one can choose each eigenvalue of $S_v$, $v\neq (0,0,0)$ to any $\mathbb{Z}_N$ value, whereas $S_{(0,0,0)}$ can  have only $n_{xyz}$ kinds of eigenvalues : $\omega^{x+d_{xyz}k} \  (k=0,1,\cdots ,n_{xyz}-1)$, where $x=0,1,\cdots ,d_{xyz}-1$ is automatically determined from the constraint in Eq.~(\ref{vertexProduct}). Therefore, the subspace dimension that all vertex terms can span is $N^{L_xL_yL_z-1}n_{xyz}$.

Next, we discuss the local cube type constraints of plaquette terms. For each cube, there is a relationship between the six plaquette terms:
\begin{equation}
    \begin{aligned}
        C_{(m_1+\frac{1}{2},m_2+\frac{1}{2},m_3+\frac{1}{2})}\equiv &
        B_{(m_1,m_2+\frac{1}{2},m_3+\frac{1}{2})}^{-a_x}
        B_{(m_1+\frac{1}{2},m_2,m_3+\frac{1}{2})}^{-a_y}
        B_{(m_1+\frac{1}{2},m_2+\frac{1}{2},m_3)}^{-a_z}\\
        &B_{(m_1+1,m_2+\frac{1}{2},m_3+\frac{1}{2})}
        B_{(m_1+\frac{1}{2},m_2+1,m_3+\frac{1}{2})}
        B_{(m_1+\frac{1}{2},m_2+\frac{1}{2},m_3+1)} = 1,  
    \end{aligned}\label{cube_relation}
\end{equation}

where $m_1=0,\cdots,L_x-1$, $m_2=0,\cdots,L_y-1$ and $m_3=0,\cdots,L_z-1$.

Note that we have $L_xL_yL_z$ relationships, but not all of them can reduce the freedom of one plaquette term by $N$.
In fact, if we multiply all local constraints which satisfy that $(m_1,m_2,m_3)\neq(L_x-1,L_y-1,L_z-1)$ together with some exponents
\begin{equation}
    \begin{aligned}
        &\prod_{(m_1,m_2,m_3)\neq(L_x-1,L_y-1,L_z-1)}C_{(m_1+\frac{1}{2},m_2+\frac{1}{2},m_3+\frac{1}{2})}^{-n_{xyz}a_x^{L_x-1-m_1}a_y^{L_y-1-m_2}a_z^{L_z-1-m_3}}=
        (B_{(L_x-1,L_y-\frac{1}{2},L_z-\frac{1}{2})}^{-a_x}
        B_{(L_x-\frac{1}{2},L_y-1,L_z-\frac{1}{2})}^{-a_y}\\
        &B_{(L_x-\frac{1}{2},L_y-\frac{1}{2},L_z-1)}^{-a_z}
        B_{(L_x,L_y-\frac{1}{2},L_z-\frac{1}{2})}^{a_x^{L_x}}B_{(L_x-\frac{1}{2},L_y,L_z-\frac{1}{2})}^{a_y^{L_y}}B_{(L_x-\frac{1}{2},L_y-\frac{1}{2},L_z)}^{a_z^{L_z}})^{n_{xyz}}
        =(B_{(L_x-1,L_y-\frac{1}{2},L_z-\frac{1}{2})}^{-a_x}\\
        &B_{(L_x-\frac{1}{2},L_y-1,L_z-\frac{1}{2})}^{-a_y}B_{(L_x-\frac{1}{2},L_y-\frac{1}{2},L_z-1)}^{-a_z}
        B_{(L_x,L_y-\frac{1}{2},L_z-\frac{1}{2})}B_{(L_x-\frac{1}{2},L_y,L_z-\frac{1}{2})}
        B_{(L_x-\frac{1}{2},L_y-\frac{1}{2},L_z)})^{n_{xyz}}\\
        &=C_{(L_x-\frac{1}{2},L_y-\frac{1}{2},L_z-\frac{1}{2})}^{n_{xyz}}=1,
    \end{aligned}
\end{equation}
where we used the fact that $(a_i^{L_i}-1)n_{xyz}=0\mod N$ for $i=x,y,z$.
This equation means that the local cube constraints $C_{(L_1-\frac{1}{2},L_2-\frac{1}{2},L_3-\frac{1}{2})}=1$ cannot reduce the freedom of stabilizers by $N$ like others,
because before we apply it, the freedom is not $N$ but $n_{xyz}$, already reduced by the other $L_xL_yL_z-1$ local cube constraints.
Thus $C_{(L_1-\frac{1}{2},L_2-\frac{1}{2},L_3-\frac{1}{2})}=1$ reduces the freedom by $n_{xyz}$.

Now we find the global constraints for the plaquette terms, which are independent to the local cube constraints Eq.~(\ref{cube_relation}). There are three independent global constraints for plaquette terms for $x=0$, $y=0$ and $z=0$ planes :

\begin{align}
    \prod_{m_1=0}^{L_x-1}\prod_{m_2=0}^{L_y-1}B_{(m_1+\frac{1}{2},m_2+\frac{1}{2},0)}^{n_{xy}a_x^{L_x-1-m_1}a_y^{L_y-1-m_2}}=\prod_{m_2=0}^{L_y-1}Z_{(0,m_2+\frac{1}{2},0)}^{-n_{xy}(a_x^{L_x}-1)a_y^{L_y-1-m_2}}\prod_{m_1=0}^{L_x-1}Z_{(m_1+\frac{1}{2}0,0)}^{n_{xy}(a_y^{L_y}-1)a_x^{L_x-1-m_1}}=1\label{z0_plane}\\
    \prod_{m_2=0}^{L_y-1}\prod_{m_3=0}^{L_z-1}B_{(0,m_2+\frac{1}{2},m_3+\frac{1}{2})}^{n_{yz}a_y^{L_y-1-m_2}a_z^{L_z-1-m_3}}=\prod_{m_3=0}^{L_z-1}Z_{(0,0,m_3+\frac{1}{2})}^{-n_{yz}(a_y^{L_y}-1)a_z^{L_z-1-m_3}}\prod_{m_2=0}^{L_y-1}Z_{(0,m_2+\frac{1}{2},0)}^{n_{yz}(a_z^{L_z}-1)a_y^{L_y-1-m_2}}=1\label{x0_plane}\\
    \prod_{m_3=0}^{L_z-1}\prod_{m_1=0}^{L_x-1}B_{(m_1+\frac{1}{2},0,m_3+\frac{1}{2})}^{n_{zx}a_z^{L_z-1-m_3}a_x^{L_x-1-m_1}}=\prod_{m_1=0}^{L_x-1}Z_{(m_1+\frac{1}{2},0,0)}^{-n_{zx}(a_z^{L_z}-1)a_x^{L_x-1-m_1}}\prod_{m_3=0}^{L_z-1}Z_{(0,0,m_3+\frac{1}{2})}^{n_{zx}(a_x^{L_x}-1)a_z^{L_z-1-m_3}}=1\label{y0_plane}
\end{align}

where $n_{ij}\equiv \frac{N}{d_{ij}}$, $d_{ij}=\gcd(a_i^{L_i}-1, a_j^{L_j}-1,N)$ and $(i,j)$ is an ordered pair, which belongs to $\{ (x,y), (y,z), (z,x) \}$.
These three relations reduce degree of freedom of three plaquettes : $B_{(0,L_y-\frac{1}{2},L_z-\frac{1}{2})}$ to $n_{yz}$-fold, $B_{(L_x-\frac{1}{2},0,L_z-\frac{1}{2})}$ to $n_{zx}$-fold, and $B_{(L_x-\frac{1}{2},L_y-\frac{1}{2},0)}$ to $n_{xy}$-fold. Yet, we will see that the three plaquette terms already have the freedom less than $N$-fold due to the local cube relations Eq.~\eqref{cube_relation} before we apply the plane constraints Eq.~\eqref{x0_plane}, Eq.~\eqref{y0_plane}, and Eq.~\eqref{z0_plane}.

Before to move on, let's clarify the point that there are many other planes in the lattice that are parallel to these three planes, and they have also similar global constraints of plaquette terms. However, they are not independent relations to these three plane constraints and local cube constraints. For example, the global constraint along $z=1$ plane becomes 

\begin{align}
    \prod_{m_1=0}^{L_x-1}\prod_{m_2=0}^{L_y-1}B_{(m_1+\frac{1}{2},m_2+\frac{1}{2},1)}^{n_{xy}a_x^{L_x-1-m_1}a_y^{L_y-1-m_2}}=\prod_{m_1=0}^{L_x-1}\prod_{m_2=0}^{L_y-1}B_{(m_1+\frac{1}{2},m_2+\frac{1}{2},0)}^{n_{xy}a_x^{L_x-1-m_1}a_y^{L_y-1-m_2}a_z}\prod_{m_1=0}^{L_x-1}\prod_{m_2=0}^{L_y-1}C_{(m_1+\frac{1}{2},m_2+\frac{1}{2},\frac{1}{2})}^{n_{xy}a_x^{L_x-1-m_1}a_y^{L_y-1-m_2}} =1,\nonumber
\end{align}

Therefore, only the three planes(which we picked as $x=0$, $y=0$, and $z=0$) can have independent relations from the local cube constraints.

Products of all local cube constraints with proper exponents becomes 

\begin{align}
    \prod_{m_1=0}^{L_x-1}\prod_{m_2=0}^{L_y-1}\prod_{m_3=0}^{L_z-1} & C_{(m_1+\frac{1}{2},m_2+\frac{1}{2},m_3+\frac{1}{2})}^{{n'} a_x^{L_x-1-m_1}a_y^{L_y-1-m_2}a_z^{L_z-1-m_3}}=\prod_{m_1=0}^{L_x-1}\prod_{m_2=0}^{L_y-1}B_{(m_1+\frac{1}{2},m_2+\frac{1}{2},0)}^{{-n'}(a_z^{L_z}-1)a_x^{L_x-1-m_1}a_y^{L_y-1-m_2}} \nonumber \\  &\prod_{m_2=0}^{L_y-1}\prod_{m_3=0}^{L_z-1}B_{(0,m_2+\frac{1}{2},m_3+\frac{1}{2})}^{{-n'}(a_x^{L_x}-1)a_y^{L_y-1-m_2}a_z^{L_z-1-m_3}}  \prod_{m_3=0}^{L_z-1}\prod_{m_1=0}^{L_x-1}B_{(m_1+\frac{1}{2},0,m_3+\frac{1}{2})}^{{-n'}(a_y^{L_y}-1)a_z^{L_z-1-m_3}a_x^{L_x-1-m_1}}=1. \label{cube_plane_relation}
\end{align}

When we put $n'=-n_{xy}$, then it becomes the $z=0$ plane constraint Eq.~(\ref{z0_plane}) with additional exponent $a_z^{L_z}-1$. 

\begin{align}
    \prod_{m_1=0}^{L_x-1}\prod_{m_2=0}^{L_y-1}\prod_{m_3=0}^{L_z-1} C_{(m_1+\frac{1}{2},m_2+\frac{1}{2},m_3+\frac{1}{2})}^{{-n_{xy}} a_x^{L_x-1-m_1}a_y^{L_y-1-m_2}a_z^{L_z-1-m_3}}=\prod_{m_1=0}^{L_x-1}\prod_{m_2=0}^{L_y-1}B_{(m_1+\frac{1}{2},m_2+\frac{1}{2},0)}^{{n_{xy}}(a_z^{L_z}-1)a_x^{L_x-1-m_1}a_y^{L_y-1-m_2}}=1 \nonumber
\end{align}

Bezout's identity~\ref{Bezout_identity} states that integer $\alpha$ and $\beta$ exist such as

\begin{align}
    \alpha d_{xy}+ \beta (a_z^{L_z}-1)=\gcd(d_{xy},a_z^{L_z}-1)=\gcd(a_x^{L_x}-1,a_y^{L_y}-1,a_z^{L_z}-1,N)=d_{xyz}, \nonumber
\end{align}

which indicates that when we put $n'=-n_{xy}\beta$,

\begin{align}
    \prod_{m_1=0}^{L_x-1}\prod_{m_2=0}^{L_y-1}\prod_{m_3=0}^{L_z-1} C_{(m_1+\frac{1}{2},m_2+\frac{1}{2},m_3+\frac{1}{2})}^{{-n_{xy}\beta} a_x^{L_x-1-m_1}a_y^{L_y-1-m_2}a_z^{L_z-1-m_3}}=\left(\prod_{m_1=0}^{L_x-1}\prod_{m_2=0}^{L_y-1}B_{(m_1+\frac{1}{2},m_2+\frac{1}{2},0)}^{n_{xy}a_x^{L_x-1-m_1}a_y^{L_y-1-m_2}}\right)^{\beta (a_z^{L_z}-1)} \nonumber \\
    =\left(\prod_{m_1=0}^{L_x-1}\prod_{m_2=0}^{L_y-1}B_{(m_1+\frac{1}{2},m_2+\frac{1}{2},0)}^{n_{xy}a_x^{L_x-1-m_1}a_y^{L_y-1-m_2}}\right)^{\alpha d_{xy}+\beta (a_z^{L_z}-1)} =\left(\prod_{m_1=0}^{L_x-1}\prod_{m_2=0}^{L_y-1}B_{(m_1+\frac{1}{2},m_2+\frac{1}{2},0)}^{n_{xy}a_x^{L_x-1-m_1}a_y^{L_y-1-m_2}}\right)^{d_{xyz}}=1   \nonumber
\end{align}
Therefore, before we apply the $z=0$ plane constraint Eq.~(\ref{z0_plane}), $B_{(L_x-\frac{1}{2}, L_y-\frac{1}{2},0)}$ already has automatically reduced freedom $n_{xy}d_{xyz}$, divisor of $N$ determined by the local cube constraints. Then the $z=0$ plane constraint reduces the freedom of $B_{(L_x-\frac{1}{2}, L_y-\frac{1}{2},0)}$ from $n_{xy}d_{xyz}$ to $n_{xy}$-fold, $1/d_{xyz}$ times smaller. Under the same arguments, each $x=0$ plane constraint Eq.~(\ref{x0_plane}) and $y=0$ plane constraint Eq.~(\ref{y0_plane}) also reduces the freedom of plaquette terms by $d_{xyz}$.

The dimension of total Hilbert space  is known as $N^{3L_xL_yL_z}$, and the freedom of stabilizers is 
\begin{align}
    D_{\text{local terms}}&=N^{L_xL_yL_z-1}n_{xyz}\frac{N^{3L_xL_yL_z}}{N^{L_xL_yL_z-1}n_{xyz}d_{xyz}^3}=\frac{N^{3L_xL_yL_z}}{d_{xyz}^3},\nonumber
\end{align}

therefore the ground state degeneracy could be calculated as
\begin{align}
    N_{\text{deg}}=\frac{N^{3L_xL_yL_z}}{D_{\text{local terms}}}=d_{xyz}^3.
\end{align}
Since $d_{xyz}= \gcd(a_x^{L_x}-1,a_y^{L_y}-1,a_z^{L_z}-1,N)\le N$, $N_{\text{deg}}\le N^3=N_{\text{deg}}(a_x=a_y=a_z=1)$.
    
\end{widetext}
\section{Haah's code}\label{Appendix : gHC}

\subsection{Derivation of the exponent sum rule Eq.~(\ref{eqn : generalized Pascal's sum rule})}\label{App_sum_rule}
In this section, we derive the $\mathbb{Z}_N$ generalized Pascal's tetrahedron sum rule Eq.~(\ref{eqn : generalized Pascal's sum rule}).
To begin, We extract $Z\otimes I$ component from $S_c^Z$, which is centered at the dual lattice $\vec{R}=\vec{r}+(1/2,1/2,1/2)$, 
\begin{align}
    Z^{a_x}_{\vec{r}+(0,1,1)}Z^{a_y}_{\vec{r}+(1,0,1)}Z^{-a_z}_{\vec{r}+(1,1,0)}Z^{-1}_{\vec{r}+(1,1,1)} ,\nonumber
\end{align}
where we have replaced $Z\otimes I$ with $Z$ for brevity.

Consequently, the commuting relation between $X^{s_{\vec{r}+(1,0,1)}}X^{s_{\vec{r}+(1,1,0)}}X^{s_{\vec{r}+(0,1,1)}}X^{s_{\vec{r}+(1,1,1)}}$ in $\prod_{\vec{r}} (X\otimes I)^{s_{\vec{r}}}$ and $S_c^Z$ becomes
\begin{align}
    a_xs_{\vec{r}+(0,1,1)}+a_ys_{\vec{r}+(1,0,1)}-a_zs_{\vec{r}+(1,1,0)}=s_{\vec{r}+(1,1,1)} \text{ mod } N,\nonumber
\end{align}
which is exactly the sum rule given in Eq.~(\ref{eqn : generalized Pascal's sum rule}).

\begin{widetext}
    \subsection{Proof of \texorpdfstring{${p^k \choose x}=0$}{} for prime \texorpdfstring{$p$}{}}\label{Appendix : gHC_prove}
In this section we prove the core property Eq.~(\ref{eqn : gHC_combinatorics}) for the fracton multipole in the $\mathbb{Z}_N$ generalized Haah's code.
\begin{equation}
    {p^k \choose x}=\frac{p^k!}{x!(p^k-x)!}=\frac{{p^k}{(p^k-1)!}}{x(x-1)!(p^k-1-(x-1))!}\rightarrow x{p^k \choose x}=p^k{p^k-1 \choose x-1}\nonumber
\end{equation}
\end{widetext}
For $1\le {x}\le {p^k-1}$, $p$ divides $x$ at most $k-1$ times, and thus ${p^k \choose x}=0$ mod $p$. This it not true for a composite integer $N=\prod {p_i}^{r_i}$, since $x$ can be of the form $N^{k-1}p_i$, which means we cannot generally confirm ${N^k \choose x}=0$ mod $N$.

\subsection{Details on the excitation of tetrahedron Eq.~(\ref{eqn : gHC_tetrahedron_bulk})}\label{Appendix : gHC_tetrahedron}

In this section, we provide a detailed explanation of the structure of the trinomial exponents in the tetrahedron operator given by Eq.~(\ref{eqn : gHC_tetrahedron_bulk}). We will verify directly that this bulk commutes with $S_c^Z$ and also demonstrate how it violates the condition $S_c^Z=1$ at the vertex and on the base plane. 

\begin{figure}
\includegraphics[width=\columnwidth]{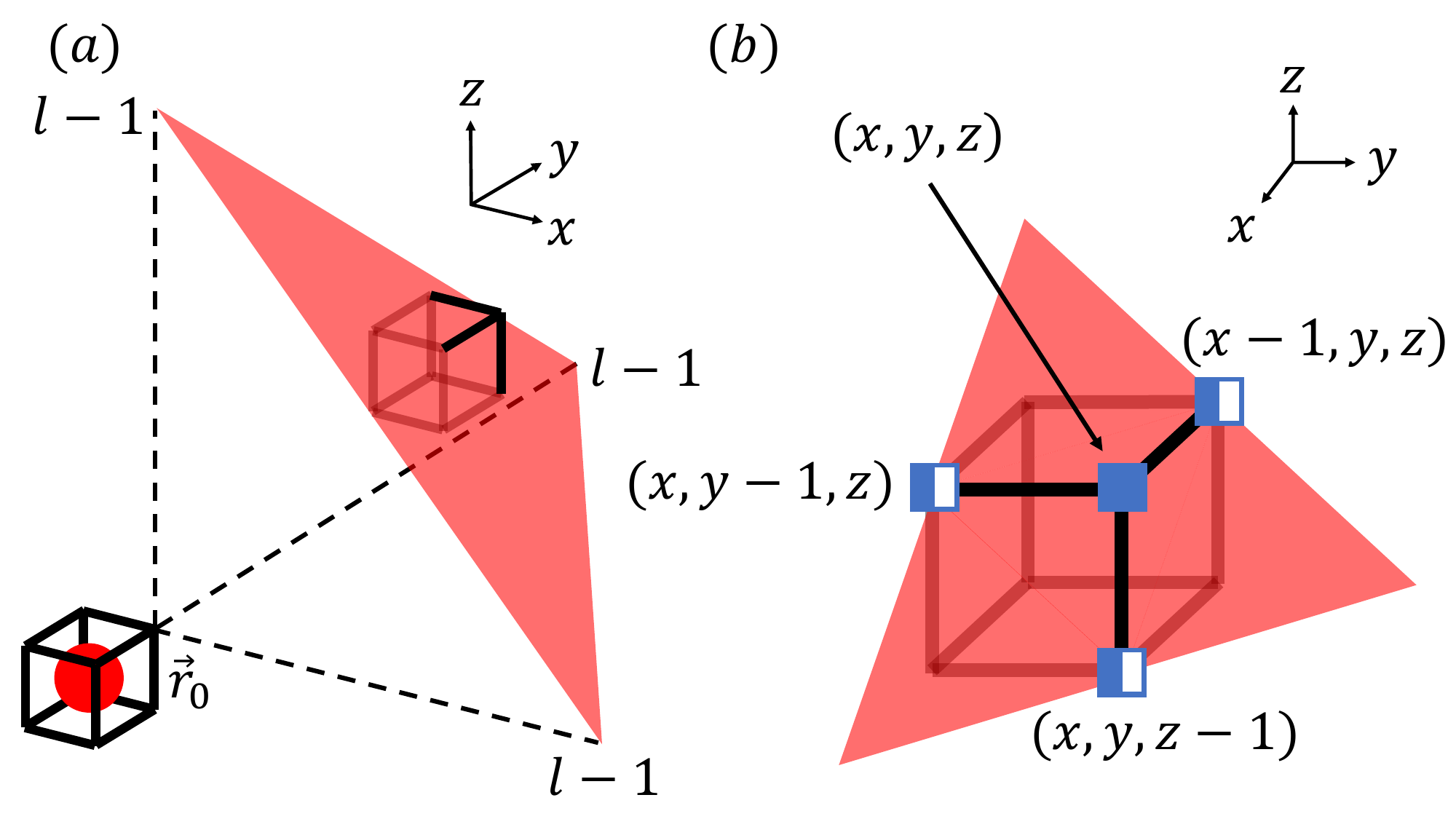}
\caption{\label{fig : gHC_tetrahedron_surface} \textbf{Pictorial representation of the fracton excitations of the tetrahedron operator Eq.~(\ref{eqn : gHC_tetrahedron_bulk}). } Translucent red surface depicts the plane $(x-x_0)+(y-y_0)+(z-z_0)=l-1$. (a) A fracton at the vertex $\vec{r}_0$ and surface excitation violating $S_c^Z=1$ of the cube terms straddling on the plane $(x-x_0)+(y-y_0)+(z-z_0)=l-1$. (b) Zoom-in of the single straddling $S_c^Z$ cube term, whose three vertices having $Z\otimes I$ operator are exactly on the surface $(x-x_0)+(y-y_0)+(z-z_0)=l-1$, while the vertex at $(x,y,z)$ having $Z^{-1} \otimes Z^{-1}$ is on the next layer $(x-x_0)+(y-y_0)+(z-z_0)=l$ }
\end{figure}

We can rewrite the trinomial coefficients in the following convenient form as  
\begin{align}
    \binom{x+y+z}{x+y}\binom{x+y}{x}=\binom{x+y+z}{x \ |\ y\ |\ z}\nonumber.
\end{align}
This expression emphasizes that it represents the number of ways to divide $x+y+z$ into three groups with sizes $x$, $y$, and $z$, respectively.  
We can now express the properties of trinomial factors as follows :
\begin{align}
    \binom{x+y+z}{x\ |\ y\ |\ z} & =\binom{x+y+z-1}{x-1\ |\ y\ |\ z}  \nonumber \\
    & +\binom{x+y+z-1}{x\ |\ y-1\ |\ z}+\binom{x+y+z-1}{x\ | \ y\ |\ z-1}. \nonumber
\end{align}
This identity demonstrates that adding exponents $(a_x,a_y,-a_z)$ to the relation above results in a form identical to the sum rule Eq.~(\ref{eqn : generalized Pascal's sum rule}). Consequently, the tetrahedron operator $O_{\{\vec{r}_0,l\}}^{XI}$ in Eq.~(\ref{eqn : gHC_tetrahedron_bulk}) commutes with all $S_c^Z$ cube terms that are entirely contained within the tetrahedron and touch the side plane at $x=x_0$, $y=y_0$ and $z=z_0$.
However, because the tetrahedron is truncated abruptly along the base plane defined by $(x-x_0)+(y-y_0)+(z-z_0)=l-1$, the tetrahedron operator violates $S_c^Z=1$ for the most cube terms located on the surface where $(x-x_0)+(y-y_0)+(z-z_0)=l-1$. Additionally it violates $S_c^Z=1$ at the vertex $\vec{r}_0$, as shown in [Fig.~\ref{fig : gHC_tetrahedron_surface} (a)].

Let's consider the point $(x,y,z)$ on the plane $(x-x_0)+(y-y_0)+(z-z_0)=l$, the neighboring upper layer of the base plane of the tetrahedron, as depicted in [Fig.~\ref{fig : gHC_tetrahedron_surface} (b)].
We can use the exponent $s_{(x,y,z)}$ to diagnose a violation of $S_c^Z=1$ at the point $(x-1/2,y-1/2,z-1/2)$ for the plane $(x-x_0)+(y-y_0)+(z-z_0)=l$.
More precisely, if $s_{(x,y,z)}=0$, it means that the product of three $X\otimes I$ operators at $(x-1,y,z)$, $(x,y-1,z)$ and $(x,y,z-1)$ also commutes to $S_{(x-1/2,y-1/2,z-1/2)}^Z$, indicating that there is no excitation. On the other hand, if $s_{(x,y,z)}\ne 0$, then the products of the three $X\otimes I$ operators on the surface $(x-x_0)+(y-y_0)+(z-z_0)=l-1$ cannot commute with $S_{(x-1/2,y-1/2,z-1/2)}^Z$, because there is no $X\otimes I$ operator at the point $(x,y,z)$, which is outside the tetrahedron operator $O_{\{\vec{r}_0,l\}}^{XI}$ in Eq.~\eqref{eqn : gHC_tetrahedron_bulk}.

Now let's find the cases of when $s_{(x,y,z)}=0$ on the surface $(x-x_0)+(y-y_0)+(z-z_0)=l$. If $l\ne N^k$, there are some stripe regions on the surface without excitation, by satisfying the conditions $(x-x_0)+(y-y_0)=N^r$, $(y-y_0)+(z-z_0)=N^r$ and $(z-z_0)+(x-x_0)=N^r$ for {$N^r<l$}. Let's check the exponents on the layer $(x-x_0)+(y-y_0)+(z-z_0)=l$ to see if it becomes zero under the conditions above.
When $x+y+z=l\ne N^k$ and $(x-x_0)+(y-y_0)=N^r$, the exponent at $(x,y,z)$ is given by
\begin{align}
    s_{(x,y,z)}=\binom{l}{N^r}\binom{N^r}{x} {a_x}^{x} {a_y}^{N^r-x} (-a_z)^{l-N^r}. \nonumber
\end{align}
According to the property Eq.~(\ref{eqn : gHC_combinatorics}), $s_{(x,y,z)}=0$ for $1\le x\le N^r-1$, along the intersection between two planes, $(x-x_0)+(y-y_0)=N^r$ and $(x-x_0)+(y-y_0)+(z-z_0)=l$. Rotational symmetry ensures the existence other stripes without excitation, such as $(y-y_0)+(z-z_0)=N^r$ and $(z-z_0)+(x-x_0)=N^r$ for $N^r<l$ [Fig.~\ref{fig : gHC_local_excitation} (c)].

When $l=N^k$, the exponent coefficient on the surface $(x-x_0)+(y-y_0)+(z-z_0)=N^k$ is
\begin{align}
    s_{(x,y,z)}=\binom{N^k}{x+y}\binom{x+y}{x} {a_x}^{x} {a_y}^{y} (-a_z)^{N^k-x-y}. \nonumber
\end{align}

It is always 0 to modulo $N$ due to Eq.~(\ref{eqn : gHC_combinatorics}), except for the three cases: (i) $x+y=0$, (ii) $x+y=N^k, x=0$, and (iii) $x+y=N^k, x=N^k$. Combining the constraint $(x-x_0)+(y-y_0)+(z-z_0)=l$, each case corresponds to a single point as: (i) $(x,y,z)=\vec{r}_0+(0,0,N^k)$, (ii) $(x,y,z)=\vec{r}_0+(0,N^k,0)$, and (iii) $(x,y,z)=\vec{r}_0+(N^k,0,0)$. Consequently, a fracton quadrupole is created, with eigenvalues 
\begin{align}
    S_{\vec{r}_0-\frac{1}{2}(1,1,1)}^Z=\omega^{-1}, \nonumber \\
    S_{\vec{r}_0+N^k\hat{x}-\frac{1}{2}(1,1,1)}^Z=\omega^{a_x},  \nonumber \\
    S_{\vec{r}_0+N^k\hat{y}-\frac{1}{2}(1,1,1)}^Z=\omega^{a_y},  \nonumber \\
    S_{\vec{r}_0+N^k\hat{z}-\frac{1}{2}(1,1,1)}^Z=\omega^{-a_z}, \nonumber
\end{align}
as depicted in [Fig.~\ref{fig : gHC_local_excitation} (b)]. For example, let's calculate the eigenvalue of the fracton at $\vec{r}_0+N^k\hat{x}-\frac{1}{2}(1,1,1)$.
\begin{align}
    S_{\vec{r}_0+N^k\hat{x}-\frac{1}{2}(1,1,1)}^Z O_{\{\vec{r}_0,N^k\}}^{XI} =\omega^{{a_x}^{N^k}} O_{\{\vec{r}_0,N^k\}}^{XI} S_{\vec{r}_0+N^k\hat{x}-\frac{1}{2}(1,1,1)}^Z . \nonumber \\
    \because (Z\otimes I)^{a_x} (X\otimes I)^{{a_x}^{N^k-1}}=\omega^{{a_x}^{N^k}} (X\otimes I)^{{a_x}^{N^k-1}} (Z\otimes I)^{a_x}, \nonumber
\end{align}
and applying Euler's theorem which states ${a_x}^{N-1}=1$ mod $N$ for prime $N$ in Section~\ref{subsec : number theory}, we get
\begin{align}
    {a_x}^{N^k}=({a_x}^{(N-1)}{a_x})^{N^{(k-1)}}={a_x}^{N^{(k-1)}}={a_x} \ \text{mod} \ N. \nonumber
\end{align}

Therefore, the fracton at $\vec{r}_0+N^k\hat{x}-\frac{1}{2}(1,1,1)$ has an eigenvalue of $S_c^Z$ of $\omega^{a_x}$, and the eigenvalues of the other fractons composing the quadrupole can be calculated in the same way.

\subsection{Details on the fracton multipole}\label{Appendix : multipole}

\begin{figure*}
\includegraphics[width=2\columnwidth]{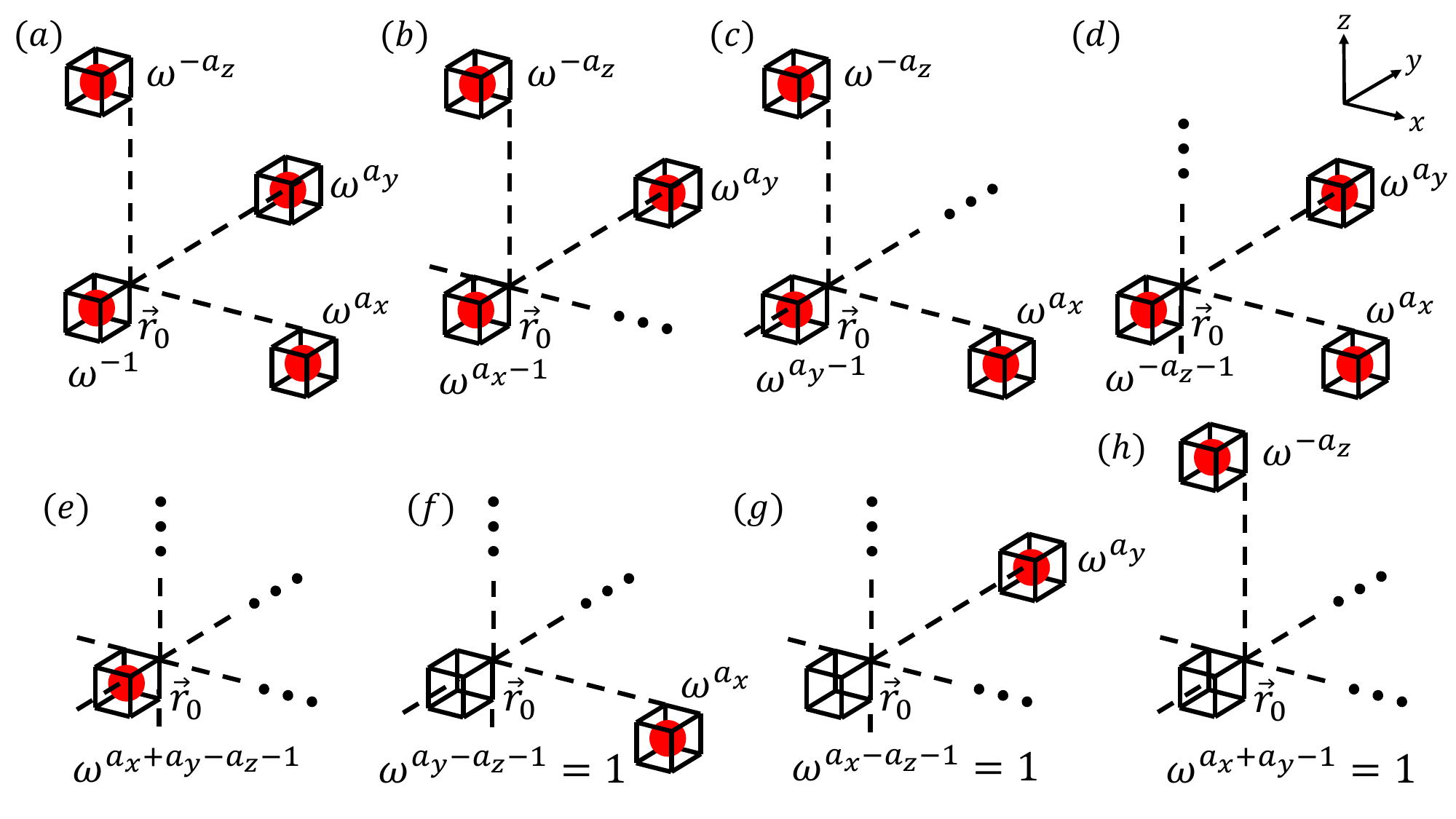}
\caption{\label{fig : App_gHC_multipole} \textbf{Fracton multipole configurations of $\mathbb{Z}_N$ generalized Haah's code.} Eigenvalues of $S_c^Z$ on each fracton are denoted, with $\omega =e^{2\pi i/N}$. (a) A fracton quadrupole can be created by the tetrahedron operator $O$ Eq.~\eqref{eqn : gHC_tetrahedron_bulk} with $l=N^k$, when $L_i \neq N^{k}$ for all $i=x,y,z$. (b) A fracton tripole can be created by the tetrahedron operator $O$ Eq.~\eqref{eqn : gHC_tetrahedron_bulk} with $l=L_x=N^k$, $L_x\ne L_y$, $L_x\ne L_z$ and $a_x\ne 1$ mod $N$. If $a_x= 1$ mod $N$, the same operator creates a dipole as in the $\mathbb{Z}_2$ case. (c) A fracton tripole can also be created by the tetrahedron operator $O$ Eq.~\eqref{eqn : gHC_tetrahedron_bulk} with $l=L_y=N^k$, $L_y\ne L_x$, $L_y\ne L_z$ and $a_y\ne 1$ mod $N$. (d) A fracton tripole can also be created by the tetrahedron operator $O$ Eq.~\eqref{eqn : gHC_tetrahedron_bulk} with $l=L_z=N^k$, $L_z\ne L_x$, $L_z\ne L_y$, and $-a_z\ne 1$ mod $N$. (e) A fracton monopole can be created by the tetrahedron operator $O$ Eq.~\eqref{eqn : gHC_tetrahedron_bulk} with $l=L_x=L_y=L_z=N^k$ and $a_x+a_y-a_z\ne 1$ mod $N$. If $a_x+a_y-a_z = 1$ mod $N$, the same operator creates no excitation. (f) A fracton monopole can also be created by the tetrahedron operator $O$ Eq.~\eqref{eqn : gHC_tetrahedron_bulk} with $l=L_y=L_z=N^k$ and $a_y-a_z=1$ mod $N$. If $a_y-a_z\ne 1$ mod $N$, the same operator creates a dipole as in the $\mathbb{Z}_2$ case. (g) A fracton monopole can also be created by the tetrahedron operator $O$ Eq.~\eqref{eqn : gHC_tetrahedron_bulk} with $l=L_x=L_z=N^k$ and $a_x-a_z=1$ mod $N$. (h) A fracton monopole can also be created by the tetrahedron operator $O$ Eq.~\eqref{eqn : gHC_tetrahedron_bulk} with $l=L_x=L_y=N^k$ and $a_x+a_y=1$ mod $N$.}
\end{figure*}

In this section, we discuss the fracton multipole of $\mathbb{Z}_2$ and $\mathbb{Z}_N$ Haah's code in detail and explicitly provide the conditions for constructing dipoles, tripoles and monopoles beyond the quadrupole.

We will investigate the $\mathbb{Z}_2$ case first. When the tetrahedron operator $O_{\{\vec{r}_0,l\}}^{XI}$ Eq.~(\ref{Z2HaahO}) has the length $l=2^k$, we know that a fracton quadrupole is created as follows :
\begin{align}
    S_{\vec{r}_0-\frac{1}{2}(1,1,1)}^Z=-1, \nonumber\\
    S_{\vec{r}_0+2^k\hat{x}-\frac{1}{2}(1,1,1)}^Z=-1,\nonumber\\
    S_{\vec{r}_0+2^k\hat{y}-\frac{1}{2}(1,1,1)}^Z=-1, \nonumber\\
    S_{\vec{r}_0+2^k\hat{z}-\frac{1}{2}(1,1,1)}^Z=-1. \nonumber
\end{align}

However, when the system size $L_y$ along the $\hat{y}$-direction is equal to $2^k$, then $S_{\vec{r}_0+L_y\hat{y}-\frac{1}{2}(1,1,1)}^Z=-1$ creates a fracton at the same position as $S_{\vec{r}_0-\frac{1}{2}(1,1,1)}^Z=-1$ and the pair of fractons cancels each other. Therefore, only the fracton dipole located at $\vec{r}_0+2^k\hat{x}-\frac{1}{2}(1,1,1)$ and $\vec{r}_0+2^k\hat{z}-\frac{1}{2}(1,1,1)$ remains [Fig.~\ref{fig : Z2_HC_new} (e)]. Due to rotational symmetry, such a fracton dipole can also be created when $l=2^k=L_x$, or $l=2^k=L_z$. Since the fracton in the $\mathbb{Z}_2$ model has eigenvalue $S_c^Z=-1$, the two fracton at the same position always cancel each other. As a result, only quadrupoles and dipoles, which consist of even number of fractons can be created. However, we will see that different configurations, such as fracton tripoles and monopoles are also possible in the $\mathbb{Z}_N$ generalized model. 

As mentioned in the main text Section~\ref{sec : gHC}, the overall behavior of $\mathbb{Z}_N$ model is the same as the $\mathbb{Z}_2$ case, if at least one of the followings is true: (i) $L_i \neq N^{k}$ for all $i=x,y,z$, (ii) $L_y=N^k$ for some integer $k$ and $a_y= 1$ mod $N$, (iii) $L_z=N^k$ for some integer $k$  and $-a_z= 1$ mod $N$, (iv) $L_x=N^k$ for some integer $k$ and $a_x= 1$ mod $N$. 
When the condition (i), $L_i \neq N^{k}$ for all $i=x,y,z$ is true, then one cannot create fracton multipole except quadrupole from the tetrahedron operator $O_{\vec{r}_0,N^k}^{XI}$, since one cannot superpose several fractons belonging to the same quadrupole. Therefore both $\mathbb{Z}_N$ and $\mathbb{Z}_2$ case exhibits the fracton quadrupole and the surface excitation, which are mainly discussed in~\ref{Appendix : gHC_tetrahedron}.
When the condition (ii) is true, i.e. $L_y=N^k$ for some integer $k$ and $a_y= 1$ mod $N$, a fracton multipole can be created from the tetrahedron operator $O_{\vec{r}_0,N^k}^{XI}$ with size $L_y=N^k$. This allows merging of the two fractons, $S_{\vec{r}_0-\frac{1}{2}(1,1,1)}^Z=\omega^{-1}$ and $S_{\vec{r}_0+L_y\hat{y}-\frac{1}{2}(1,1,1)}^Z=\omega^{a_y}$, but they cancel each other out because $S_{\vec{r}_0-\frac{1}{2}(1,1,1)}^Z=\omega^{a_y-1}=1$. Eventually, the remaining fractons form a dipole, as in the $\mathbb{Z}_2$ model. Similary, the $\mathbb{Z}_N$ model under the other conditions (iii) or (iv) also exhibits a $\mathbb{Z}_2$-like fracton dipole. Thus, the overall behavior of the $\mathbb{Z}_N$ model satisfying at least the one of the four conditions is the same as the $\mathbb{Z}_2$ case.

If all four conditions above are violated, $\mathbb{Z}_N$ generalized Haah's code Eq.~\eqref{eqn : H_gHC} exhibits unique excitations, where the tetrahedron operator can create the tripole or monopole of the fracton. Such excitations are impossible in the $\mathbb{Z}_2$ model Eq.~\eqref{eqn : Z2_HC_Hamiltonian} due to its $\mathbb{Z}_2$ structure of eigenvalues $\pm 1$. First, we list the all conditions which exhibit the tripole configurations : (i) $a_x \neq 1$ mod $N$ and $L_x = N^k$ with $L_x\ne L_y,\ L_x\ne L_z$, (ii) $a_y \neq 1$ mod $N$ and $L_y = N^k$ with $L_y\ne L_x,\ L_y\ne L_z$, (iii) $-a_z \neq 1$ mod $N$ and $L_z = N^k$ with $L_z\ne L_x,\ L_z\ne L_y$ [Fig.~\ref{fig : App_gHC_multipole} (b-d)].
We explicitly write the fractons created by the tetrahedron operator $O$ Eq.~(\ref{eqn : gHC_tetrahedron_bulk}) with $l=N^k$ under the condition (i), 
\begin{align}
    S_{\vec{r}_0-\frac{1}{2}(1,1,1)}^Z=\omega^{a_x-1}, \nonumber \\
    S_{\vec{r}_0+N^k\hat{y}-\frac{1}{2}(1,1,1)}^Z=\omega^{a_y},  \nonumber \\
    S_{\vec{r}_0+N^k\hat{z}-\frac{1}{2}(1,1,1)}^Z=\omega^{-a_z}, \nonumber
\end{align}
where the eigenvalue of the fracton at $\vec{r}_0-\frac{1}{2}(1,1,1)$ is changed, as the tetrahedron operator now shares two points with $S_{\vec{r}_0-\frac{1}{2}(1,1,1)}^Z$, specifically $\vec{r}_0$ and $\vec{r}_0+(L_x-1)\hat{x}$. Precisely, the $X\otimes I$ operator at $\vec{r}_0$ contributes $\omega ^{-1}$ to the eigenvalue of $S_c^Z$, while the $X\otimes I$ operator at $\vec{r}_0+(L_x-1)\hat{x}$ contributes $\omega ^{a_x}$. Therefore, $S_c^Z$ at $\vec{r}_0-\frac{1}{2}(1,1,1)$ has an eigenvalue of $\omega ^{a_x-1}$. If $a_x=1$ mod $N$, then $S_{\vec{r}_0-\frac{1}{2}(1,1,1)}=1$, and there is no excitation. In this case, the configuration becomes a dipole, consisting of two fractons at $\vec{r}_0+N^k\hat{y}-\frac{1}{2}(1,1,1)$ and $\vec{r}_0+N^k\hat{z}-\frac{1}{2}(1,1,1)$. The tripole configuration is available only when both $a_x \neq 1$ mod $N$ and $L_x = N^k$ with $L_x\ne L_y,\ L_x\ne L_z$ are simultaneously satisfied. Rotational symmetry ensures that Other conditions also provide the tripole configuration [Fig.~\ref{fig : App_gHC_multipole} (c--d)].

Next, we list the all the conditions that exhibit the monopole configurations: (i) $L_x=L_y=L_z=L = N^{k}$, $a_x+a_y-a_z\ne 1\ \text{mod}\ N$ (ii) $L_x=L_y=N^k$, $a_X+a_y=1$ mod $N$, (iii) $L_y=L_z=N^k$, $a_y-a_z=1$ mod $N$, (iv) $L_z=L_x=N^k$, $a_x-a_z=1$ mod $N$  
[Fig.~\ref{fig : App_gHC_multipole} (e--h)]. Under condition (i), the tetrahedron operator $O_{\{\vec{r}_0,l\}}^{XI}$ Eq.~(\ref{eqn : gHC_tetrahedron_bulk}) with $l=N^k$ creates a single fracton
\begin{align}
    S_{\vec{r}_0-\frac{1}{2}(1,1,1)}^Z=\omega^{a_x+a_y-a_z-1}. \nonumber 
\end{align}
The tetrahedron operator now shares four points with $S_{\vec{r}_0-\frac{1}{2}(1,1,1)}^Z$, which are $\vec{r}_0$, $\vec{r}_0-\hat{x}$, $\vec{r}_0-\hat{y}$ and $\vec{r}_0-\hat{z}$. At each points, the $X\otimes I$ operators contribute to the eigenvalue of $S_c^Z$ as $\omega ^{-1}$, $\omega ^{a_x}$, $\omega ^{a_y}$, and $\omega ^{-a_z}$, respectively. Therefore $S_c^Z$ at $\vec{r}_0-\frac{1}{2}(1,1,1)$ has the eigenvalue $\omega ^{a_x+a_y-a_z-1}$. If $a_x+a_y-a_z=1$ mod $N$, then $S_{\vec{r}_0-\frac{1}{2}(1,1,1)}=1$ and there is no excitation.
Under condition (ii), the tetrahedron operator $O_{\{\vec{r}_0,l\}}^{XI}$ Eq.~(\ref{eqn : gHC_tetrahedron_bulk}) with $l=N^k$ creates a fracton configuration as follows:
\begin{align}
    S_{\vec{r}_0-\frac{1}{2}(1,1,1)}^Z=\omega^{a_x+a_y-1}=1, \nonumber \\
    S_{\vec{r}_0+N^k\hat{z}-\frac{1}{2}(1,1,1)}^Z=\omega^{-a_z}. \nonumber
\end{align}
The eigenvalue of the fracton at $\vec{r}_0-\frac{1}{2}(1,1,1)$ is 1, so there is no excitation at this point. Consequently, the only remaining fracton is $S_{\vec{r}_0+N^k\hat{z}-\frac{1}{2}(1,1,1)}^Z=\omega^{-a_z}$ which represents a monopole configuration. The tetrahedron operator shares three points with $S_{\vec{r}_0-\frac{1}{2}(1,1,1)}^Z$, $\vec{r}_0$, $\vec{r}_0-\hat{x}$ and $\vec{r}_0-\hat{y}$. The $X\otimes I$ operator at each point contributes to the eigenvalue of $S_{\vec{r}_0-\frac{1}{2}(1,1,1)}^Z$ as $\omega ^{-1}$, $\omega ^{a_x}$ and $\omega ^{a_y}$, respectively. Therefore the eigenvalue of $S_c^Z$ at $\vec{r}_0-\frac{1}{2}(1,1,1)$ is $\omega ^{a_x+a_y-1}$. If $a_x+a_y\ne 1$ mod $N$, then $S_{\vec{r}_0-\frac{1}{2}(1,1,1)}\ne 1$, and a fracton is created, resulting in a dipole configuration. This is similar to the behavior observed in the $\mathbb{Z}_2$ model. Rotational symmetry ensures that conditions (iii) and (iv) also yield the monopole configuration [Fig.~\ref{fig : App_gHC_multipole} (g--h)].

\subsection{Minimal model of \texorpdfstring{$\mathbb{Z}_N$}{} generalized Haah's code}
\begin{figure}
\includegraphics[width=\columnwidth]{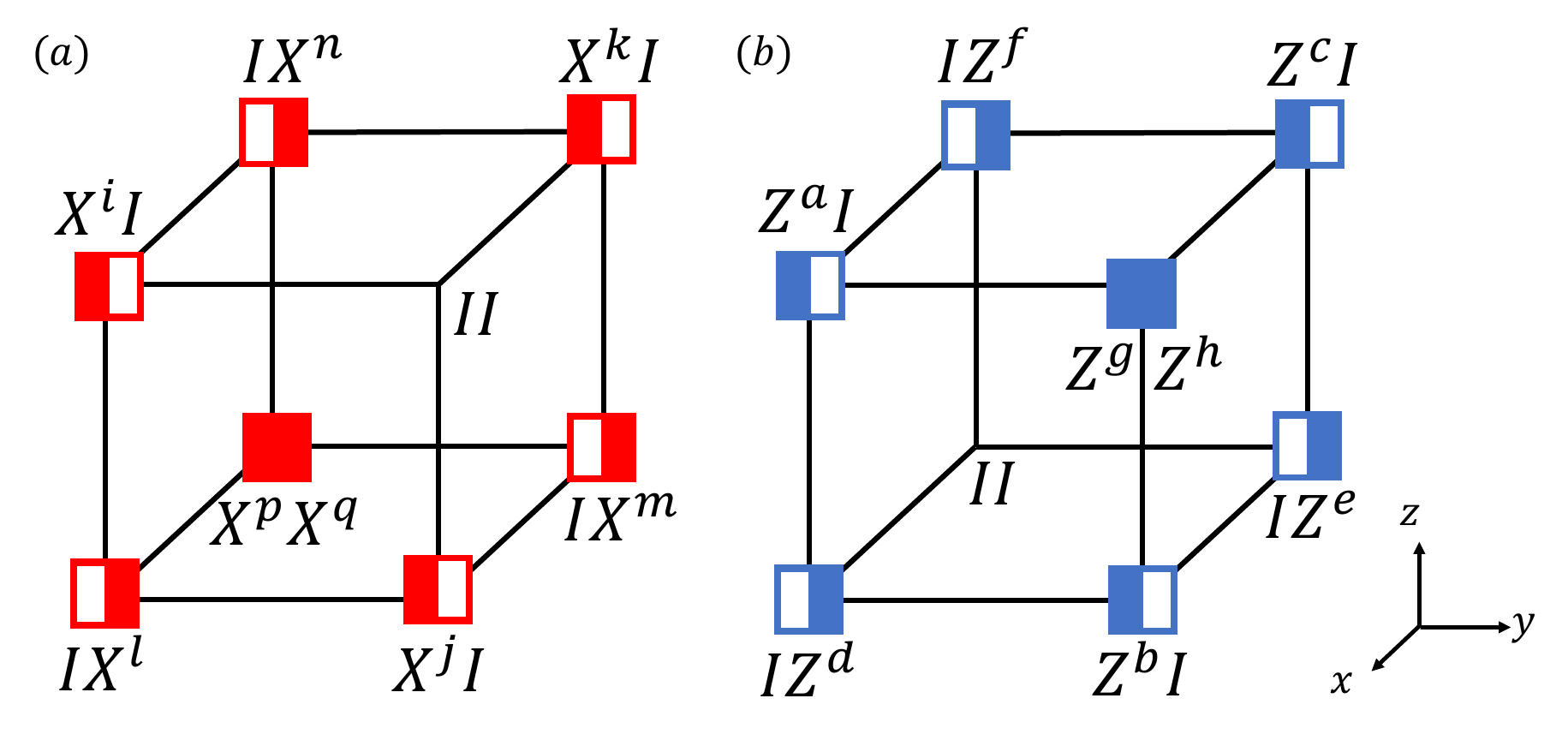}
\caption{\label{fig : gHC_most_generalized} \textbf{The most generalized forms of $\mathbb{Z}_N$ generalized Haah's code with full redundant exponents }}
\end{figure}

\begin{figure}
\includegraphics[width=\columnwidth]{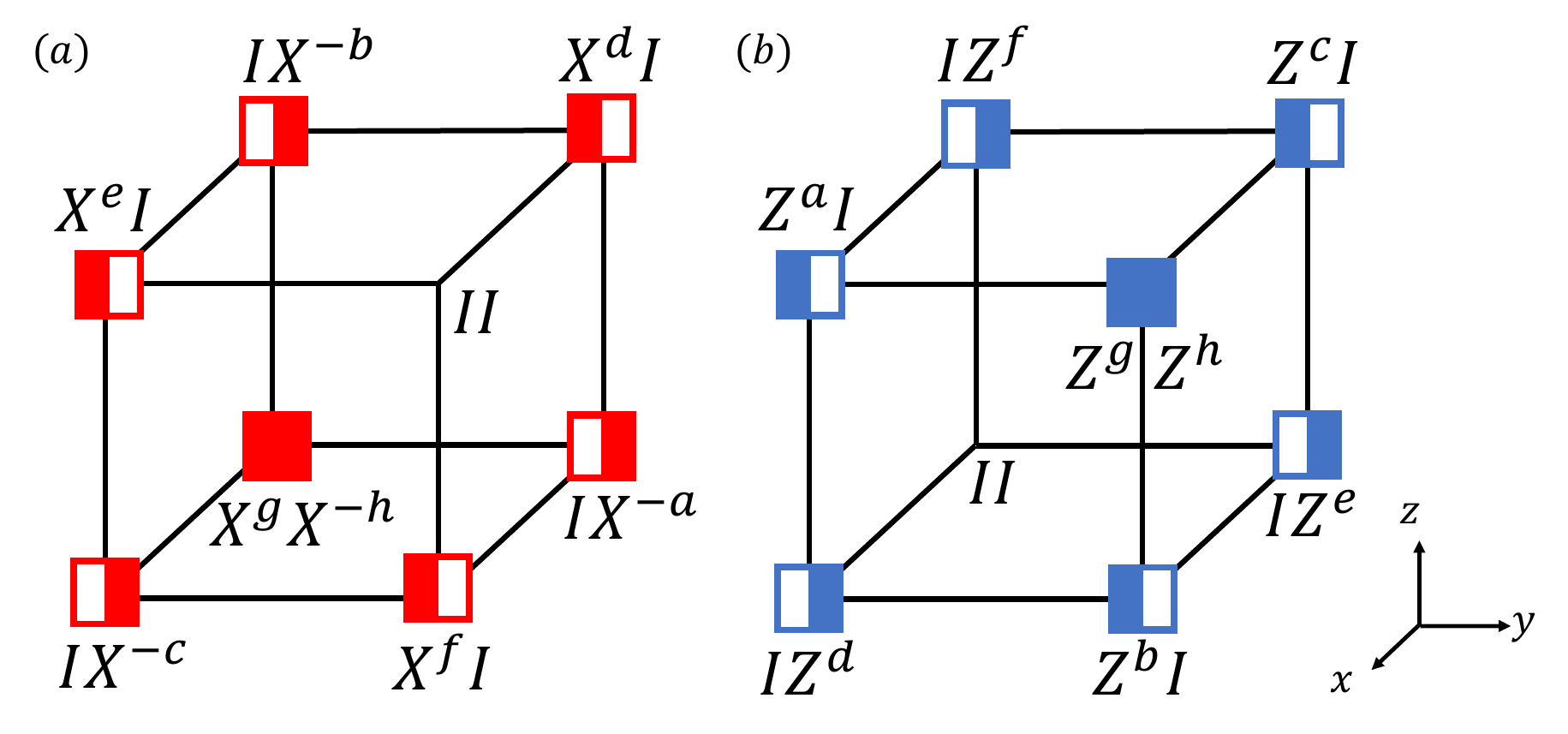}
\caption{\label{fig : gHC_first_reduced} \textbf{The first reduced stabilizer terms of gHC}}
\end{figure}

In this section, we demonstrate that the model in the main text Eq.~\eqref{eqn : H_gHC} is the minimal model capturing the essential physics, starting from the most generalized form [Fig.~\ref{fig : gHC_most_generalized} (a--b)] and then reducing the redundant exponents as much as possible. Due to the commuting relations between $S_c^X$ and $S_c^Z$, several parameter relationships emerge: $hp+gq=0$ from the sharing a vertex, $ap+gm=0$, $lf+cj=0$, $bp+gn=0$, $ak+dm=0$, $bi+en=0$, $cp+gl=0$, $aj+fm=0$, $el+ci=0$, $dn+bk=0$ from sharing a single edge, $hi+eq=0$, $hj+fq=0$, $hk+dq=0$ from sharing a face, and $ai+fn+ck+dl+bj+em=0$ from sharing a whole cube. By solving the equations simultaneously, we obtain the ratio $x$, which is defined as: 
\begin{equation}
    x\equiv \frac{p}{g}=\frac{i}{e}=\frac{j}{f}=\frac{k}{d}=-\frac{q}{h}=-\frac{m}{a}=-\frac{n}{b}=-\frac{l}{c}.
\end{equation}
After setting the ratio $x=1$, the first reduced version of the model emerges [Fig.~\ref{fig : gHC_first_reduced} (a--b)]. Setting $x$ as other integers corresponds to replacing $S_c^X$ to $(S_c^X)^x$, so setting $x=1$ does not lose any generality.

In the first reduced cube terms, generalized exponent sum rules, as given by Eq.~(\ref{eqn : generalized Pascal's sum rule}) for each $X^{s_{\vec{r}}} \otimes I$ and $I \otimes X^{s_{\vec{r}}}$ operators become:
\begin{widetext}
\begin{equation}
    as_{\vec{r}+(1,0,1)}+bs_{\vec{r}+(0,1,1)}+cs_{\vec{r}+(1,1,0)}+hs_{\vec{r}+(1,1,1)}=0 \;\text{mod}\; N \nonumber
\end{equation}

\begin{equation}
    ds_{\vec{r}+(1,0,0)}+es_{\vec{r}+(0,1,0)}+fs_{\vec{r}+(0,0,1)}+gs_{\vec{r}+(1,1,1)}=0 \;\text{mod}\; N \nonumber
\end{equation}
\end{widetext}

Since we are interested in the prime $N$, an arbitrary integer is always coprime to $N$, and thus it always has a modular multiplicative inverse $i\times i^{-1}=1 \ \text{mod}\ N$ due to the B\'ezout's identity~\ref{Bezout_identity}. Therefore, if we multiply $g^{-1}$ and $h^{-1}$ on each of the exponent sum rules, the remaining 6 variables(which are renamed as $(a_x,a_y,a_z,a_{xy}',a_{yz}',a_{zx}')$ in the main text) are sufficient to capture the nontrivial physics of the $\mathbb{Z}_N$ generalized Haah's code.

\subsection{\texorpdfstring{$\mathbb{Z}_2$}{Z\_2} Haah's code with zero parameters}\label{appendix : Z2_HC_zeros}
In this section, we examine in detail the case where the parameters of the $\mathbb{Z}_2$ Haah's code include zeros. For brevity, we consider only the $\{a_x,a_y,a_z\}$ parameter set. The $\{a_{xy}',a_{yz}',a_{zx}'\}$ set can be treated in the same way as in Section~\ref{sec : gHC_local_cube_excitation}. For $N=2$, the parameters can take only the values 0 and 1. Additionally, we compare the $\mathbb{Z}_2$ Haah's code with zero parameters to other previously known models with similar structures.

First, when only one parameter is 0, fracton multipoles and fracton lines can be generated using local triangle operators, as discussed in Section~\ref{sec : multoipoles_quasi_fracton}. However, since the number of $Z\otimes I$ and $I\otimes X$ terms at the eight corners of the Z and X cube terms are three—an odd number—there is no longer an identity relation when multiplying all cube terms together.

When two parameters are 0, there always exists a stabilizer relation that makes the product of all cube terms equal to the identity, just like in the original model. However, for the same reason as in Section~\ref{sec : gHC_local_cube_excitation}, the cube excitations now become lineons. This case corresponds to the Sierpinski fractal spin liquid (SFSL) model introduced in Ref.~\cite{PhysRevB.88.125122,PhysRevB.100.155137}. When all three parameters are 0, the cube excitations behave as free particles for the same reason discussed in Section~\ref{sec : gHC_local_cube_excitation}.

Consequently, the $\mathbb{Z}_2$ Haah's code with zero parameters not previously considered a viable quantum memory, as it fails to simultaneously satisfy the two key conditions as 18 types of cubic codes proposed in Ref.~\cite{PhysRevA.83.042330}: the presence of immobile excitations and a ground state degeneracy greater than 1 for all system sizes. However, in this work, we numerically confirm that even when a single parameter is set to 0 in the $\mathbb{Z}_2$ Haah's code, the GSD oscillates between 1 and values greater than 1 [Fig.~\ref{fig : gHC_GSD_N2}]. This finding establishes that such cases also belong to the fracton phase. 

In conclusion, we find that when the $\mathbb{Z}_2$ model has zero parameters, it realizes a fracton order distinct from previously known $\mathbb{Z}_2$ models, such as SFSL and other Haah cubic codes, provided that at most one parameter in each of the two parameter sets, $\{a_x,a_y,a_z\}$ and $\{a_{xy}',a_{yz}',a_{zx}'\}$, is set to zero.

\bibliography{main}
\end{document}